\documentclass{ametsoc}

\usepackage{xspace} % Incorporates correct optional space after commands

\newcommand{\be}{\begin{equation}}
\newcommand{\ee}{\end{equation}}
\newcommand{\xxi}{\vec{\xi}}

\newcommand{\ave}[1]{\left\langle {#1} \right\rangle}

\renewcommand{\vec}[1]{{\mathchoice
                     {\mbox{\boldmath$\displaystyle{#1}$}}
                     {\mbox{\boldmath$\textstyle{#1}$}}
                     {\mbox{\boldmath$\scriptstyle{#1}$}}
                     {\mbox{\boldmath$\scriptscriptstyle{#1}$}}}}

        \newcommand{\argmin}{\arg\!\min}
        
        \newcommand{\MIT}{{\sc MITgcm}\xspace}

\journal{mwr}

\bibpunct{(}{)}{;}{a}{}{,}

\title{Polynomial Chaos-based Bayesian Inference of K-Profile Parametrization in a General Circulation Model of the Tropical Pacific}

    \authors{Ihab Sraj\correspondingauthor{Ihab Sraj, 
     Division of Physical Sciences and Engineering, 
     King Abdullah University for Science and Technology, Thuwal, Saudi Arabia.}}

     \affiliation{
     King Abdullah University for Science and Technology, Thuwal, Saudi Arabia}

\email{ihab.sraj@kaust.edu.sa}
\extraauthor{Sarah E. Zedler }
    \extraaffil{
    University of Texas at Austin, %201 E 24th ST. Stop C0200, 
    Austin, TX, USA 
    }
    \extraauthor{Omar M. Knio}
    \extraaffil{
    Duke University, 
    Durham, NC, USA;
    King Abdullah University for Science and Technology, Thuwal, Saudi Arabia% 27708, USA
     }

    \extraauthor{Charles S. Jackson }
    \extraaffil{
    University of Texas at Austin, %201 E 24th ST. Stop C0200, 
    Austin, TX, USA 
    }
     
     \extraauthor{Ibrahim Hoteit}
    \extraaffil{
    King Abdullah University of Science and Technology, Thuwal, Saudi Arabia}

\abstract{The authors present a Polynomial Chaos (PC)-based Bayesian inference method for quantifying the uncertainties 
of the K-Profile Parametrization (KPP) within the MIT General Circulation Model (\MIT) of the tropical pacific. The 
inference of the uncertain parameters is based on a Markov Chain Monte Carlo (MCMC) scheme that utilizes a 
newly formulated test statistic taking into account the different components representing the structures of 
turbulent mixing on both daily and seasonal timescales in addition to the data quality, and filters for the 
effects of parameter perturbations over those due to changes in the wind. 
To avoid the prohibitive computational cost of integrating the \MIT model at each MCMC iteration, we build a 
surrogate model for the test statistic using the PC method. 
To filter out the noise in the model predictions and avoid related convergence issues, we resort to a 
Basis-Pursuit-DeNoising (BPDN) compressed sensing approach to determine the PC coefficients of a representative surrogate model. The PC surrogate is then used to evaluate the test statistic in the MCMC step for sampling the posterior of the uncertain parameters. Results of the posteriors
indicate good agreement with the default values 
for two parameters of the KPP model namely the critical bulk and gradient Richardson numbers;  
while the posteriors of the remaining parameters were barely informative.} 

\begin{document}

%% Necessary!
\maketitle

%%%%%%%%%%%%%%%%%%%%%%%%%%%%%%%%%%%%%%%%%%%%%%%%%%%%%%%%%%%%%%%%%%%%%
% MAIN BODY OF PAPER
%%%%%%%%%%%%%%%%%%%%%%%%%%%%%%%%%%%%%%%%%%%%%%%%%%%%%%%%%%%%%%%%%%%%%
%
\section{Introduction}

\label{sec:intro}

The present work seeks to calibrate the model parameters of the K-Profile Parameterization (KPP) model~\citep{Large:94,Large:97}
as implemented in the ocean model  MIT General Circulation Model (\MIT)~\citep{Ferreira200686,mar97b} of the tropical pacific.
The KPP model relies on a number of parameters whose default values are set 
based on a combination of theory, laboratory experiments, and atmospheric/oceanic boundary layer 
observations~\citep{Large:94,Large:97}. Our goal here is to quantify the uncertainties in these parameters 
where the ocean resolves the chaotic behavior of fluid dynamic models.
The chaos in the response of the deterministic \MIT model to the perturbation of these parameters leads to internal noise
that in turn results in low signal to noise challenges as will be discussed in more detail below.

An inverse modelling approach is adopted for the objective stated above
in which a set of temperature, salinity and horizontal current measurements
are used to estimate the KPP parameters. Specifically, we employ a Bayesian approach to 
inverse problems that provides complete posterior statistics
and not just a single value for the quantity of interest~\citep{Tarantola:2005}. Traditionally, local model-data 
misfit of short-term turbulent mixing events are used to construct a cost function
and then Bayesian inference is employed for the estimation of the uncertain parameters~\citep{sivia}. 
Here, instead, we use a test statistic for KPP parameters' estimation 
that was introduced in~\citet{wag14} and in~\citet{Zedler2015}. This statistic seeks to formulate 
a total cost function of different components representing 
the structures of turbulent mixing on both daily and seasonal timescales, takes data quality into account, 
and filters for the effects of parameter perturbations over those due to changes in the wind.
At both timescales, the model and data are filtered before taking differences to capture the time integrated response, rather than  
individual mixing events.

The end result of the Bayesian inference formulation is a multi-dimensional posterior that
can be directly sampled via Markov Chain Monte Carlo (MCMC). This, however, requires a prohibitive 
number of simulations of the forward model, one for every proposed set 
of parameters of the Markov chain~\citep{Malinverno2002}. This practice
renders Bayesian methods computationally prohibitive for large-scale models such as the \MIT where one model evaluation takes $22$ hours in computing time
using $256$ processors. To overcome this issue, we construct a surrogate model 
that approximates the forward model and can be used in the sampling MCMC.
More precisely, we use the Polynomial Chaos (PC) method to construct the surrogate model from an ensemble of 
\MIT model runs~\citep{MarzoukNajmRahn:2007,MarzoukNajm2009}. This approach further offers additional advantages such as computing model output sensitivities and additional statistics~\citep{LeMaitreKnio2010}.

The PC method has been extensively investigated in the literature, and its suitability  
for large-scale models has been recently demonstrated in various settings. 
\citet{Alexanderian2012} implemented a sparse spectral projection PC approach to propagate parametric uncertainties of three KPP parameters 
in addition to the wind drag coefficient during a hurricane event. The study demonstrated
the possibility of building a representative surrogate model for a realistic ocean model; however the inverse problem was not tackled. 
\citet{winokur:2012} followed up on \citet{Alexanderian2012} work and implemented an adaptive strategy to design sparse ensembles of oceanic simulations for the purpose of constructing a PC surrogate with even less computational effort. 
\citet{sraj:2013a} extended the previous work and combined a spectral projection PC approach with Bayesian inference to estimate parametrized wind drag coefficient using temperature data collected during a typhoon event. The same problem was also solved using gradient based search method~\citep{sraj:2013b}.
\citet{MatternFennelDowd2012} have similarly exploited the virtues of such polynomial expansions for
examining the response of ecosystem models to finite perturbations of their uncertain parameters. Tsunami~\citep{sraj:2014,Cheung2011}, climate~\citep{OlsonEtAl2012} and subsurface flow modeling~\citep{Elsheikh2014515}
were also investigated using similar PC approaches. 

What is common in the aforementioned PC applications is that the processes studied occurred over
short timescales of few days, so that internal noise in the model was small. This enabled
a successful PC expansion construction using traditional spectral projection~\citep{sraj:2013a,Reagan:2003,Alexanderian2012}.
In this study, however, the major hurdle of constructing a PC surrogate model was the internal noise present in the \MIT model due to the perturbation of the chosen KPP parameters, which was amplified over time by non-linear interactions. As explained in Section~\ref{sec:uqpce}\ref{sec:pccs}(\ref{sec:nisp}), a spectral projection technique failed to construct a PC expansion that faithfully represent the model. 
Instead, we resort to a compressed sensing technique
namely the Basis-Pursuit-DeNoising (BPDN) method~\citep{Doostan:2014} to determine the PC expansion coefficients.
This technique first seeks to estimate the noise in the model output, filter it out and then solve an optimization problem 
assuming sparsity in the PC expansion to determine the non-zero PC coefficients efficiently.
The BPDN method offers an additional advantage that a smaller number of model runs compared to the spectral projection 
method is required to determine the PC coefficients as shown in Section~\ref{sec:uqpce}\ref{sec:size}.
BPDN was recently employed to build a proxy model for an integral oil-gas plume model~\citep{wang2015}
and for an ocean model with initial and wind forcing uncertainties~\citep{li2015}.
In the former, the model output was noisy due to the iterative solver used in the double-plume calculation~\citep{Socolofsky2008}
and BPDN proved to be successful in filtering the noise and thus building a representative PC model.
In the latter, no noise was present in the model output, however, the model was unable to produce realistic simulations
for pre-specified sets of parameters as required by the spectral projection method. BPDN was therefore used as an alternative
approach as it does not have this requirement, a random sample of model parameters can be modeled instead~\citep{Doostan2011} .
To our best knowledge, BDPN has not previously been applied to a noisy, large scale ocean model.

The structure of the paper is as follows: Section~\ref{sec:model} describes the \MIT, presents the choice of the uncertain parameters 
and describes the observation and cost function used in the estimation process.
Section~\ref{sec:inference} introduces the Bayesian inference and its application to our specific problem.
Section~\ref{sec:uqpce} discusses the PC method and presents several error studies to show the convergence of the constructed PC expansion.
Section~\ref{sec:mcmc} presents the results of the inference of KPP parameters and Section~\ref{sec:conc} 
summarizes our findings.

\section{Model, Uncertain parameters and Observations}
\label{sec:model}

\subsection{{MITcgm} Model}
%%%%%%%%%%%%%%%%%%%%%%%%%%%%%%%%%%%%%%%%%%%%%%%%%%%%%%%%%%%%%%%%%%%%%%%%%%%%%%%%%%%%%%%%%%%%%%%%%%%%%%%%%%%%^M
%%%%%%%%%%%%%%%%%%%%%%%%%%%%%%%%%%%%%%%%%%%%%%%%%%%%%%%%%%%%%%%%%%%%%%%%%%%%%%%%%%%%%%%%%%%%%%%%%%%%%%%%%%%%^M

\begin{figure}
\begin{center}
\includegraphics[width=0.9\textwidth]{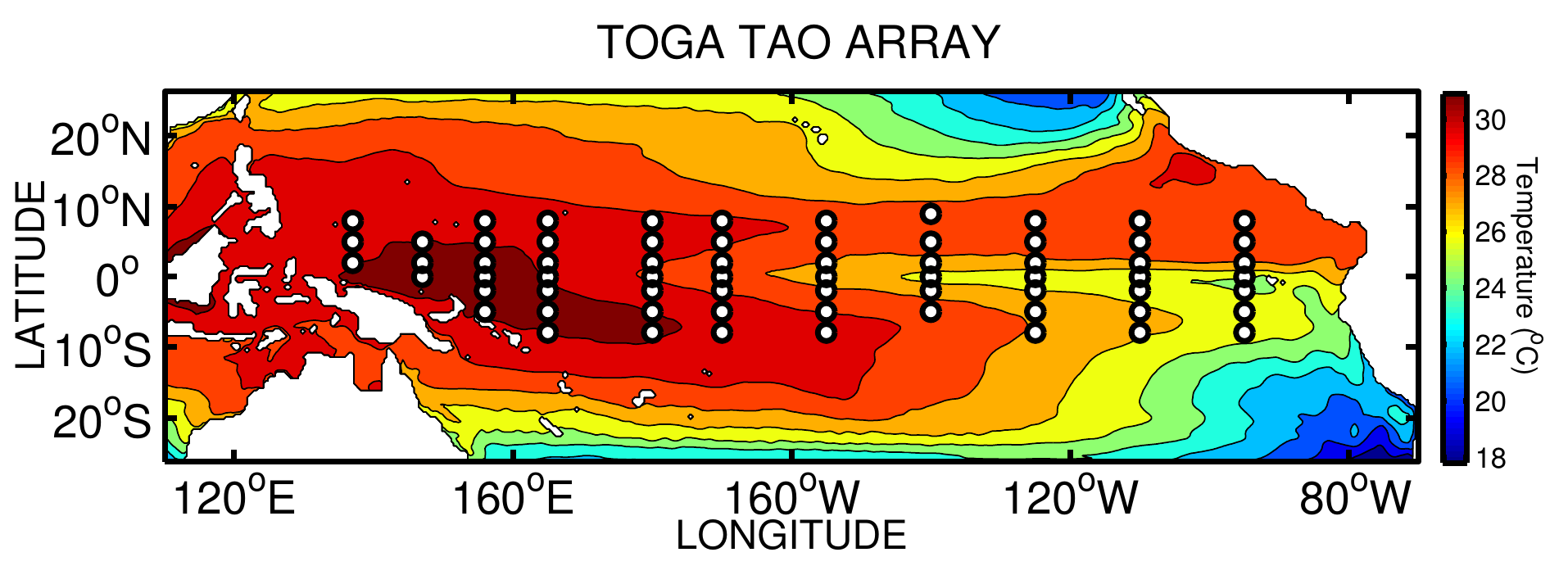}
\caption{Model domain showing 2004-2007 averaged sea surface temperature using default \MIT KPP parameters
and the TOGA/TAO mooring array.}
\label{fig:obs}
\end{center}
\end{figure}

The {\MIT} employed in this work is based on the primitive Navier Stokes equations
implemented in spherical coordinates with an implicit non-linear 
free surface. The {\MIT} implements the K-Profile Parameterization (KPP) turbulent mixing 
scheme~\citep{adc95,mar97b,mar97a,lar94} (see below) in a regional configuration
based on that of \cite{hot08} and \cite{hot10} for the simulation
of oceanic flow. In particular, the domain chosen covers the region with latitudes 
from 26$^\circ$S to 30$^\circ$N and longitudes from 104$^\circ$E to 70$^\circ$W (Figure~\ref{fig:obs}). The time period of our model simulation is 2004-2007.
The initial and lateral boundary conditions are provided by the OCean Comprehensible Atlas (OCCA) reanalysis that was
developed for the 2004-2007 time period using data assimilation in the {\MIT} of available temperature and salinity
ocean data sets~\citep{for10}.
The lateral boundary conditions for our model are implemented with a 
sponge layer (with a thickness of 9 grid cells, and inner and outer boundary relaxation timescales 
of 20 and 1 days, respectively). For the lateral boundary conditions, the 
OCCA data assimilation product is interpolated at the model resolution of $1/3^\circ$ and
with a time step of one day. Therefore, at the initial timestep, the boundary temperature and salinity 
conditions are approximately in equilibrium with the interior fields.   
Once our higher resolution simulation starts, the velocity field quickly adjusts to the pressure gradient forces and 
establishes a realistic Equatorial circulation.  We note that our model runs on $256$ processors
and takes about $22$ hours for a single simulation.  
As described below in Section~\ref{sec:uqpce}, we needed to run the model 903 times,
which required a total of about 5.5 million compute hours. 

\subsection{KPP model}
In the ocean, turbulent mixing can ensue when there is net heat released to the atmosphere at the sea surface (i.e.\  at night), producing gravitationally unstable density inversions (convective mixing) and when there is sufficient vorticity-producing (in the x-z plane) vertical shear in the horizontal currents to overturn a nominally stratified water column (shear-driven or Kelvin-Helmholz instability induced mixing).  In general terms, the intensity of convective and shear-driven mixing depend on local water column properties and surface forcing conditions.  Theoretically, the most vigorous turbulent mixing should occur when a weakly unstably stratified, strongly sheared flow is forced with strong winds and convection (i.e.\  at night).  By contrast, the flow is most likely to be laminar when a strongly stably stratified, weakly sheared flow is forced with weak winds and large net heat going into the ocean (i.e.\  during the day). The KPP embodies these basic relationships, by making the intensity of mixing a function of locally diagnosed properties of the water column that relate to the amenability to turbulent mixing (such as the bulk and gradient Richardson numbers) as well as the non-local surface wind stress and net heat flux forcing (through forcing parameters such as the friction velocity of wind and the Monin-Obukhov length).   In the KPP, mixing is more intense under unstable convective surface forcing conditions. Readers interested in the details of the KPP are referred to~\cite{lar94} and \cite{lar99}.

The KPP generates depth profiles of two quantities that are relevant for turbulent mixing, the eddy diffusivity/viscosity and a non-local term. The eddy diffusivity and viscosity at depth z can be thought of as a scale of the intensity of the turbulent mixing there, with larger values indicating more vigorous turbulence.  The role of the non-local term is to enhance turbulent fluxes of temperature and salinity (but not the horizontal velocity components) under convective forcing conditions.

There are nine parameters in the KPP that pertain to convective or shear-driven mixing.  
A list of the five uncertain parameters in the KPP model under investigation 
in this work is presented in Table~\ref{tab:param} as well as minimum and maximum values for the uniform prior assumed for each. The default 
values in \MIT are also indicated in the table. The critical bulk and gradient Richardson ($Ri_c$ and $Ri_g$, 
respectively) numbers relate directly to local water column shear/stratification properties, with larger 
values generally making turbulent mixing more intense (for the same water-column).  Convective 
mixing parameters $\phi_{s,unst}$ and $\phi_{m,unst}$ depend directly on a combination of surface forcing and 
local water column shear/stratification considerations and are zero under stable forcing conditions (during 
the day).  Increasing their value generally makes convection more vigorous (given the same water column 
properties and surface forcing).  The non-local convective mixing parameter $C^*$ is proportional to the 
non-local convective term, so increasing it will make the turbulent fluxes for temperature and salinity larger.

\subsection{Observations and test statistic}
\label{sec:cost}
The observational data for our experiment come from the TOGA-TAO mooring array  
for the November 2003--November 2007 time period. The array consists of 77 moorings, shown in Figure~\ref{fig:obs}, centered on the Equator that span the width
of the tropical Pacific in the east-west direction, in the latitude range from 8$^\circ$S to 8$^\circ$N~\citep{mcp98}.    
The data used included measurements of temperature, salinity and horizontal current components.

The test statistic used in this paper has components that operate on daily and seasonal timescales. At daily
timescales, it measures the model's ability to reproduce the observed relationship between wind forcing and
subsequent lowering of the sea surface temperature that results from shear-driven mixing. Prior to calculating
the correlation between those quantities, the model and data are filtered to remove the diurnal cycle. At
seasonal timescales, the test statistic measures the ability of the model to reproduce patterns of sensitivity in
the ocean state that would result from perturbing KPP parameters (as diagnosed from an ensemble of single
KPP parameter perturbation experiments). The patterns of sensitivity are determined separately for
temperature, salinity, east velocity, and north velocity as extracted on the TOGA/TAO sensor array.
For more details of what and how data was used for the calculation of the test statistic, the reader is referred to ~\cite{wag14} and~\cite{Zedler2015}.

\section{Bayesian Inference}
 \label{sec:inference}
 
Bayesian inference is a statistical approach to inverse problems~\citep{sivia}
that has recently gained great interest in different applications, including
ocean~\citep{Alexanderian2011a,Zedler2012,sraj:2013a}, tsunami \citep{sraj:2014}
climate~\citep{OlsonEtAl2012} and geophysical~\citep{Malinverno2002} modeling.
We briefly review this approach below.
\subsection{Formalism}
Let $\vec{d}=(d_1,...,d_n)^T$ be a vector of observation data and $\vec{p}=(p_1,...,p_m)^T$ 
be a vector of model parameters. We consider a forward model $\vec G$ that predicts 
the data as function of the parameters such that:

\begin{equation}
\vec d \approx \vec{G}( \vec{p}).
\end{equation}
Let $\pi(\vec{p})$ be the prior probability distribution of $\vec{p}$ (representing any \emph{a priori} information
on $\vec{p}$), $L(\vec d| \vec{p})$ the likelihood function (the probability of obtaining $\vec{d}$ given $\vec{p}$),
and $\pi(\vec{p}| \vec d)$ the posterior probability distribution of $\vec{p}$ (the probability of occurrence of $\vec{p}$ given $\vec d $).
In this case, the Bayes' rule governs this formulation:

\begin{equation}
 \pi(\vec{p}| \vec d) \propto 
 L(\vec d | \vec{p}) \  \pi(\vec{p}).
\label{eq:post1}
\end{equation}

The expression of the likelihood function depends on the assumptions made on
the errors $\vec \epsilon$ (discrepancies) between the model and observations 
($\vec \epsilon = \vec d - \vec{G(\vec{p})}$).
It is often assumed that the errors $\epsilon_i$ are independent
and normally distributed with a covariance $\vec \Sigma$. % (of dimensions $n \times n$).
Traditionally, a metric $E(\vec{p})$, 
called the cost function, is constructed from the sum over squared errors 
normalized by estimates of the variances: %In matrix form the metric is expressed as follows:
\footnote{We note that in this work a new test statistic is adopted to construct metric $E$
as described in Section~\ref{sec:model}\ref{sec:cost} and in~\cite{Zedler2015}.}
\begin{equation} 
E(\vec{\vec{p}}) 
= 
 \frac{1}{2}\vec \epsilon^T \vec \Sigma^{-1} \vec \epsilon.
\label{eq:cost}
\end{equation}
In this case the likelihood function can be written as: %, i.e.\ 

\begin{equation} 
L(\vec d |  \vec{p}) 
= 
\frac{1}{(2\pi)^{\frac{ke}{2}} |\Sigma| ^{\frac{1}{2}} 	 }\exp \left(  - E(\vec{p}) 	\right)
\label{eq:likelihood1}
\end{equation}
and the joint posterior in Equation~\ref{eq:post1} is then expressed as:

\begin{equation} 
\pi(\vec{p} | \vec d)
\propto  \frac{1}{(2\pi)^{\frac{ke}{2}} |\Sigma| ^{\frac{1}{2}} } \exp  \left( - E(\vec{p}) \right)  \prod_{i=1}^m \pi (p_i).
\label{eq:post2}
\end{equation}
The prior of the parameters $p_i$ is assumed non-informative, i.e.\ uniform distribution such that
$\pi(p_i) = \frac{1}{b_i-a_i}$ where $a_i$ and $b_i$ are the bounds of the prior indicated in Table~\ref{tab:param}
and $ke$ is the effective degree of freedom.

To account for missing information
about off-diagonal coefficients of the covariance matrix $\vec \Sigma$, 
the test statistic $E$ can be re-scaled by a parameter $S$.
Incorporating a scaling parameter $S$ is common in statistical inference 
as a way of scaling model fit to data given the level of
agreement of the model with the data~\citep{jac2015,jac04}.
The scaling parameter $S$ is considered as an additional parameter in the Bayesian problem.
and added to the test statistic $E$ in the above equations 
where the joint posterior becomes as follows:

\begin{equation} 
\pi(\vec{p},S | \vec d)
\propto  \frac{S^{\frac{k_e}{2}}}{(2\pi)^{\frac{ke}{2}} |\Sigma| ^{\frac{1}{2}}} \exp  \left( - SE(\vec{p}) \right) \pi(S) \prod_{i=1}^m \pi (p_i).
\label{eq:post3}
\end{equation}
The scaling parameter is treated as a hyper-parameter and its prior $\pi(S)$ is taken
as a Gamma function~\citep{Zabaras2005,GelmanEtAl:2004,gelman2006} that depends on two
constants $\alpha$ and $\beta$ as follows:

\begin{equation}
	\pi(S) = \begin{cases}\displaystyle \frac{\beta^\alpha}{\Gamma(\alpha)}S^{\alpha-1} \exp \left(  -\beta S\right), & S >0\cr
	0, & \mbox{otherwise}. \end{cases}
\end{equation}
In our case we choose $\alpha=18.18$ and $\beta=72.02$.
The prior of $S$ thus has normal-like distribution
with mean value $\frac{\alpha}{\beta}=0.252$ and variance $\frac{\alpha}{\beta^2}=0.003$.
We determined the effective degrees of freedom for our integrated test statistic to be $ke=17$.

\subsection{Sampling method}
Inferring the KPP parameters amounts to sampling the posterior in Equation~\ref{eq:post3}. 
In general, when the space of the unknown parameters is multidimensional, a suitable computational strategy is 
the Markov Chain Monte  Carlo (MCMC) method, yet MCMC requires a high number of sampling iterations.
In our case, sampling the posterior however requires sampling the \MIT model, which is computationally prohibitive.
Thus we seek to build a surrogate model for the Quantity of Interest (QoI) as described in the following section,
and use the random walk Metropolis MCMC algorithm~\citep{Gareth2009,Haario2001} to accurately and
efficiently sample the posterior. Since the scaling parameter $S$ is included in our test statistic as a scalar correction 
to the data covariance matrix, it is also included as a hyper-parameter to be estimated in addition to the model parameters
$\vec p$. We assume the priors for $p$ and $S$ are independent. This implies that for each MCMC step we can use Gibbs sampling to
iteratively generate a value of $\vec p$
conditional on $S$ and a value of $S$ conditional on $\vec p$ as follows: 
\begin{enumerate}
\item We simulate $\vec p$ conditional on $S$, 
apply sampling algorithm for $\vec p$ but for just one iteration. 
\item We simulate $S$ conditional on $\vec p$; for the informative gamma distribution, we have:

\begin{equation} 
\pi(S | \vec{p}, \vec d)
\propto  S^{\frac{k_e}{2}+\alpha-1} \exp  \left( - S\left[E(\vec{p}) + \beta \right] \right),
\label{eq:postS}
\end{equation}
which results in a gamma distribution of parameters $\frac{k_e}{2}+\alpha$
and $E(\vec p) + \beta $.
\end{enumerate} 

The two steps are repeated until convergence.
%\frac{k_e}{2}+
%where $k_e$ is the effective degrees of freedom determined
%earlier.

%!TEX root = paper.tex
\section{Accelerating Bayesian Inference}
\label{sec:uqpce}

To reduce the cost of sampling the posterior, we rely on a surrogate model of the QoI
that requires a much smaller ensemble of model runs~\citep{Malinverno2002,MarzoukNajm2009}. Here, we rely on
Polynomial Chaos expansions for representing the QoIs, which, in addition can efficiently
provide statistical properties, such as the mean, variance and sensitivities~\citep{Crestaux}. 

Due to the complexity of the \MIT, constructing a surrogate for the different model
outputs is not feasible. Instead, we construct a single surrogate for the test statistic $E$ which is the QoI in this case.
This test statistic $E$ is computed from the outputs of the model runs required for the construction of the 
surrogate as explained below. This practice simplifies the PC calculation where only one surrogate model
would be constructed that can be sampled directly in the posterior of Equation~(\ref{eq:post3}). 

\subsection{Polynomial Chaos}

Polynomial Chaos (PC) is a probabilistic methodology that expresses the 
dependencies of QoI on the uncertain model inputs
as a truncated polynomial expansion~\citep{GhanemSpanos1991,Villegas2012,Lin2009,Xiu2004}.
The PC method is briefly described  below; for more details 
the reader is referred to~\cite{LeMaitreKnio2010}.

We show here the process of constructing a PC surrogate for the test statistic $E$.
To this end, we denote by
% $E=E(\xxi)$ denote a quantity of 
%interest (QoI) that is the output of a computational model or the total cost function in our case 
$\xxi=(\xi_1,...,\xi_m)$ the canonical vector of random variables that parametrize the uncertain inputs 
i.e.\  the KPP parameters. In the case of uniform distributions
the canonical vectors are calculated as follows $\xi_{i} = \frac{2p_i-(a_i+b_i)}{(a_i-b_i)}$~\citep{LeMaitreKnio2010}.
PC expresses the dependencies of $E$ on the uncertain input
variables $\xxi$ as a truncated expansion of the following
form:

\begin{equation}
 E(\xxi) \approx \sum_{k = 0}^R e_k \psi_k(\xxi),
\label{eq:stochseries}
\end{equation} 
where $e_k$ are the polynomial coefficients, and
$\psi_k(\xxi)$ are elements of an orthogonal basis of an underlying probability
space. The total number of terms in the truncated PC expansion is
$R+1 = \frac{(m+r)! }{m!\ r!}$ where $m$ is the number of stochastic dimensions and $r$ is the highest order
polynomial retained. 

The choice of the basis is dictated by the probability density
function $\rho(\xxi)$ of the stochastic vector $\xxi$, which appears as a weight
function in the probability space's inner product:

\begin{equation}
 \left<\psi_i,\psi_j\right> = \int \psi_i(\xxi) \;\psi_j(\xxi) \; \rho(\xxi) \; \mbox{d}\xxi=\delta_{ij}\ave{\psi_i^2},
\label{eq:inner}
\end{equation}
where $\delta_{ij}$ is the Kronecker delta.
For uniform distributions (our case), the basis functions are scaled Legendre polynomials.
For multi-dimensional problems the basis functions are
tensor products of 1D basis functions~\citep{LeMaitreKnio2010}.

\subsection{Determination of PC coefficients}
\label{sec:pccs}
Various methods have been proposed for the determination of the PC coefficients $e_k$. 
They can be classified into Non-intrusive and Galerkin methods.
Non-intrusive methods rely on an ensemble of deterministic model evaluations of $E(\xxi)$, 
for particular realizations of $\xxi$ selected either at random or deterministically. 
Non-Intrusive methods include Non-Intrusive Spectral and Pseudo-Spectral Projection
~\citep{Reagan:2003,Constantine:2012,Conrad:2013}, 
Least-Square-Fit and regularized variants~\citep{Berveiller:2006,Blatman:2011,Doostan:2014}, 
Collocation (interpolation) methods~\citep{Babuska:2007,Xiu_Hesthaven,Nobile:2008a}, 
that are often combined with Sparse-Grid algorithms to reduce computational complexity. 
In the present paper, we adopt non-intrusive approaches that allow the use of the forward model as a black box
with no code modifications required. PC expansion coefficients are determined
based on a set of response simulations for a specified set of the uncertain parameters. 

\subsubsection{Non-intrusive Spectral Projection}
\label{sec:nisp}
We first applied on the traditional Non-Intrusive Spectral Projection (NISP) method that 
exploits the orthogonality of the basis and applies a Galerkin projection to find the PC expansion coefficients as follows:

\begin{equation}
e_k = \frac{\left< E, \psi_k \right>}{\left< \psi_k, \psi_k \right>} = 
 \frac{1}{\left< \psi_k, \psi_k \right>} 
 \int E \psi_k(\xxi) \rho(\xxi) \mbox{ d}\xxi.
\end{equation}
This orthogonal projection minimizes the $L_2$ error on the space spanned by the basis.
The stochastic integrals are then approximated using a numerical quadrature to obtain:

\begin{equation}
  \left< E, \psi_k \right> 
%\approx \left< E, \Psi_k \right>_{ Q}
\approx \sum_{q=1}^{Q} E(\xxi_q) \psi_k(\xxi_q) \omega_q,
\end{equation}
where $\xxi_q$ and $\omega_q$ are multi-dimensional quadrature points and weights, 
respectively, and $Q$ is the number of nodes in the multi-dimensional quadrature. The quadrature order should be commensurate with the
truncation order, and should be high enough to avoid aliasing artifacts.
The choice of quadrature rule is hence critical to the performance
of the PC.

The computation of the $e_k$ can be finally expressed as a matrix-vector product of the form:

\begin{equation} 
 e_k=\sum_{q=1}^{Q} {\cal P}_{kq} E(\xxi_q),\;\;\;
 {\cal P}_{kq}=\frac{\psi_k(\xxi_q)\omega_q}{\left< \psi_k, \psi_k \right>},
\label{eq:nisp}
\end{equation} 
where ${\cal P}_{kq}$ is called the NISP projection matrix (can be pre-computed) and $E(\xxi_q)$ is obtained
from an ensemble of the deterministic model realizations with the uncertain parameters set at
the quadrature values $\xxi_q$. 

\begin{figure}[h]
\centering
\includegraphics[width=0.6\textwidth]{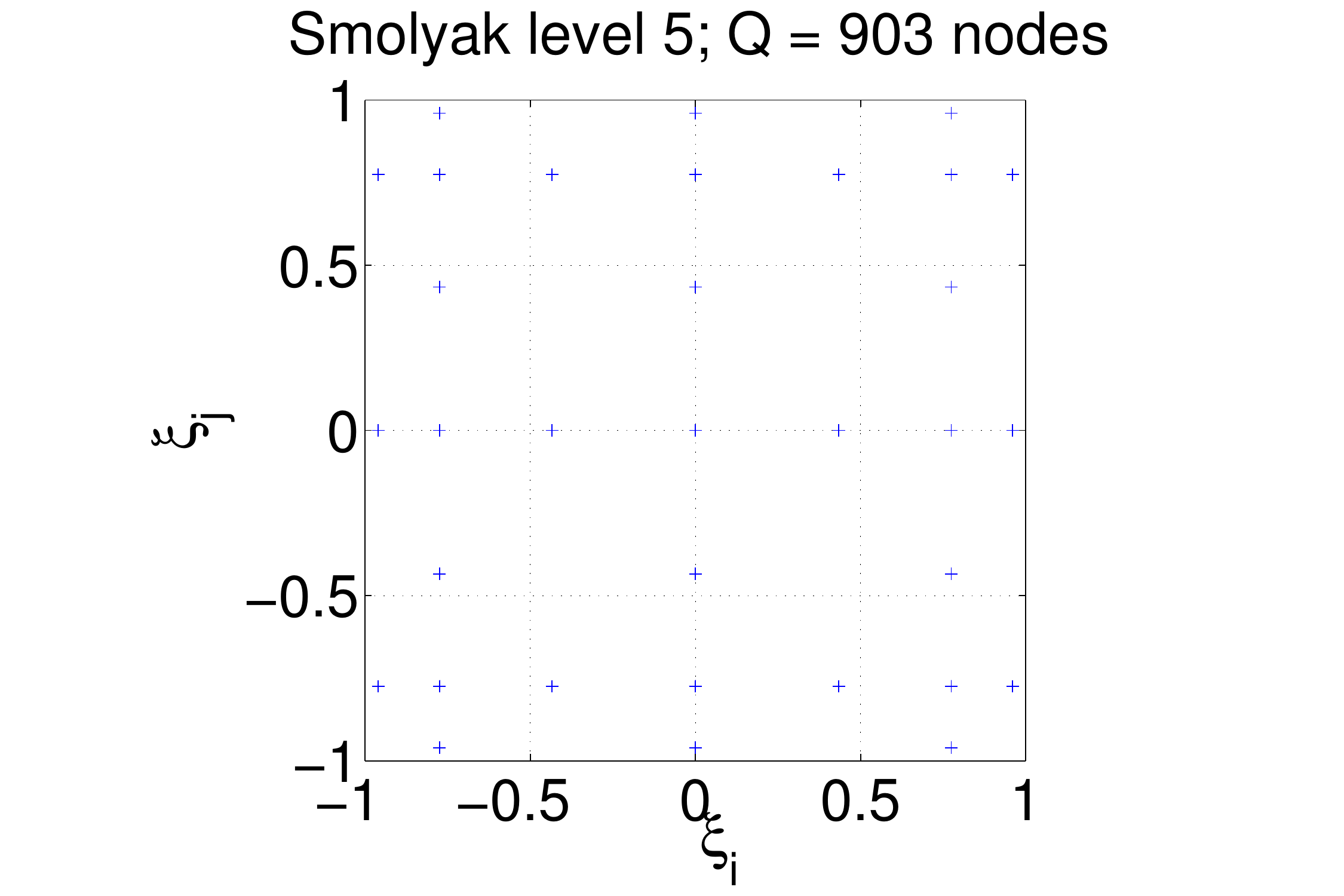}
\caption{Projection of the Smolyak Sparse quadrature nodes corresponding to level 5 on 2D plane.}  
\label{fig:s_quad}
\end{figure}

In our present work, a quadrature was built based on Smolyak sparse nested grid~\citep{Petras:2000,Petras:2001,Petras:2003,Gerstner:2003,Smolyak:1963}
to reduce the number of expensive deterministic \MIT runs. 
For a PC expansion of order $r = 5$ (total number of terms in the truncated PC expansion
$R+1 = 252$) and uncertain parameters $m=5$, a total number of $Q = 903$ quadrature nodes were
needed corresponding to Smolyak level 5. A projection of the quadrature nodes is shown in Figure~\ref{fig:s_quad} on two-dimensional plane.

\begin{figure}[h]
\centering
\begin{tabular}{clc}
{\includegraphics[width=0.45\textwidth]{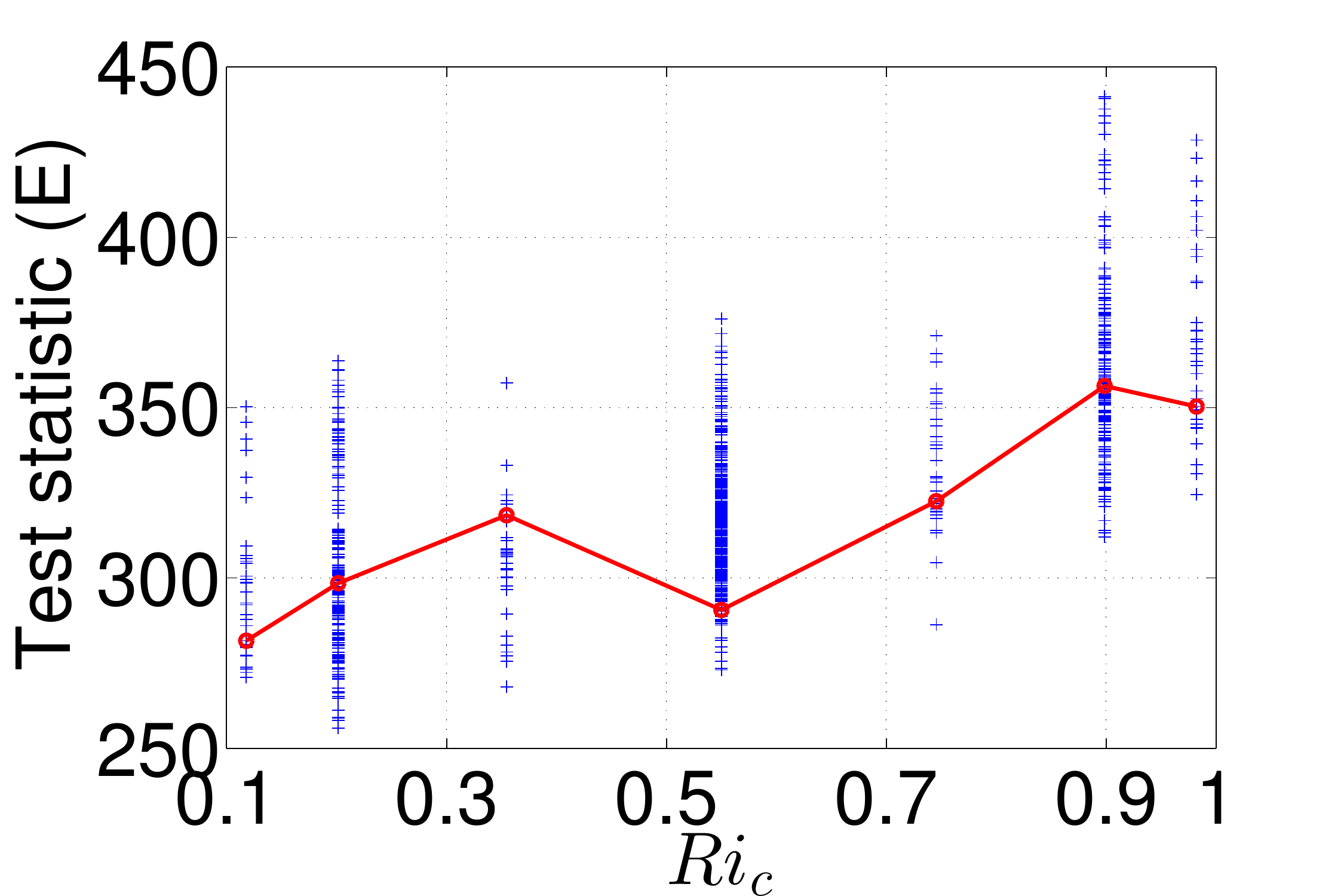}}\,
{\includegraphics[width=0.45\textwidth]{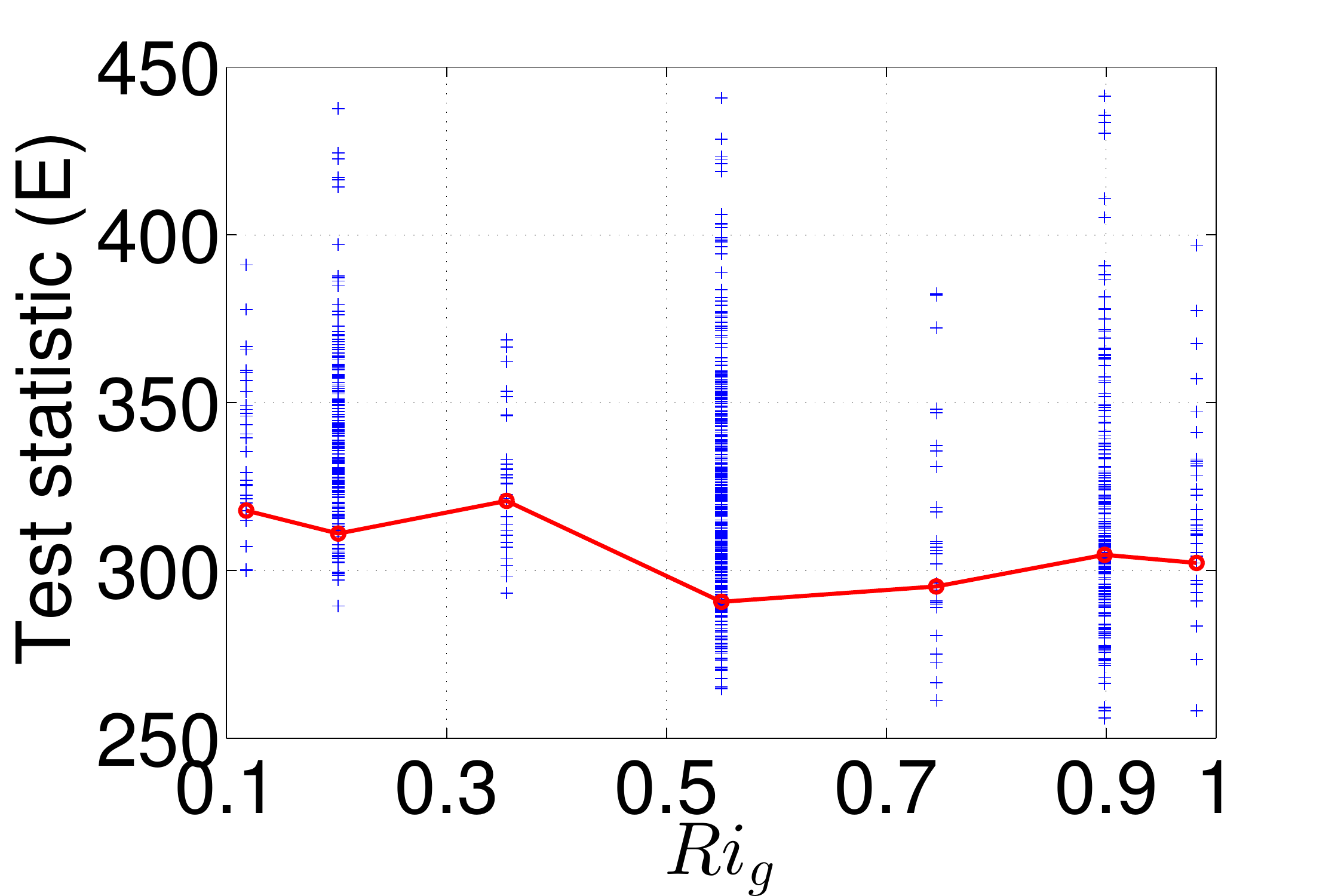}}\,
\\
{\includegraphics[width=0.45\textwidth]{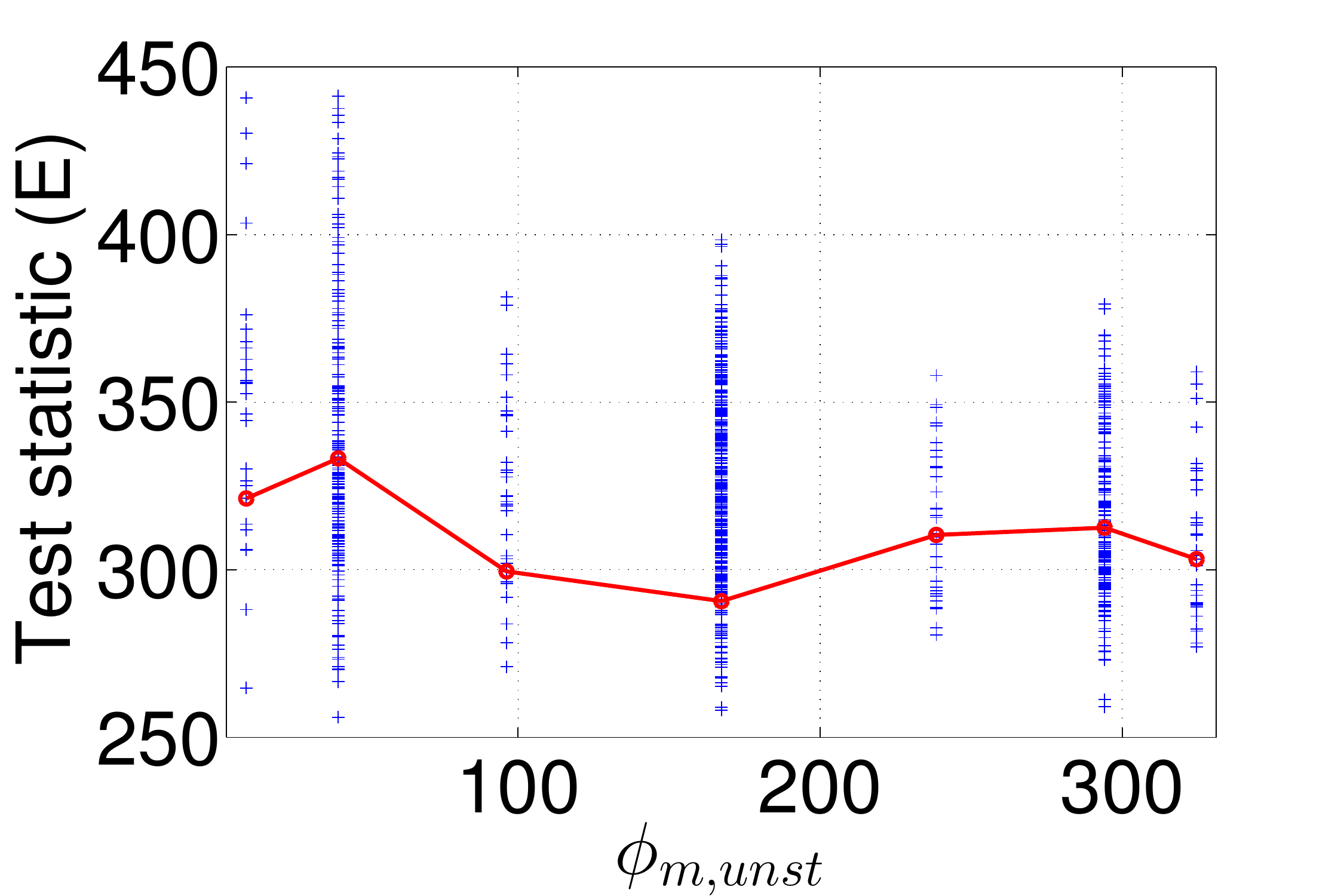}} 
{\includegraphics[width=0.45\textwidth]{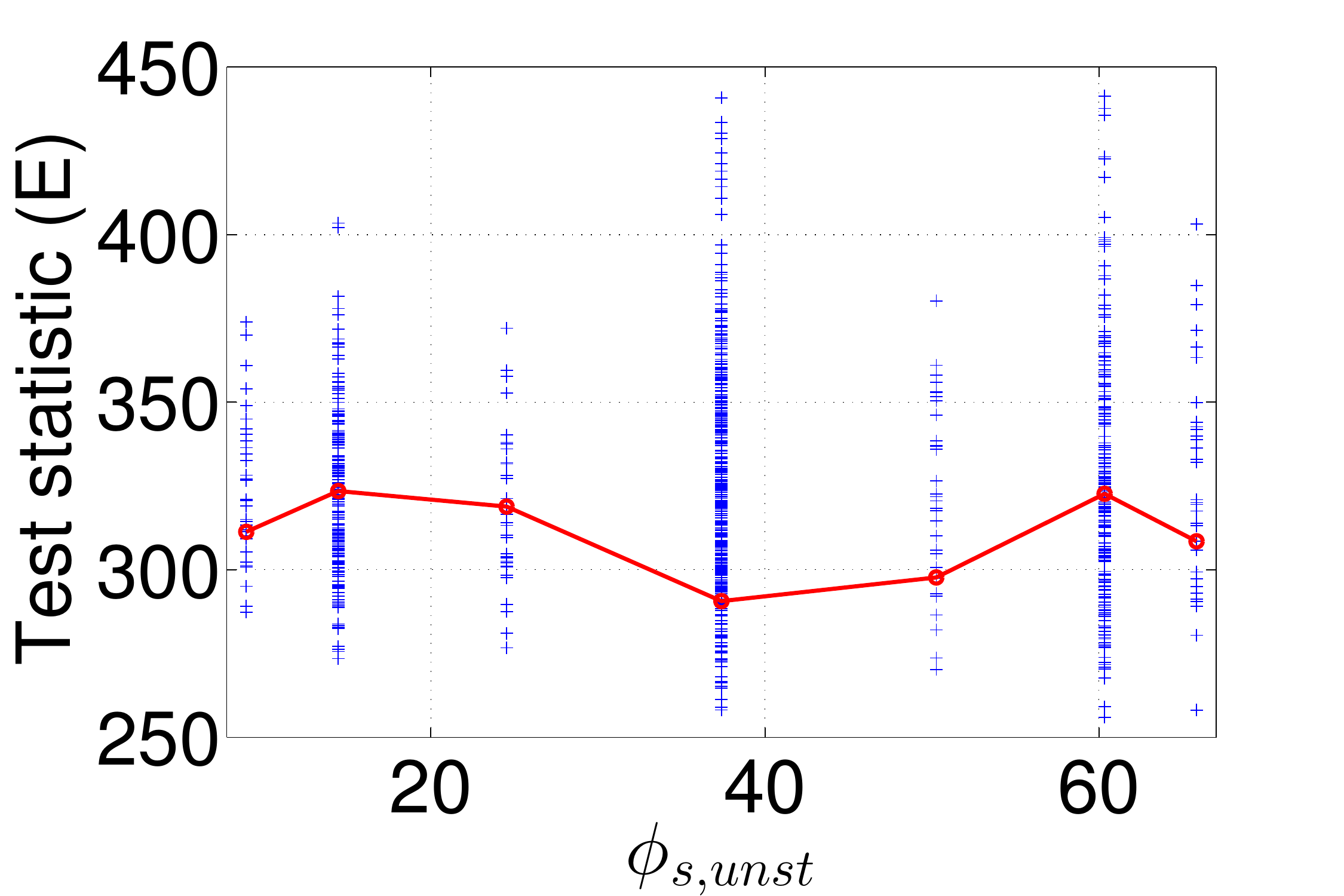}}\,
\\
{\includegraphics[width=0.45\textwidth]{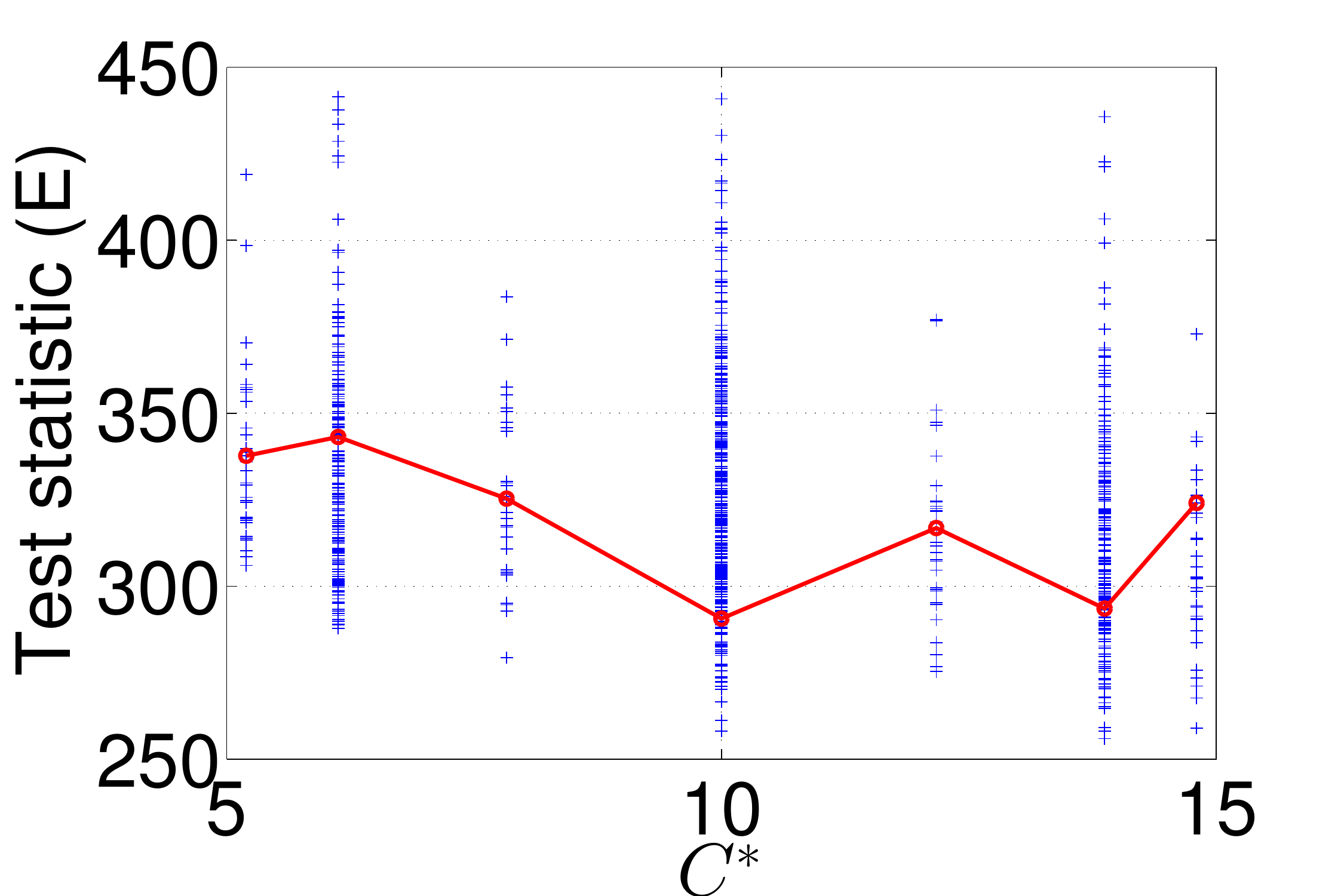}}
\end{tabular}
\caption{Test statistic ($E$) versus KPP parameters at the 903 sparse quadrature nodes. 
Each panel plots $E$ against one the uncertain parameters as indicated.
The red dashed line in each panel corresponds to the case when the other parameters are fixed to the midpoint of their uniform prior value i.e.\  $\xi_i=0$.}  
\label{fig:quad}
\end{figure}

We therefore ran \MIT 903 times as per the quadrature and calculated the test statistic
from the different model outputs. The test statistic corresponding to the sparse quadrature is shown in Figure~\ref{fig:quad}
function of different parameters' spaces as indicated in each panel. The red dashed line in each panel represents the cases where the 
other parameters are set to the center of the corresponding priors  i.e.\  $\xi_i=0$. These figures also clearly show
the uncertainty bounds in the test statistic due to the uncertainty in the input parameters. This is true for all 
five parameters. 

\begin{figure}[ht]
\centering
\includegraphics[width=0.6\textwidth]{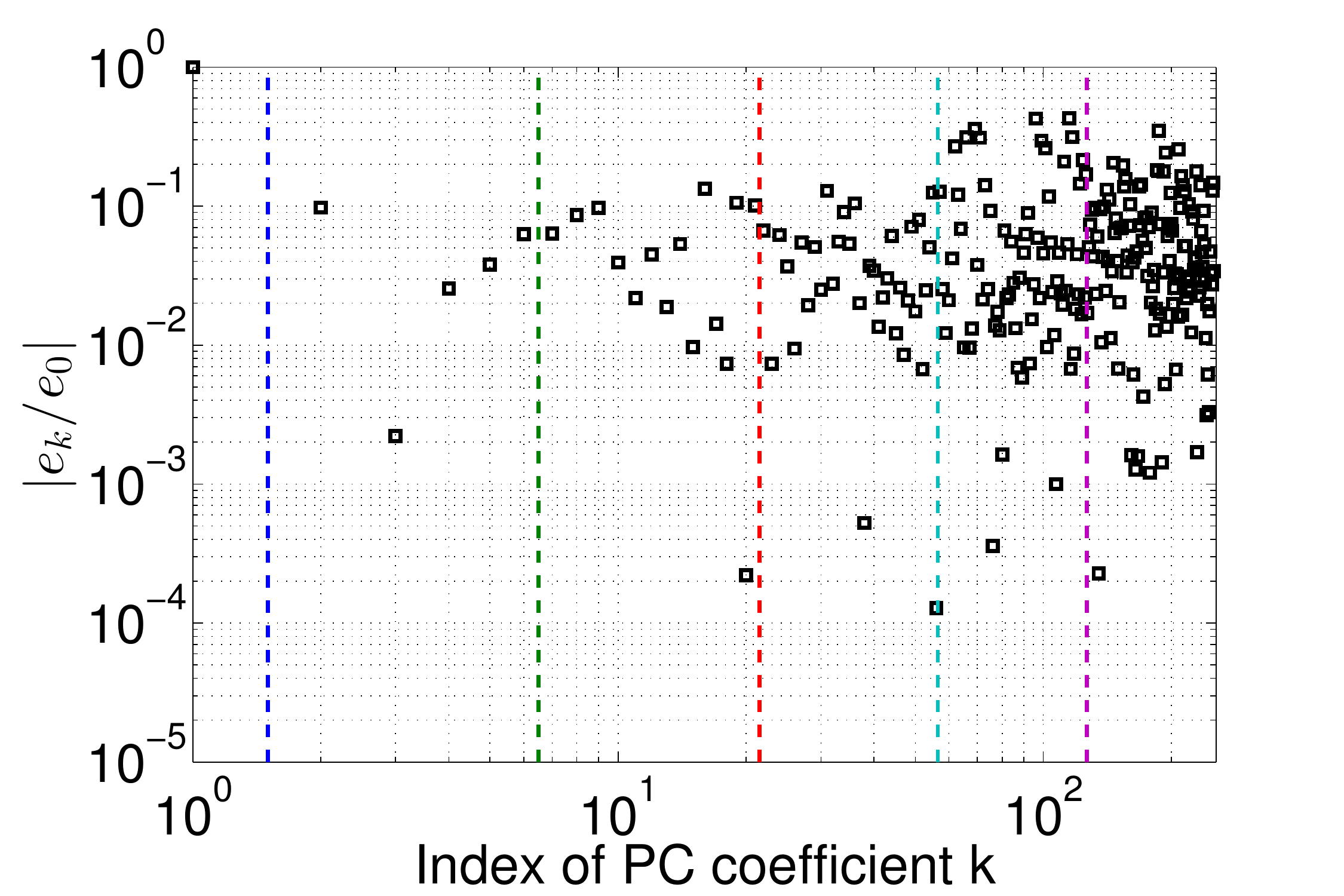} 
\caption{PC expansion normalized coefficients $|e_k/e_0|$ for PC order up to $r=5$. The dashed vertical lines separate the PC expansion terms into degrees. 
The coefficients were calculated using NISP.}  
\label{fig:pcnisp}
\end{figure}

\begin{figure}[ht]
\centering
\begin{tabular}{clc}
\includegraphics[width=0.6\textwidth]{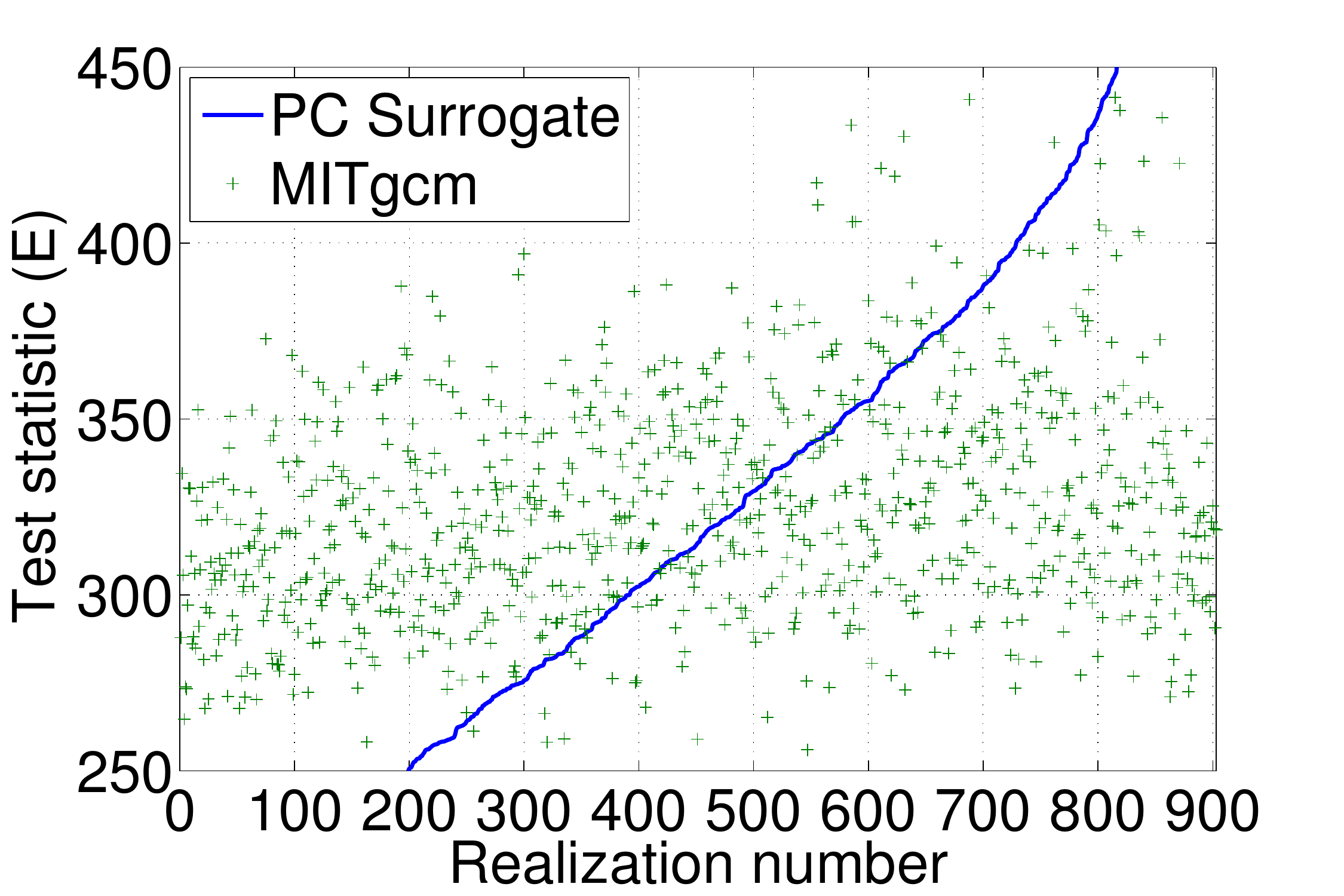} &
\end{tabular}
\caption{Comparing test statistic ($E$) from \MIT model runs superimposed with their PC surrogate counterparts constructed using NISP. 
The shown cases correspond to the sparse quadrature.% The normalized relative  error (NRE) is also indicated.
}  
\label{fig:error_nisp}
\end{figure}

\begin{figure}[h]
\centering
\begin{tabular}{clc}
\includegraphics[width=0.6\textwidth]{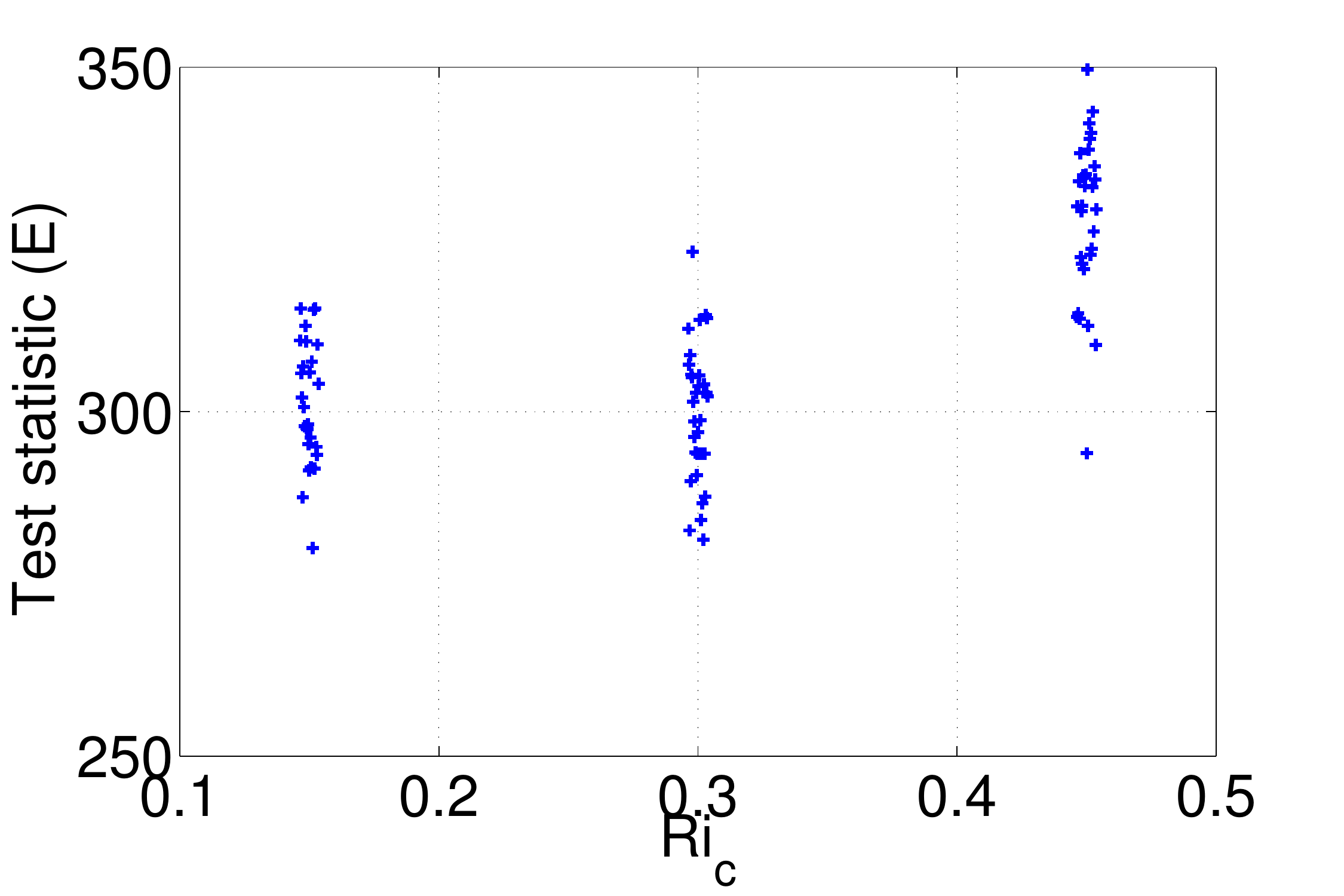} &
\end{tabular}
\caption{%Comparing test statistic ($E$) from \MIT model runs with their PC surrogate counterparts 
%(Left) Superimposed (Right) Scatter plot. 
%The shown cases correspond to $89$ model runs
Test statistic ($E$) from \MIT model runs when varying $Ri_c$ only infinitesimally}
% and PC is constructed using BPDN. }  
\label{fig:noisy}
\end{figure}

The PC expansion coefficients are computed from the output of the 903 quadrature runs
using Equation~\ref{eq:nisp}. Figure~\ref{fig:pcnisp} plots the spectrum of the normalized PC 
coefficients, $e_k/e_0$,  in absolute value. The vertical lines
separate the PC expansion terms into degrees $r={0,..,5}$.
The spectrum shows clearly that the PC suffers from convergence issues 
as the NISP-estimated PC coefficients do not decay with further increasing PC order but instead grow. 
This can be attributed to the presence of internal noise in the model 
that is not tolerated by the NISP method and thus over-fitting the model from the quadrature points with additional refinement of the PC order. 
To further asses this convergence issue, we show in Figure~\ref{fig:error_nisp} the deterministic \MIT realizations plotted 
on top of their PC expansion counterparts. The difference between the two sets of data confirms the inability of the NISP-estimated 
surrogate PC model to efficiently represent the QoI. 
In Figure~\ref{fig:noisy}, we show the test statistic $E$ corresponding to $89$ \MIT model runs
where we vary $Ri_c$ only infinitesimally as indicated on the plot. The test statistic value is highly sensitive to infinitesimal perturbations of $Ri_c$.
This likely results from internal noise in the model that is amplified over time, in part by non-linear interactions
in the model.

As a conclusion, the construction of a converging PC expansion using the NISP method 
is not successful and the surrogate model is not representative of the QoI of \MIT; therefore it can not be used for further analysis.

\subsubsection{Basis-Pursuit Denoising}
In an attempt to find a suitable PC surrogate model for the test statistic $E$, 
we resort to a different approach that tolerates noise in the model but also that is non-intrusive. 
Instead of using spectral projection, we employ a recent technique that uses Compressed Sensing (CS) for polynomial representations
which first estimates the noise in the model then determines the PC coefficients.
Let ${\vec{e}}=(e_0,...,e_R)$ be a vector of PC coefficients to be determined and let $\vec E = (E(\xxi_1),...,E(\xxi_Q) )$ be a vector of forward model evaluations function of sampled $\xxi_q$. We also define $\Psi$ as the matrix where each row corresponds to the row vector of $R+1$ PC basis functions evaluated at the sampled $\xxi_q$. CS solves the problem: 

\begin{equation}
  \vec{E} = \Psi \vec{e}\footnote{compare with Equation~(\ref{eq:stochseries})}
 \end{equation}
by exploiting the approximate sparsity of the signal i.e.\ the vector of PC coefficients ${\vec{e}}$ that necessarily converge to zero. The sparsity is set by constraining the system and minimizing its energy i.e.\  its $l_1$-norm.  CS seeks a solution with minimum number of non-zero entries by solving the optimization problem:

    \begin{equation}
    {\cal{O}}_{1,\delta} \approx \left\lbrace \argmin_{{\vec e}} || {\vec{e}} ||_1  : || \vec E - \Psi \vec{e} ||_2 \le \delta \right\rbrace
\label{eq:optim}
     \end{equation}   
where $\delta$ is the noise estimated in the signal. 
 
This $l_1$-minimization problem is referred to as Basis-Pursuit (BP) when $\delta=0$,
and to Basis-Pursuit-Denoising (BPDN) when a noise $\delta$ in the system is assumed as proposed in~\cite{Donoho:2006}.
We adapt the latter approach since we acknowledge the existence of noise in the predicted QoI. We note that in Equation~(\ref{eq:optim}) the constraint depends on selected sampled parameters $\xxi_q$ and their corresponding $E(\xxi_q)$ and not on a general sample of $\xxi$ and $E(\xxi)$. 
As a result, the coefficients ${\vec{e}}$ may be chosen to fit the input realizations, and not accurately approximate the model. 
To avoid this situation, we determine the noise $\delta$ by cross-validation as discussed in~\cite{Doostan:2014}.
To solve ${\cal{O}}_{1,\delta}$  standard $l_1$-minimization solvers may be used. In this work we use the MATLAB package SPGL1~\citep{spgl1:2007} based on the spectral projected gradient algorithm~\citep{BergFriedlander:2008}.

\begin{figure}[h]
\centering
\includegraphics[width=0.6\textwidth]{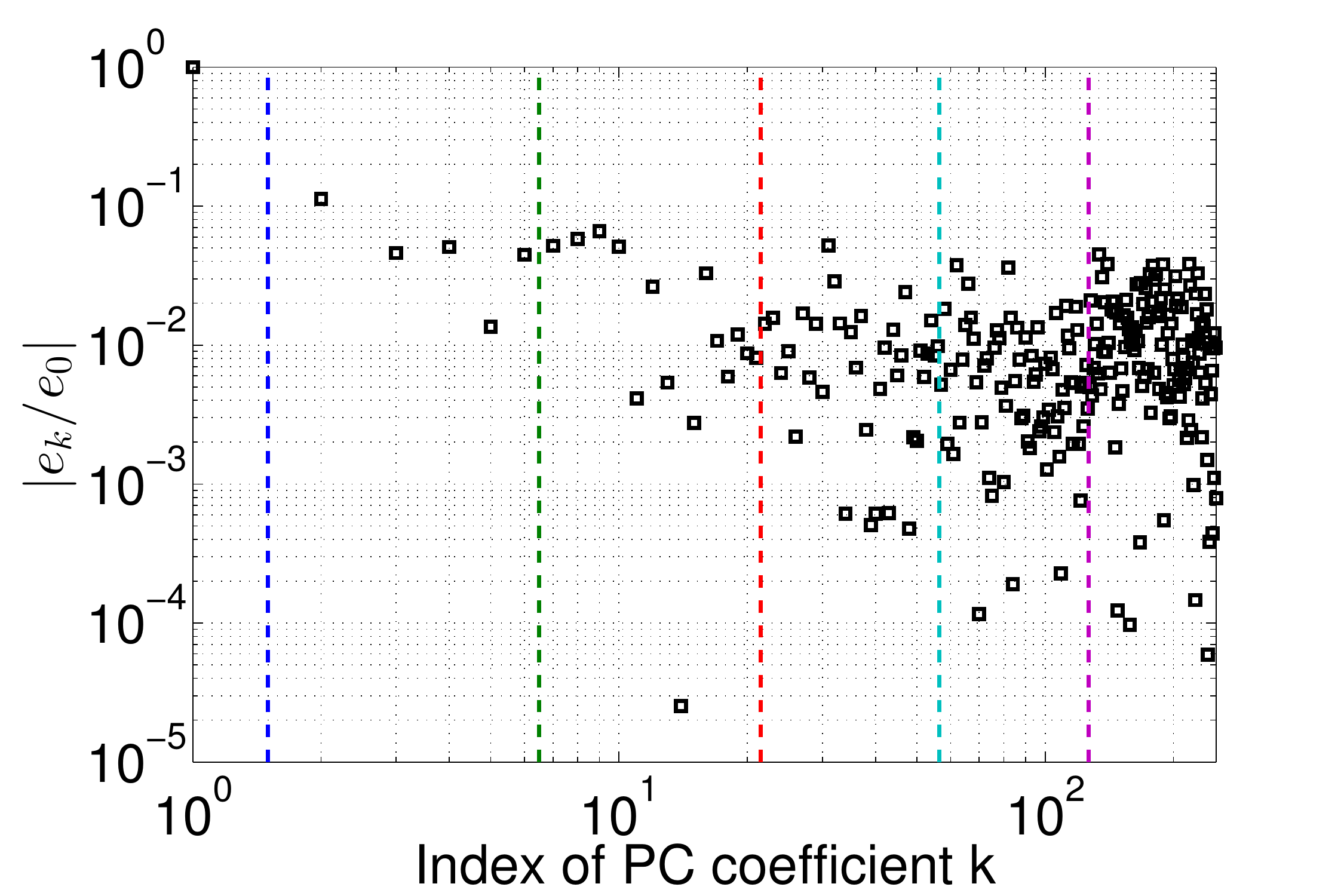} 
\caption{PC expansion normalized coefficients $|e_k/e_0|$ for PC order up to $r=5$. The dashed vertical lines separate the PC expansion terms into degrees. The coefficients were
calculated using BPDN.}  
\label{fig:pccs_bp}
\end{figure}

Here, we applied the BPDN to determine the PC coefficients for the surrogate model
of the test statistic. Instead of using Monte Carlo sampling method to generate realizations 
as in~\cite{Doostan:2014} we take
advantage of the previously simulated 903 \MIT realizations and use them to solve the
optimization problem. 
The resulting PC normalized coefficients spectrum $e_k/e_o$ is plotted in Figure~\ref{fig:pccs_bp} (in absolute value) up to order $r=5$ to asses the convergence of the PC expansion. The vertical lines
indicate the PC terms for different polynomial order $r=0,...,5$.
The spectrum shows clearly that the PC converges better with increasing PC order
using BPDN as opposed to the NISP method.

\begin{figure}[ht]
\centering
\begin{tabular}{clc}
\includegraphics[width=0.6\textwidth]{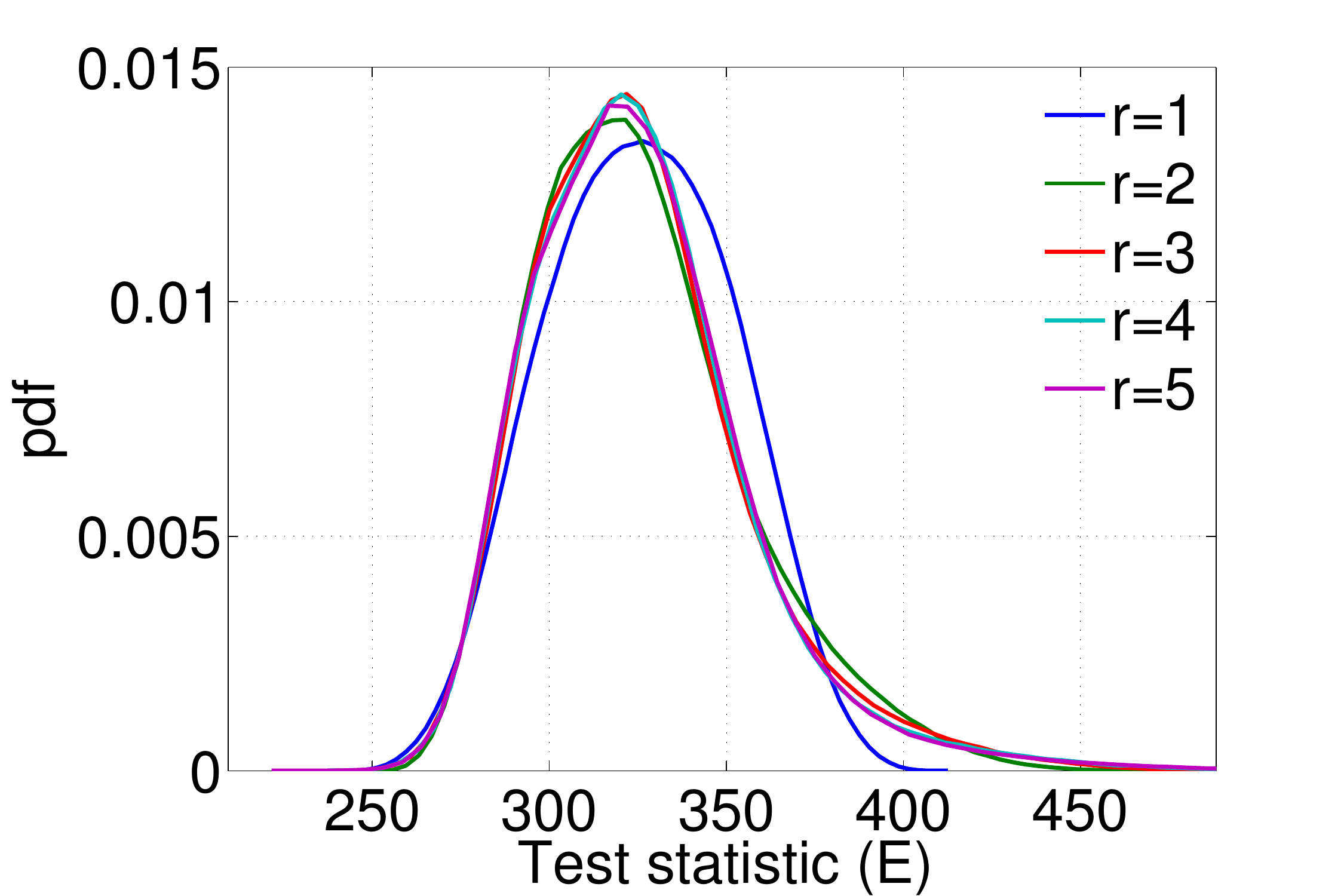}
\end{tabular}
\caption{\emph{pdfs} of test statistic $E$ with increasing order of PC constructed using BPDN-estimated PC surrogate model.}  
\label{fig:pdfc}
\end{figure}

To check the convergence of the PC expansion with PC order $r$,
the PC surrogate is sampled to find the \emph{pdf} of the test statistic $E$
using different polynomial orders.
The \emph{pdfs} are shown in Figure~\ref{fig:pdfc} where we observe that as the PC order
is increased, the \emph{pdfs} get closer to each other, indicating robust estimates.

\begin{figure}[h]
\centering
\begin{tabular}{clc}
\includegraphics[width=0.6\textwidth]{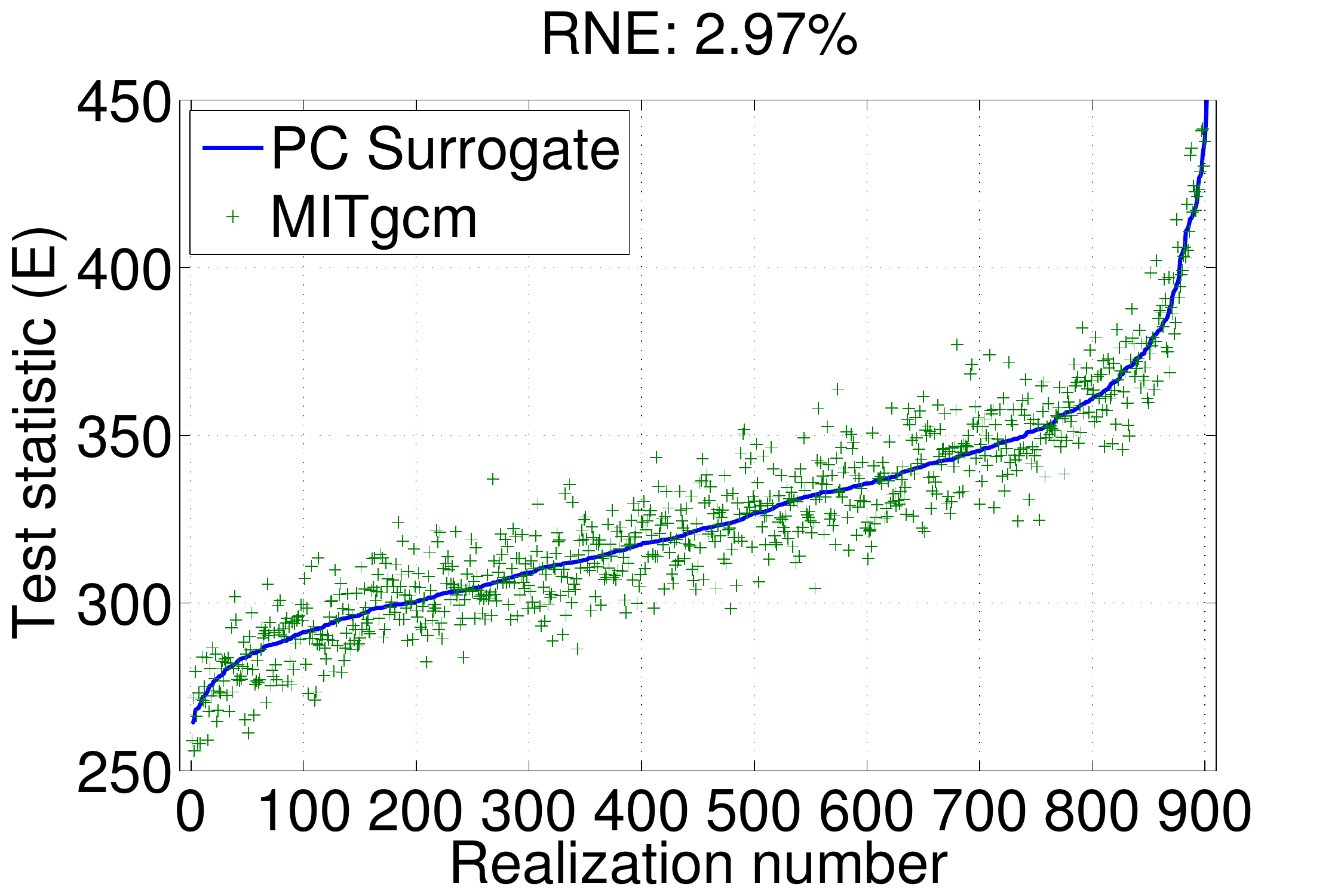} &
\hspace{-12mm}
\includegraphics[width=0.6\textwidth]{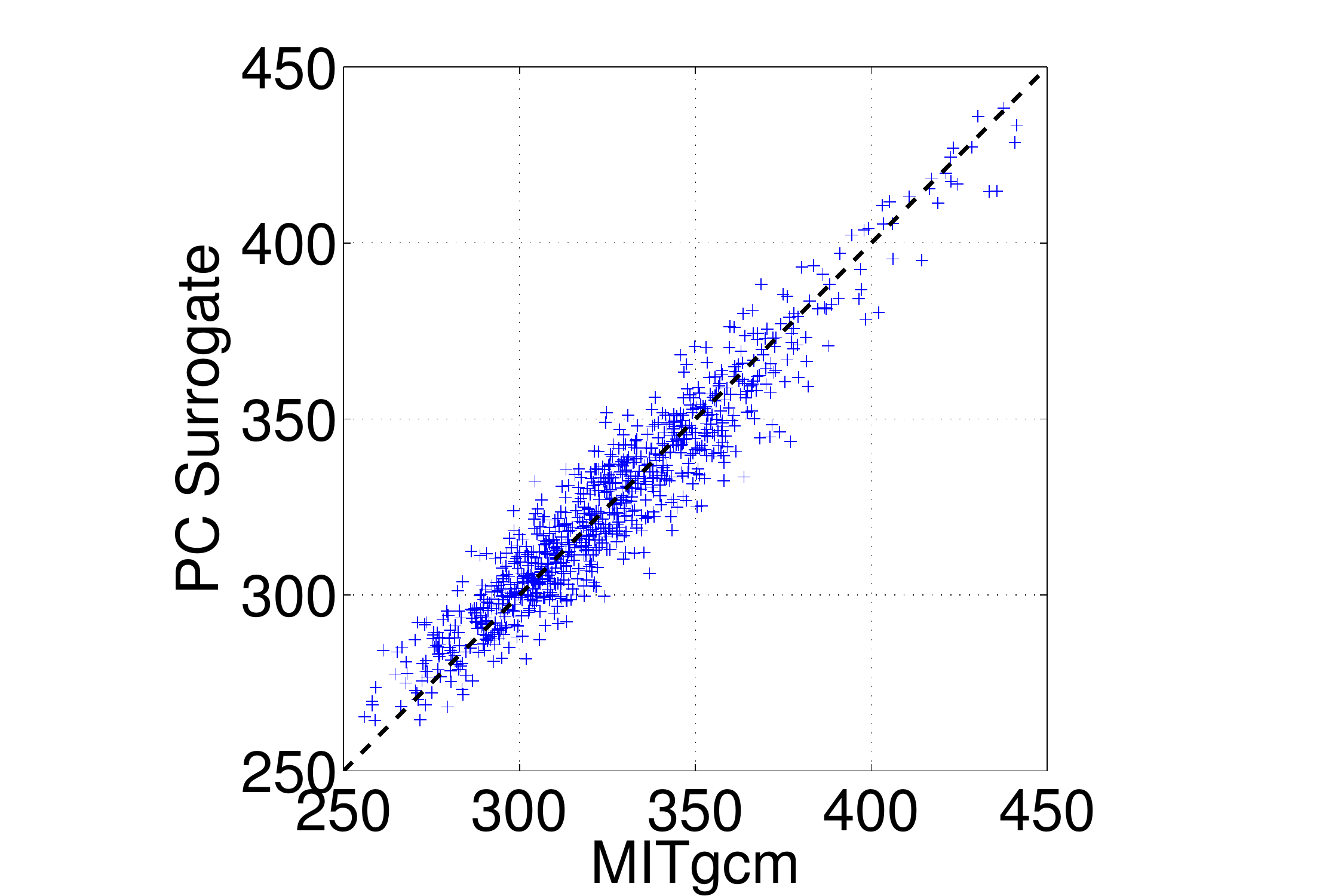} 
\end{tabular}
\caption{Comparing test statistic ($E$) from \MIT model runs with their PC surrogate counterparts 
(Left) Superimposed (Right) Scatter plot. The shown cases correspond to the sparse quadrature
and PC is constructed using BPDN. The  normalized relative error (NRE)
is also indicated.}  
\label{fig:error_bp}
\end{figure}

To confirm the ability of the constructed PC expansion to represent the QoI, we again show the \MIT realizations plotted on top of their PC counterparts
in Figure~\ref{fig:error_bp} (left). One interesting observation
is the internal noise in \MIT that appears
in the corresponding test statistic, while PC gives a smooth function for $E$.
While PC filters out the internal noise in the QoI, it clearly captures the mean of the signal.
Figure~\ref{fig:error_bp}~(right) shows the same but in a scatter plot where \MIT realizations are plotted against
their PC counterparts. 
To quantify the agreement, we use another common error metric where we calculate
the normalized relative error (NRE) between the 903 \MIT simulations and their reconstruction
using the built PC as follows:

\begin{equation} 
   NRE = \frac{\displaystyle
         \left(\sum_{q=1}^Q \left|E(\xxi_q) - \sum_{k = 0}^{R}
e_k\psi_k(\xxi_q)\right|^2
         \right)^{1/2}}
        {\displaystyle
          \left(\sum_{q=1}^Q \left|E(\xxi_q)\right|^2\right)^{1/2} 
          },
\label{eq:error}
\end{equation}
The NRE is calculated and found to be $\sim 3\%$ which is acceptable,
suggesting that BPDN is successful in constructing a surrogate that is able to accurately
simulate the QoI.

\begin{figure}[h]
\centering
\includegraphics[width=0.6\textwidth]{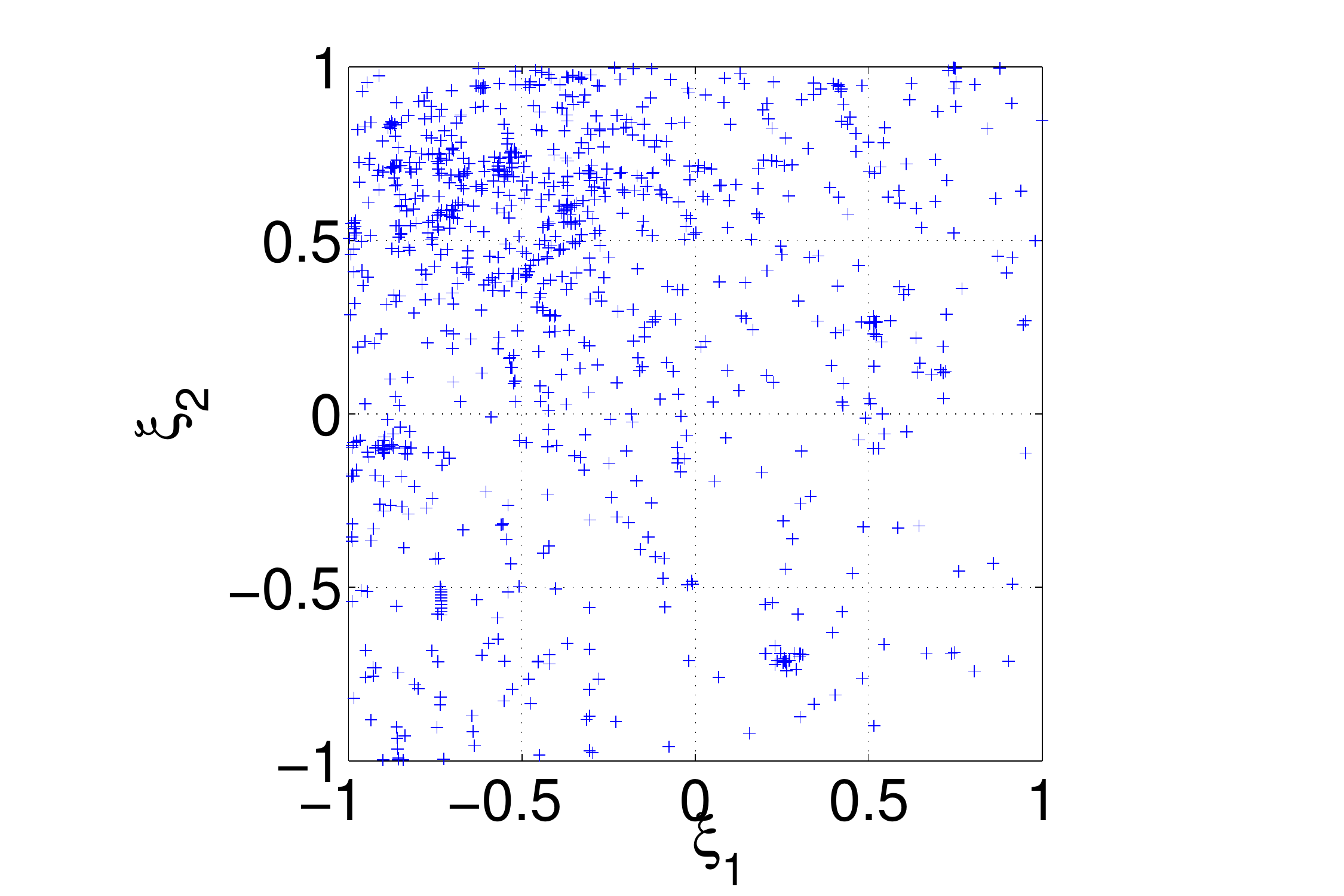}
\caption{Projection of the independent random sample on the $\xxi_1 - \xxi_2$ plane.}  
\label{fig:m_quad}
\end{figure}

\begin{figure}[h]
\centering
\begin{tabular}{clc}
{\includegraphics[width=0.45\textwidth]{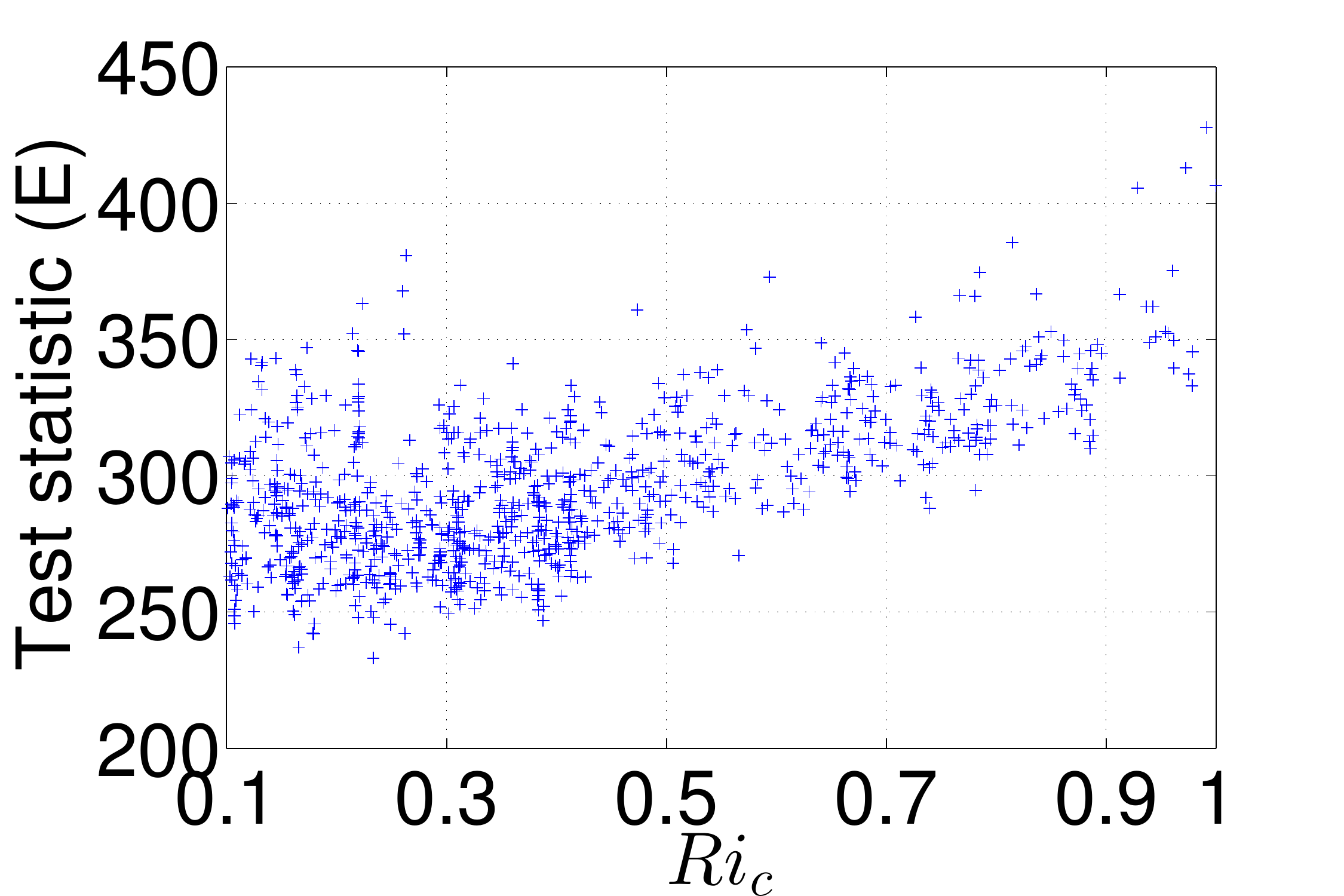}}\,
{\includegraphics[width=0.45\textwidth]{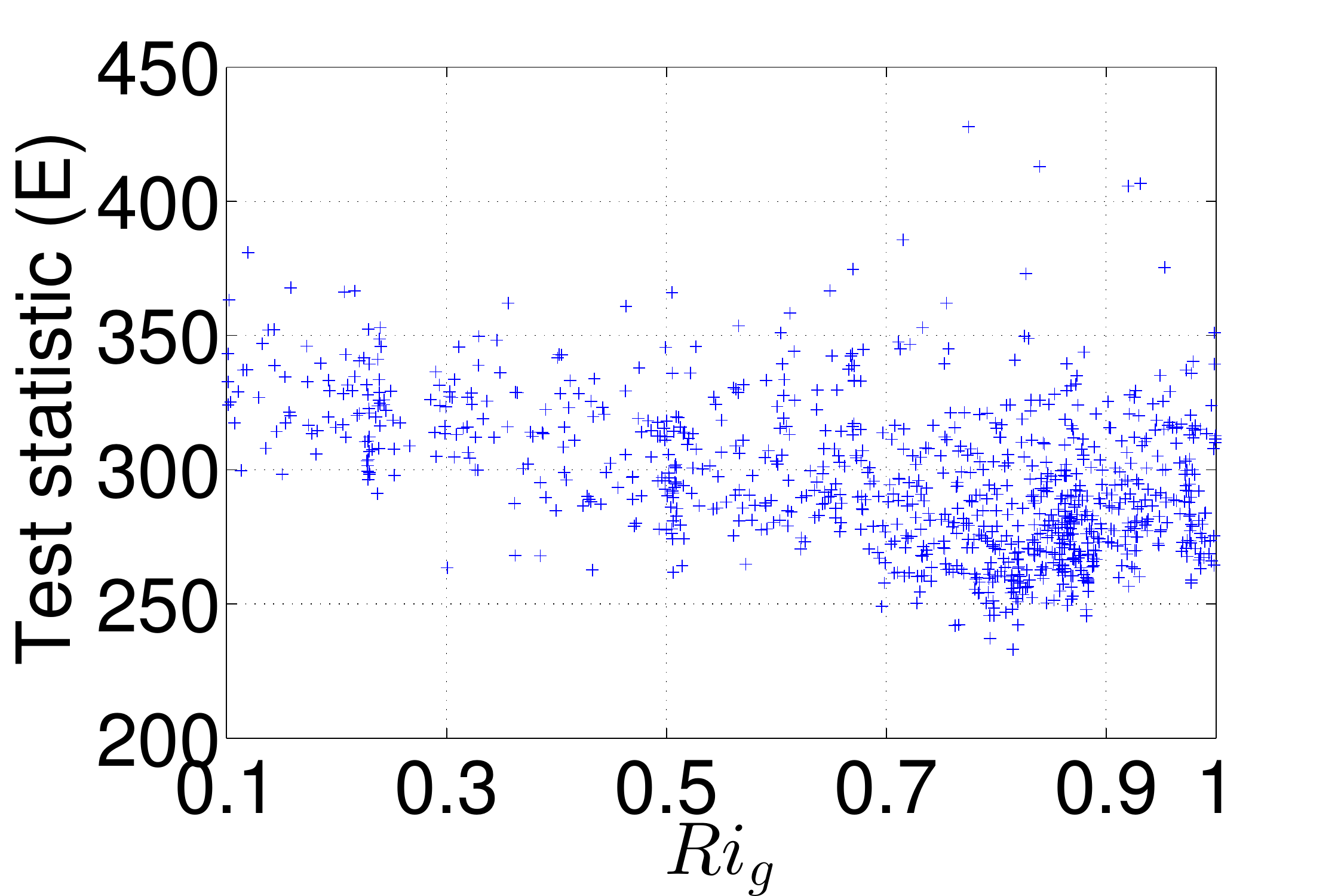}}\,
\\
{\includegraphics[width=0.45\textwidth]{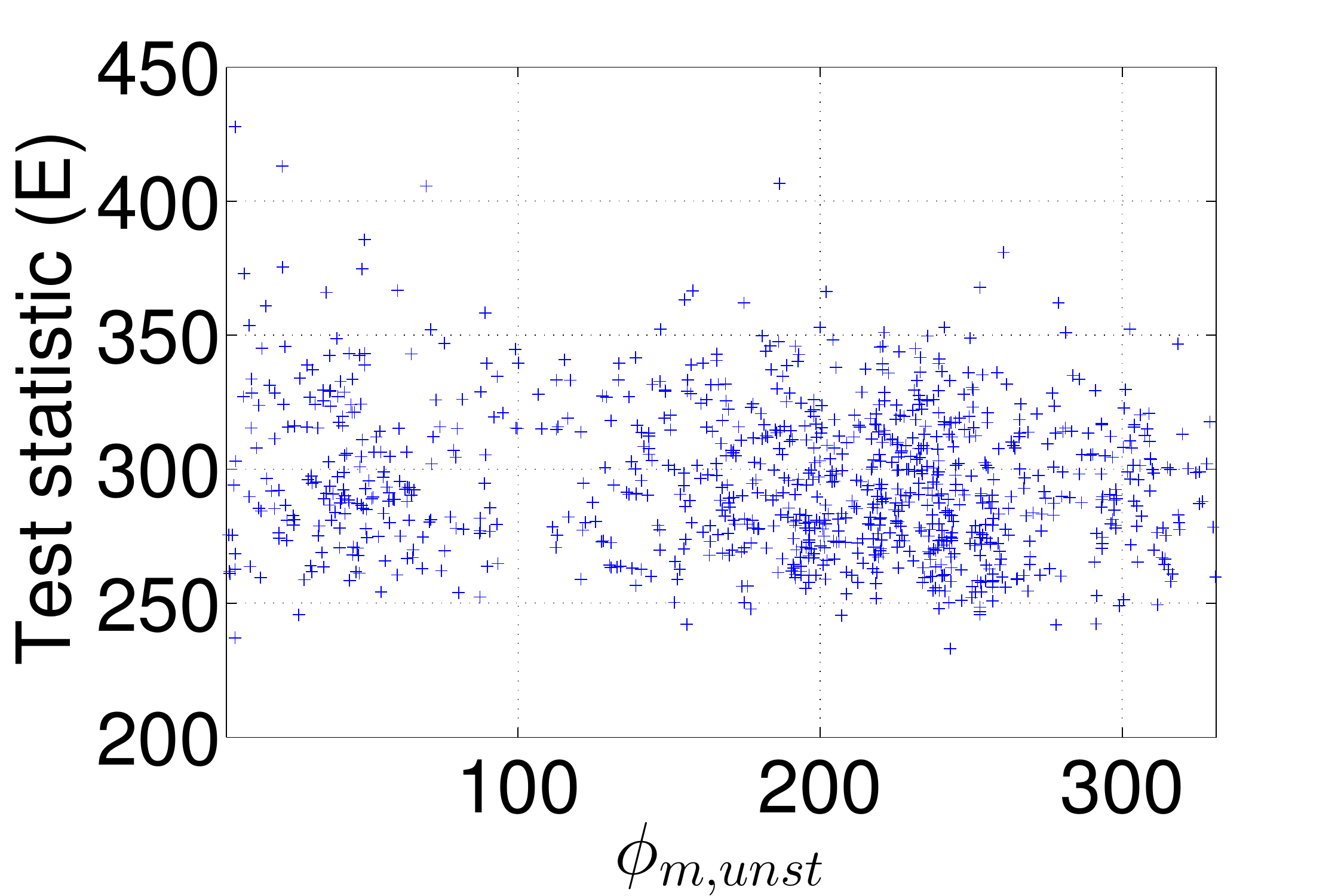}} 
{\includegraphics[width=0.45\textwidth]{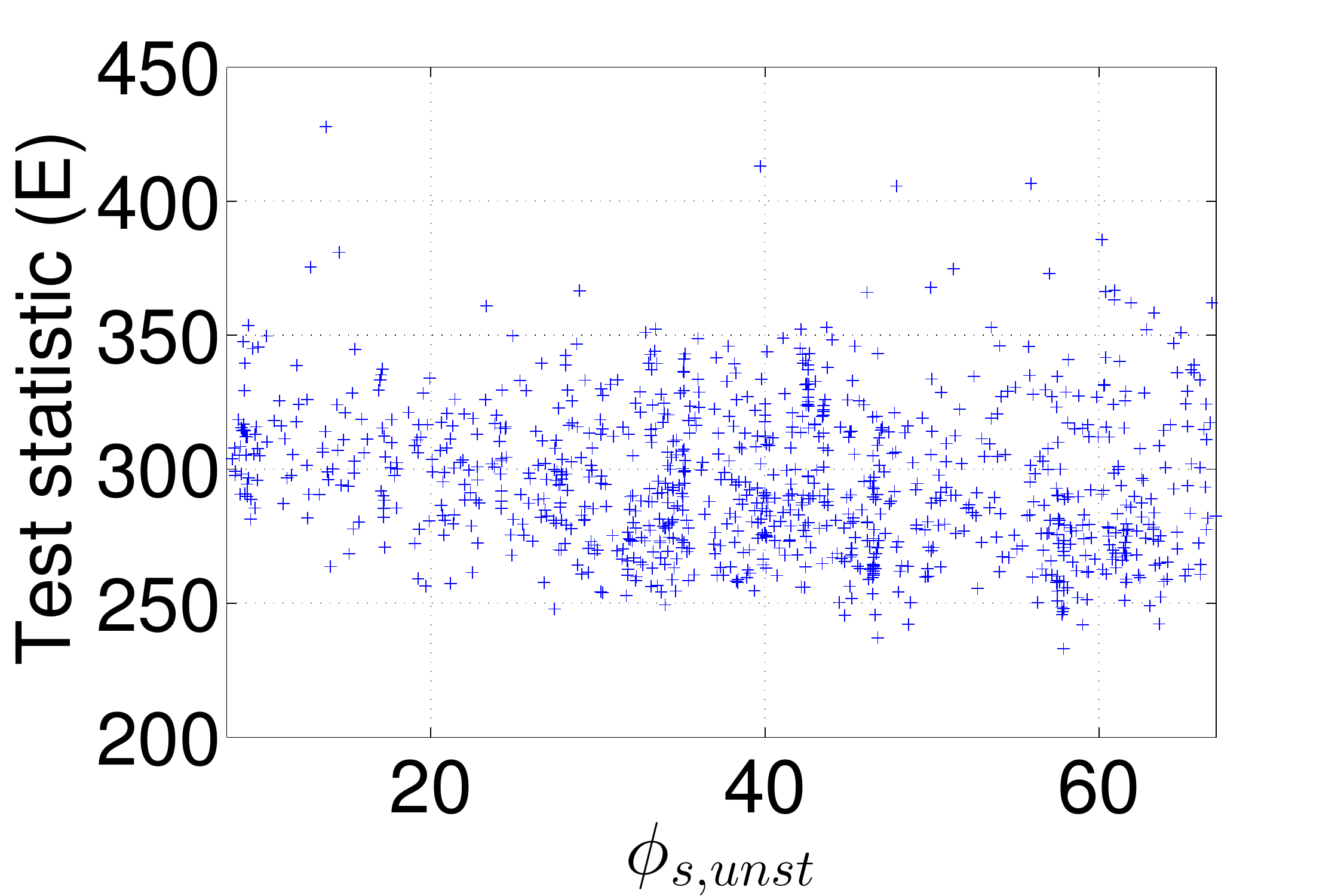}}\,
\\
{\includegraphics[width=0.45\textwidth]{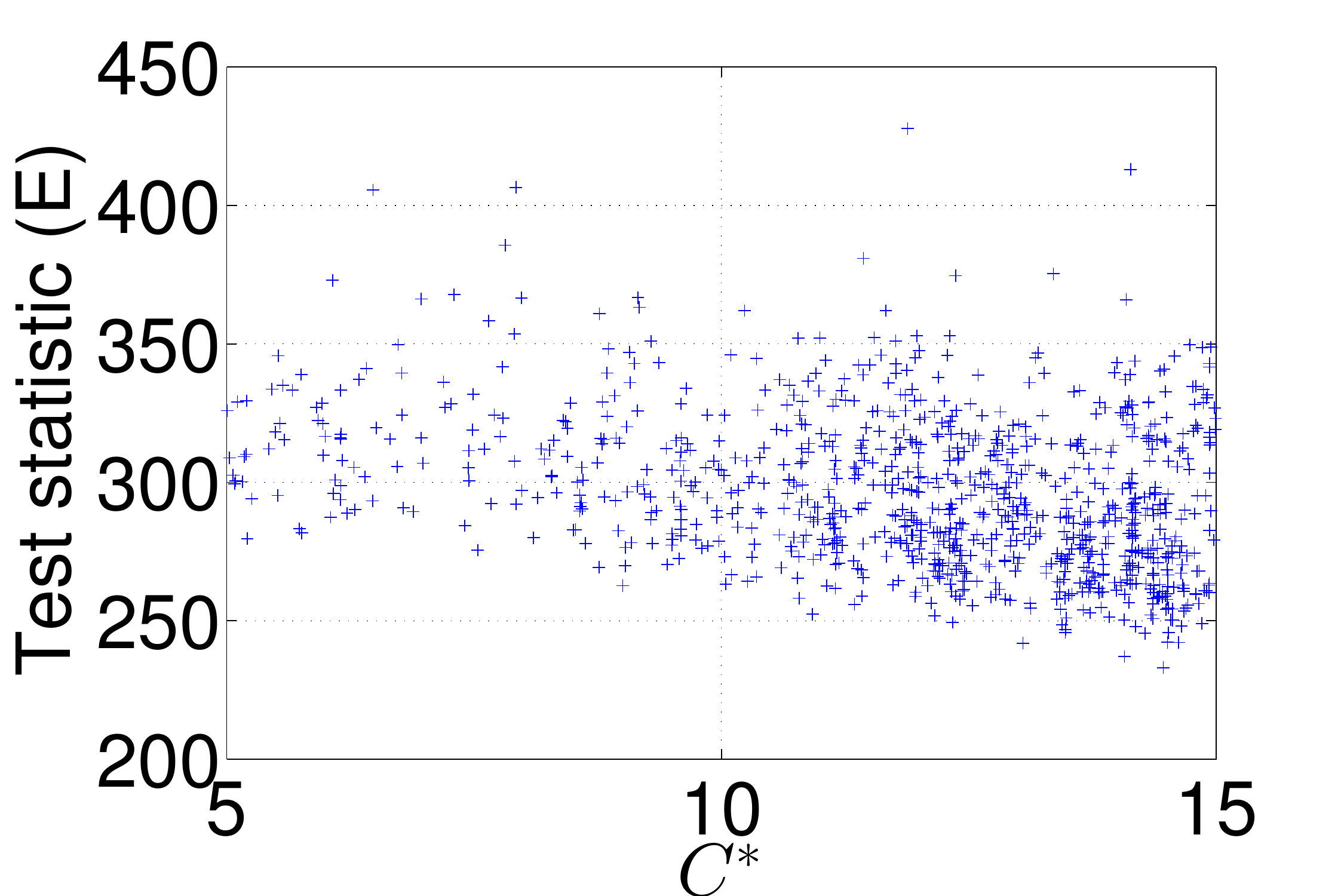}}
\end{tabular}
\caption{Test statistic (E) versus KPP parameters at the $954$ independent random sample. 
Each panel plots $E$ against one the uncertain parameters as indicated.}  
\label{fig:MVFSA}
\end{figure}

%%%%%%%%%%%%%%%%%%%%%%%%%%%%%%%%%%%%%%%%%%%%%%%%%%%%%%%%%%%%%%%%%%%%%%%%%%%%%%%%%%%%
For further validation of our surrogate model, an independent set of $954$ \MIT runs were conducted simultaneously
by varying the same set of KPP uncertain parameters in a identical model setup and using the same test statistic.
The sample is shown in Figure~\ref{fig:m_quad}
as a 2-D projection of the $\xxi_1 - \xxi_2$ plane (representing $Ri_c$ and $Ri_g$).
The corresponding test statistic $E$ is shown in Figure~\ref{fig:MVFSA} as a function of the different parameters' spaces.
The different plots in Figure~\ref{fig:MVFSA} shows a functional trend mainly for $Ri_c$ and $Ri_g$; however,
the variation in $E$ suggests an internal noise in the model outputs in addition to responding to other parameters.

%%%%%%%%%%%%%%%%%%%%%%%%%%%%%%%%%%%%%%%%%%%%%%%%%%%%%%%%%%%%%%%%%%%%%%%%%%%%%%%%%%%%%%%%%

\begin{figure}[h]
\centering
\begin{tabular}{clc}
\includegraphics[width=0.6\textwidth]{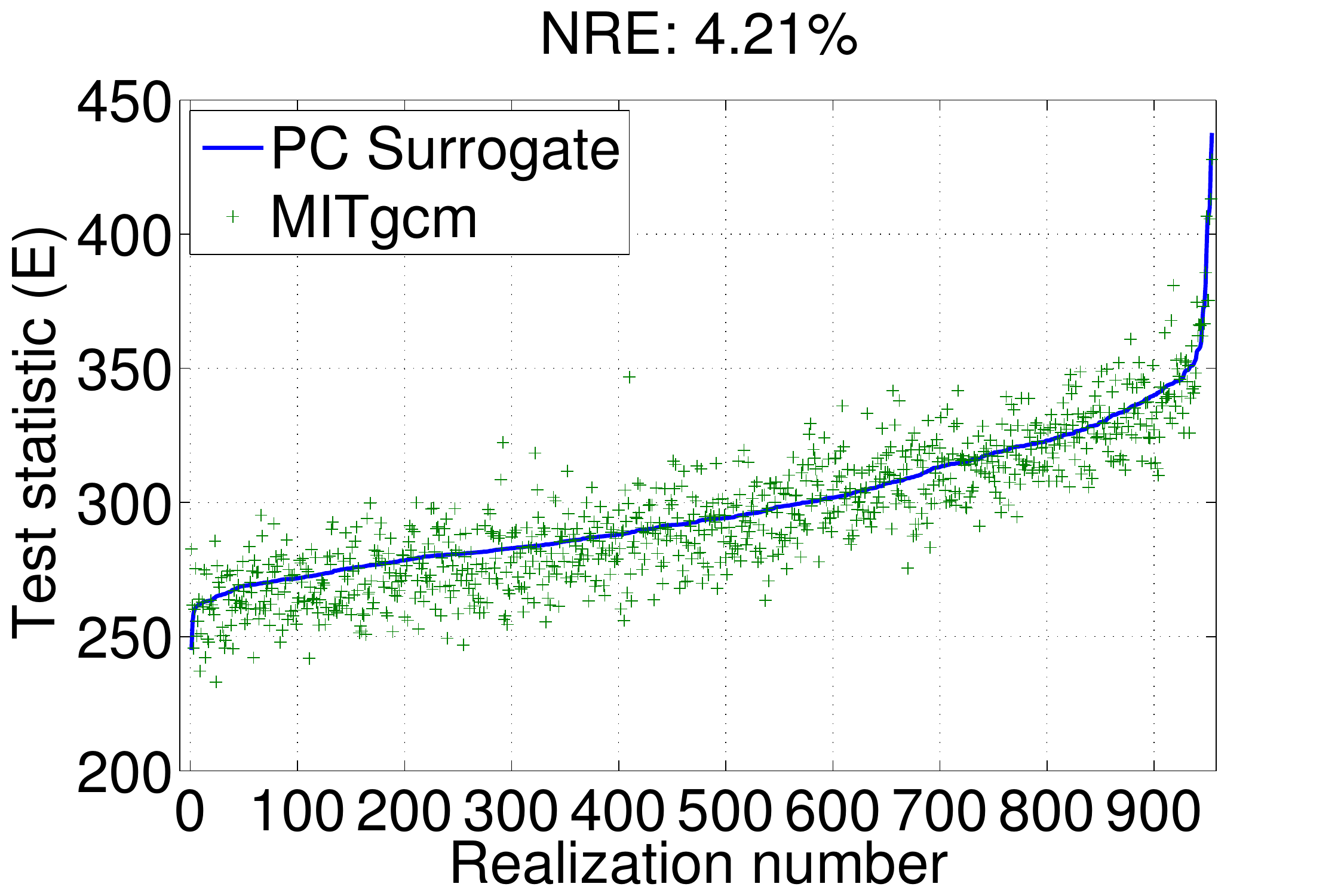} &
\hspace{-12mm}
\includegraphics[width=0.6\textwidth]{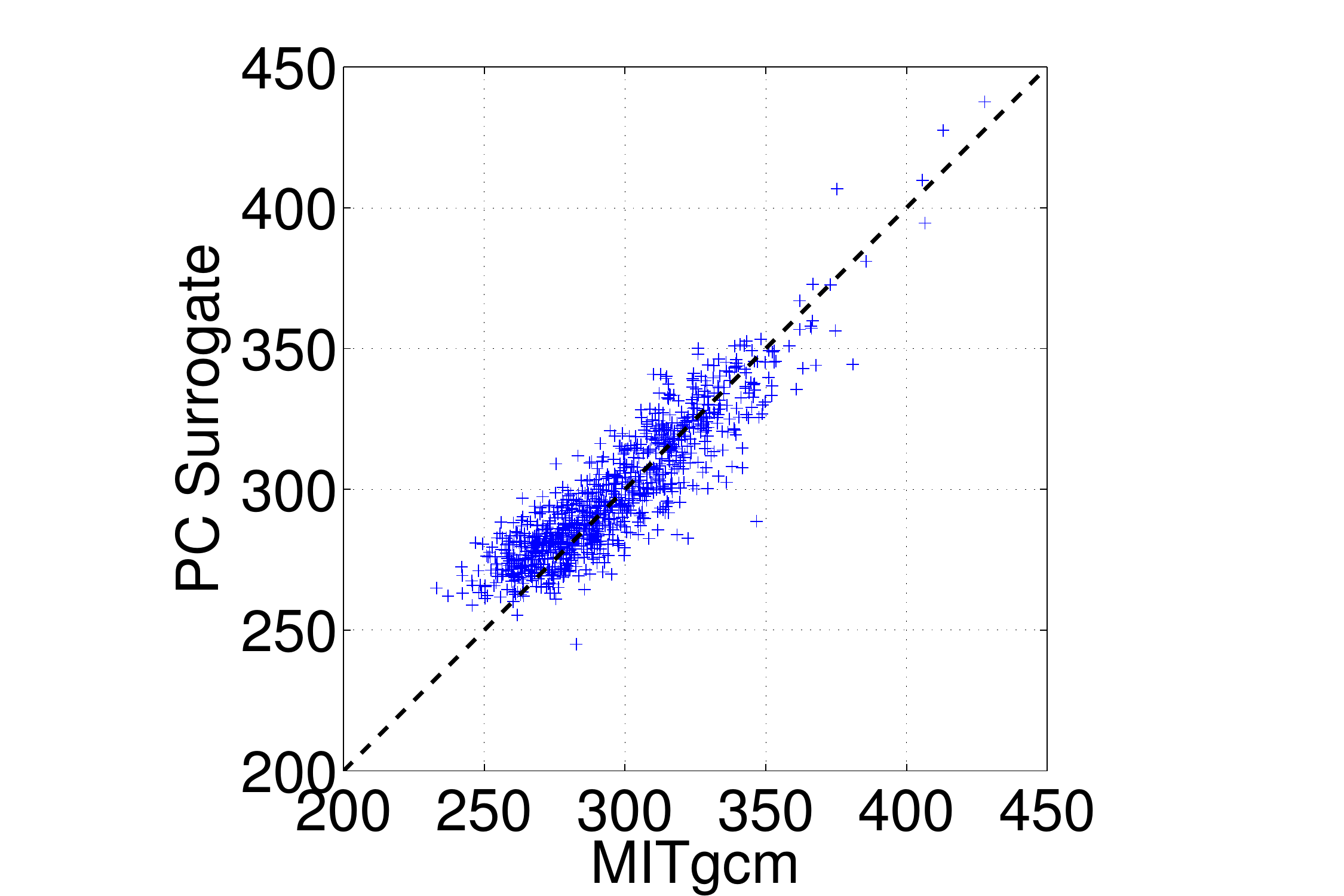}
\end{tabular}
\caption{Comparing test statistic ($E$) from \MIT model runs with their PC surrogate counterparts 
(Left) Superimposed (Right) Scatter plot. The shown cases correspond to the independent random sample
and PC is constructed using BPDN. The relative normalized error (NRE)
is also indicated.}  
\label{fig:diff_mvfsa}
\end{figure}

The PC expansion is calculated for the different sample points and is shown in Figure~\ref{fig:diff_mvfsa} (left)
versus the \MIT realization. The PC expansion appears to reproduce the mean of the deterministic model.
Figure~\ref{fig:error_bp} (right) shows the same but in a scatter plot where \MIT realizations are plotted against
their PC counterparts. The NRE is calculated and found to be $\sim 4.2\%$, which is also quite acceptable
indicating that BPDN is successful in producing a surrogate that is able to accurately 
simulate the QoI.

\subsection{PC sensitivity to ensemble size}
\label{sec:size}

\begin{figure}[h]
\centering
\begin{tabular}{clc}
\hspace{-12mm}
\includegraphics[width=0.55\textwidth]{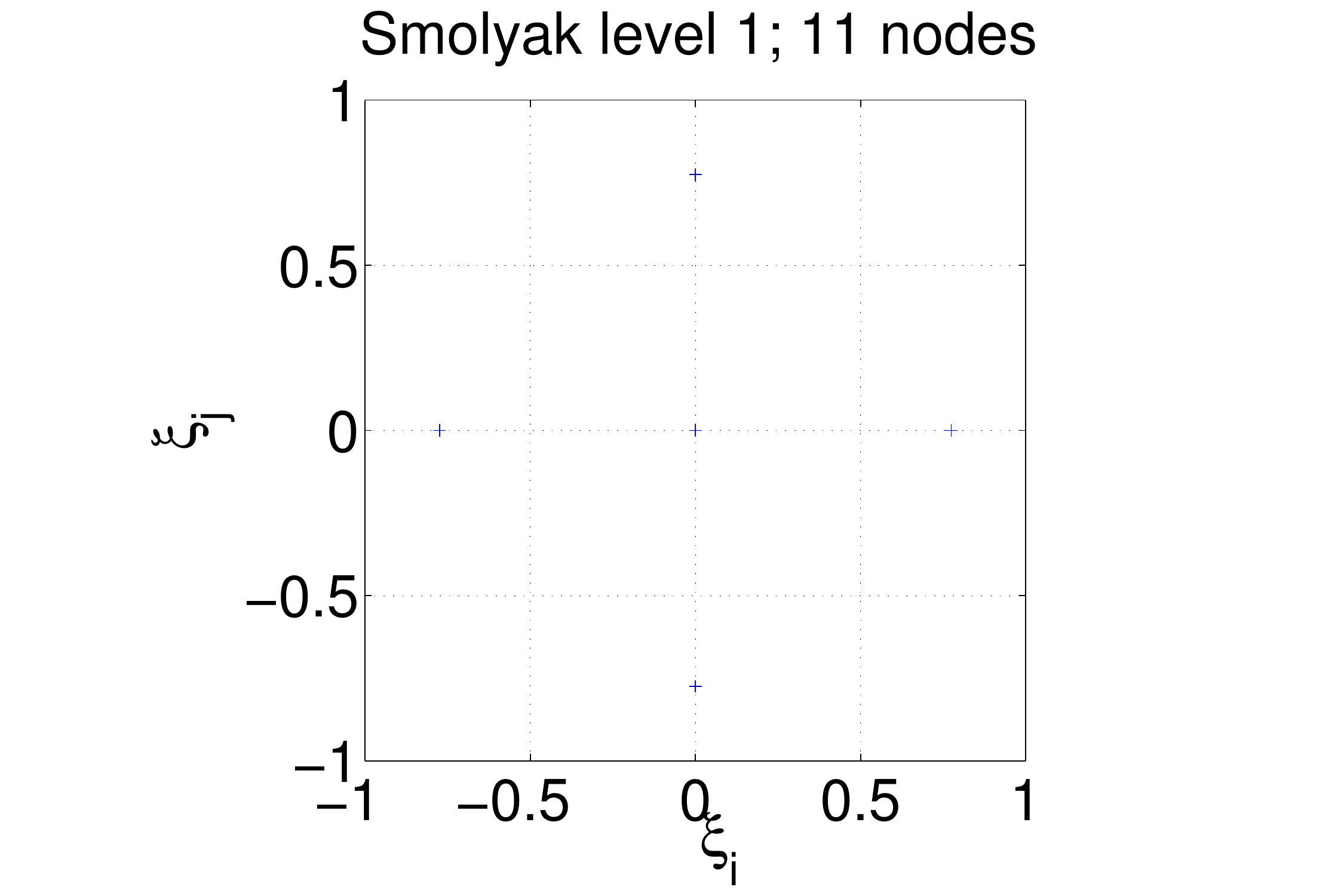} &
\hspace{-22mm}
\includegraphics[width=0.55\textwidth]{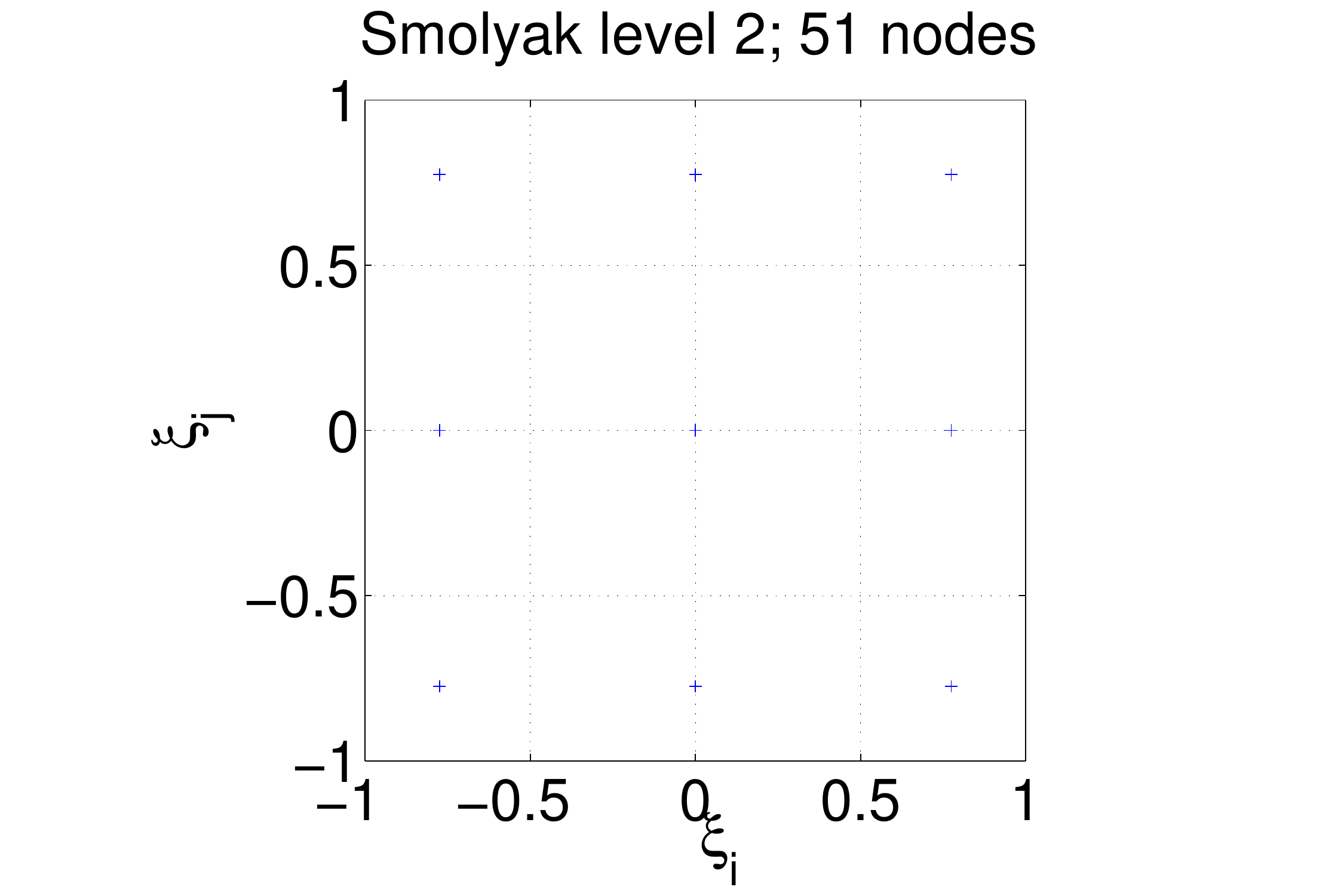} \\
\hspace{-12mm}
\includegraphics[width=0.55\textwidth]{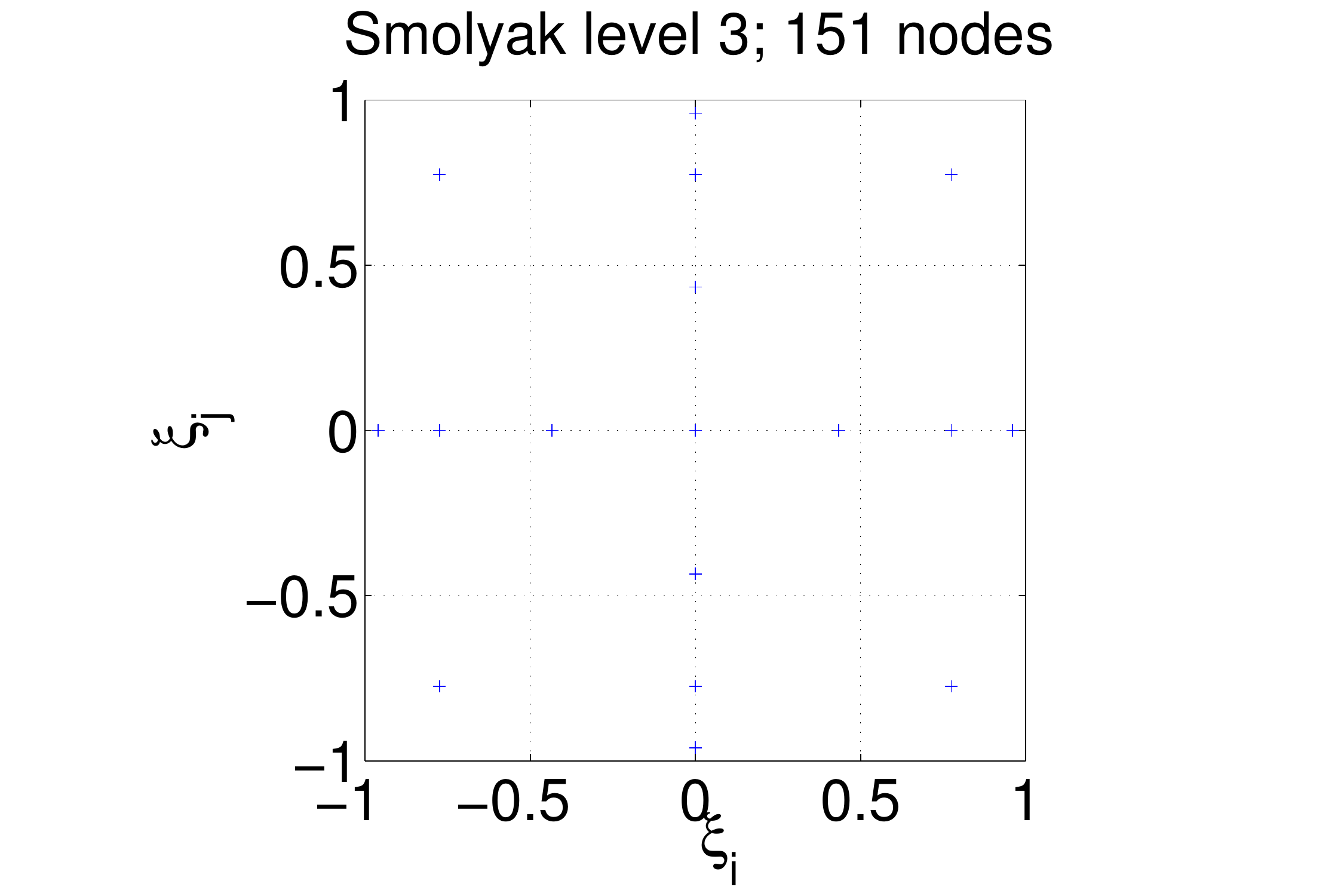} &
\hspace{-22mm}
\includegraphics[width=0.55\textwidth]{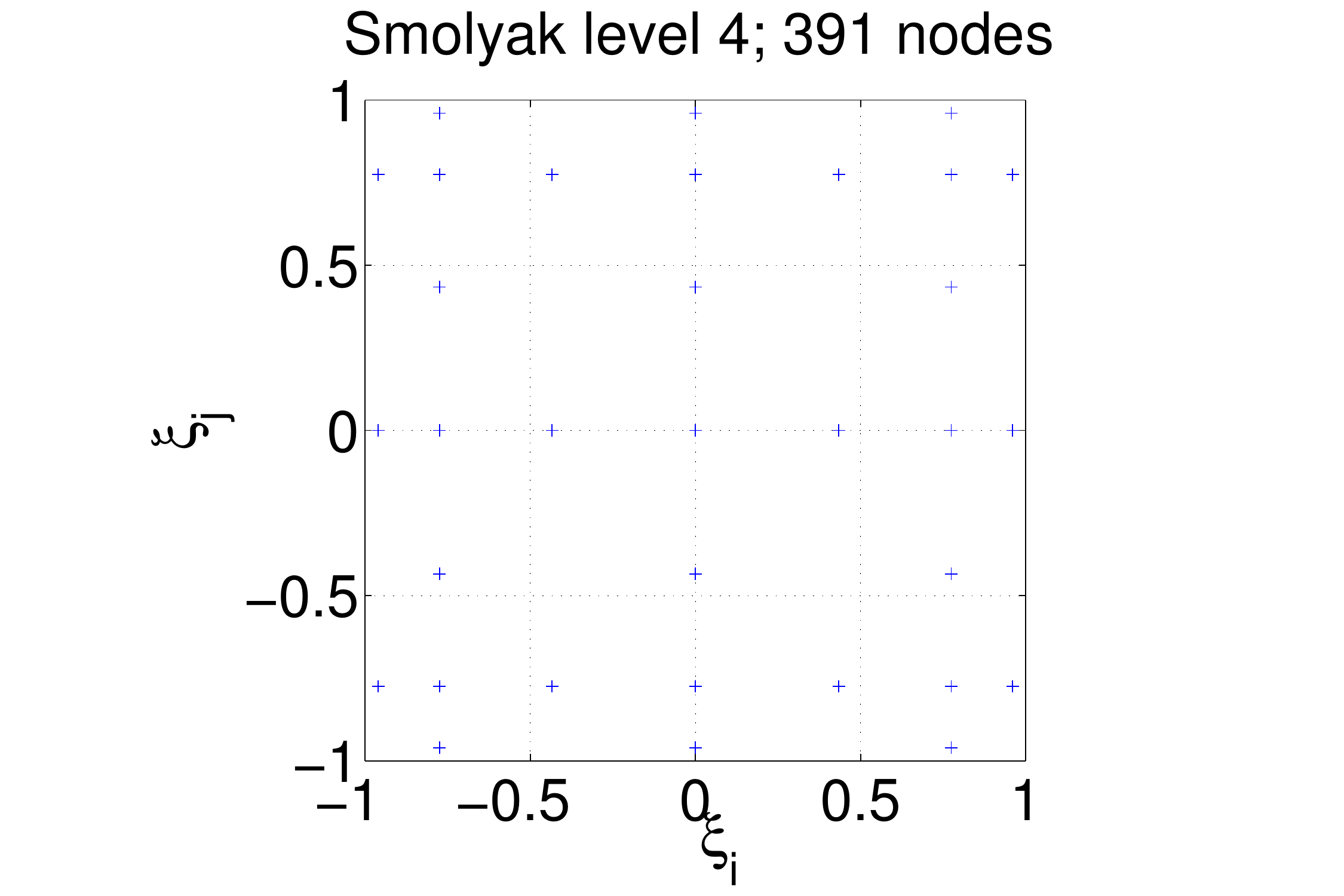}
\end{tabular}
\caption{2D Smolyak Sparse quadrature nodes corresponding to levels 1, 2, 3 and 4 in the canonical vector space.}  
\label{fig:slevels}
\end{figure}

Normally a number $N$ of random samples is required to compute the PC coefficients using BPDN;
this number is much less than the number of the unknowns i.e.\  the size of the PC expansion ($R+1 = 252$)
such that $N \ll R+1$. In the above results, we instead used an ensemble of $Q=903$ sparse quadrature \MIT model runs
corresponding to Smolyak level 5 (Figure~\ref{fig:s_quad})
to compute the PC coefficients using BPDN due to their availability (upon using NISP method) where in this case $N=Q \gg R+1$. 
Here, we explore the possibility of utilizing a small number 
of \MIT realizations to build a faithful PC expansion surrogate. To create a smaller ensemble and avoid running new expensive \MIT simulations, 
we consider lower levels of refinement
of Smolyak quadrature shown in Figure~\ref{fig:slevels} in the canonical vector
space for levels 1, 2, 3 and 4 with total number of nodes $11,~51,~151$ and $391$, respectively.
These levels are nested and therefore the nodes in each level is a subset of the higher levels. 
Thus, a small number of model runs $N$ can be extracted from the original $Q=903$ runs as per the Smolyak levels
and used to construct different PC models using BPDN.

\begin{figure}[h]
\centering
\begin{tabular}{clc}
\includegraphics[width=0.6\textwidth]{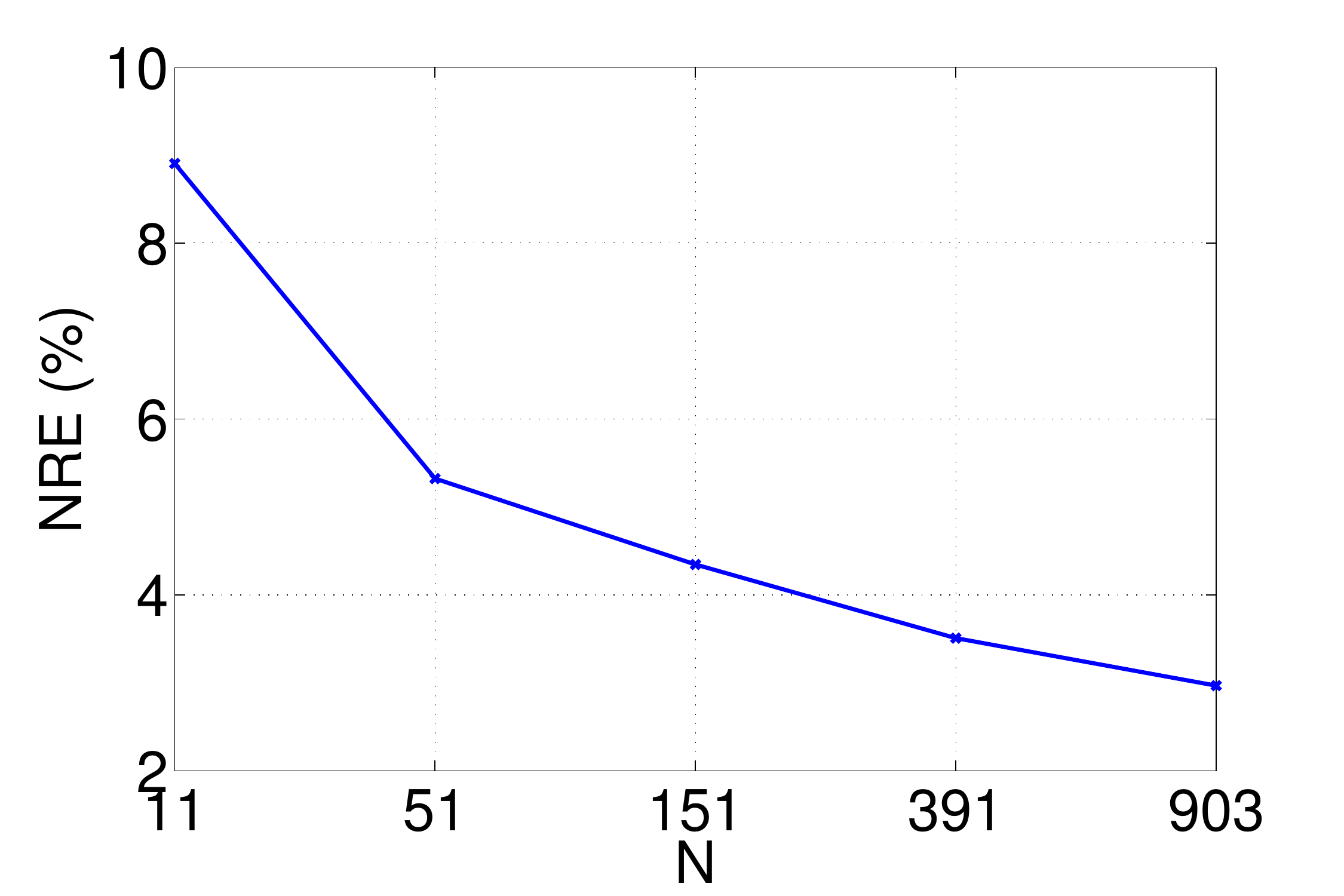}
\end{tabular}
\caption{Normalized relative error using different number of nodes corresponding to different Smolyak levels of refinement.}  
\label{fig:levels_error}
\end{figure}

In Figure~\ref{fig:levels_error}, we show the NRE of the original 903 Smolyak runs
computed using PC models built using quadrature nodes from the different Smolyak levels.
We observe that the error increases as $N$ decreases; however the increase is not significant.
In fact, even with level 2 (51 model runs) the NRE is around $5\%$.
We note that in all the results below we used the PC expansion model constructed using the level 5 Smolyak runs
with BPDN.

\subsection{Statistical moments and sensitivity analysis}
The identification of the inner product weight function
with the probability distribution of $\xxi$ simplifies the calculations of the statistical moments of $E$. 
Noting that since $\psi_0(\xxi)$ is a constant that is normalized so that 
$\left<\psi_0,\psi_0\right>=1$, the mean and variance of $E$ can be computed as:

\begin{equation}
 \mu_E = \int E \, \rho(\xxi) \, \mbox{d}\xxi \approx \left< E,\psi_0\right> = e_0,  
 \label{eq:mean}
\end{equation}

and
 
\begin{equation}
\sigma^2_E = \int (E-\mu_E)^2 \, \rho(\xxi) \, \mbox{d}\xxi \approx \sum_{k=1}^R e_k^2
 \label{eq:sigma}
\left<\psi_k,\psi_k\right>.
\end{equation}

PC representations also enable
efficient global sensitivity analysis that quantifies the
contributions of different random input parameters to the variance in the output.
This can be done by computing the so-called {\it total} 
sensitivity index $T_i$ that measures the contribution of
the $i^{th}$ random input to total model variability by
computing the fraction of the total variance due to all the terms in the
PC expansion that involve $\xi_i$
as follows:

\begin{equation} \label{eq:T-hard}
   T_i =
         \frac{\displaystyle
               \sum_{k \in K_i} e_k^2 \ave{\psi_k^2}}
              {\displaystyle\sum_{k = 1}^E e_k^2 \ave{\psi_k^2}}, \quad K_i = \left\{ k \in \{1, \ldots, R\} : \vec{\alpha}^k_i > 0 \right\},
\end{equation}
where $\vec{\alpha}^k$ is the multi-index associated with the $k^{th}$ term in the
PC expansion~\citep{LeMaitreKnio2010,Crestaux,Sudret}.

The PC expansion is used to calculate statistical moments
of $E$ due to the uncertainty in the input parameters as indicated above.
The mean of the test statistic was found to be $325.62$ with a standard deviation of $30.97$. 
The sensitivities were also computed and summarized in Table~\ref{tab:sens}. It is notable that the test statistic
is most sensitive to $Ri_c$ and then to $Ri_g$ and $\phi_{m,unst}$ as clearly reflected in the corresponding total sensitivities, $T_1-T_3$ respectively.

\subsection{Response surfaces}

\begin{figure}[h]
\centering
\begin{tabular}{clc}
\includegraphics[width=0.6\textwidth]{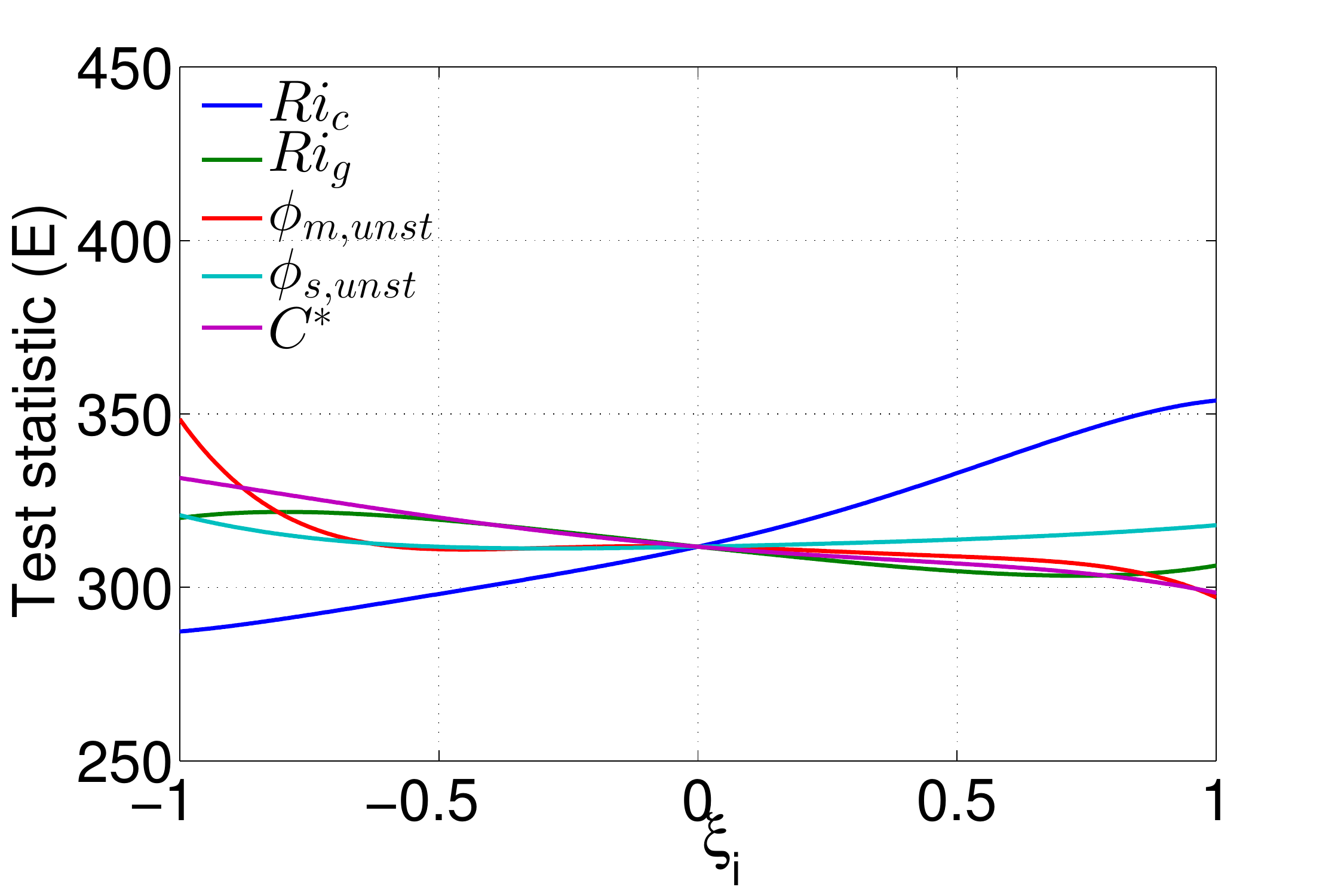} &
\end{tabular}
\caption{Response curves of test statistic ($E$) function of different parameters (in canonical space). 
For each curve, the other four parameters are set to $\xi_j=0$.}  
\label{fig:response_1d}
\end{figure}

\begin{figure}[h]
\centering
\begin{tabular}{clc}
\includegraphics[width=0.55\textwidth]{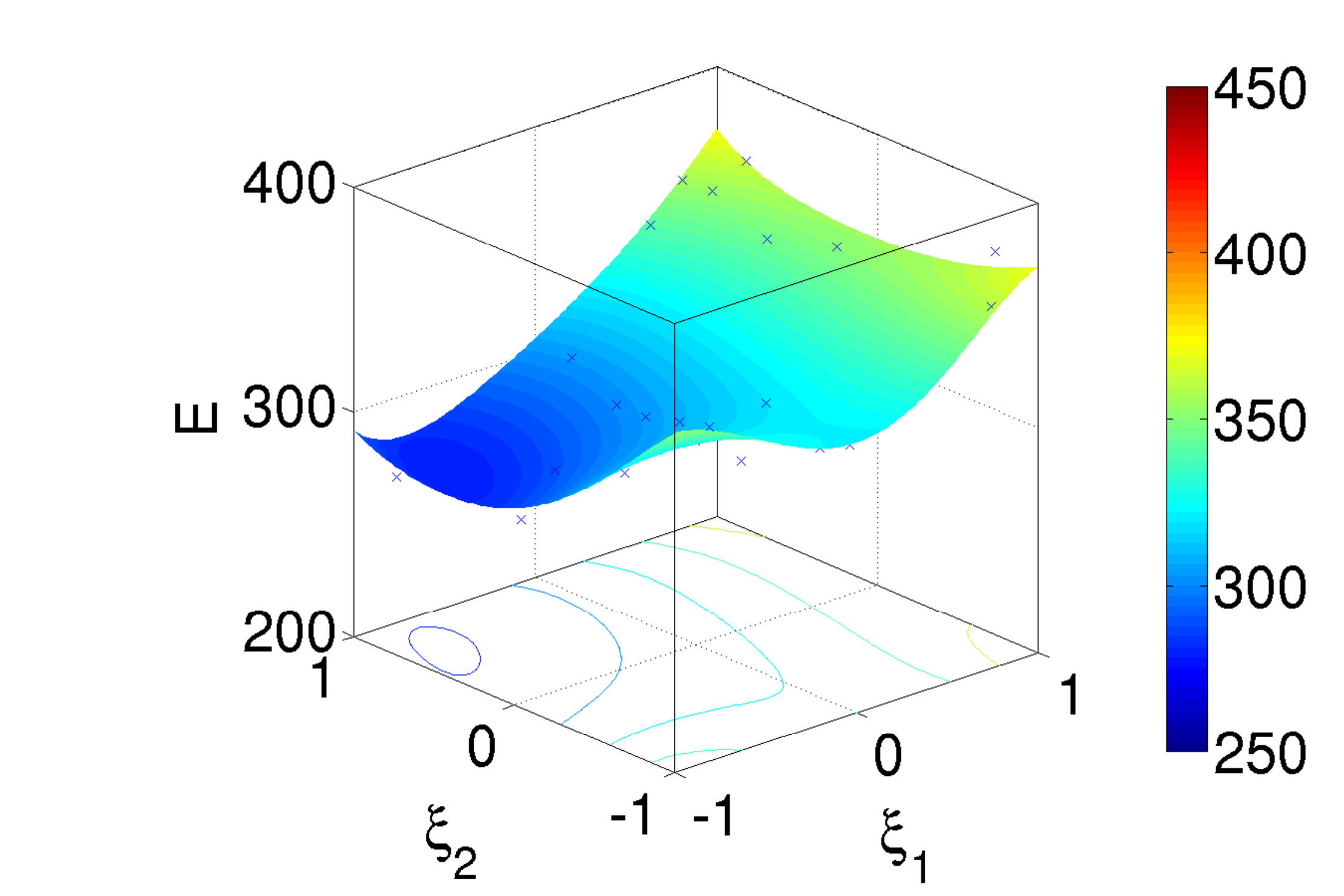} \\
%\hspace{-12mm}
\includegraphics[width=0.55\textwidth]{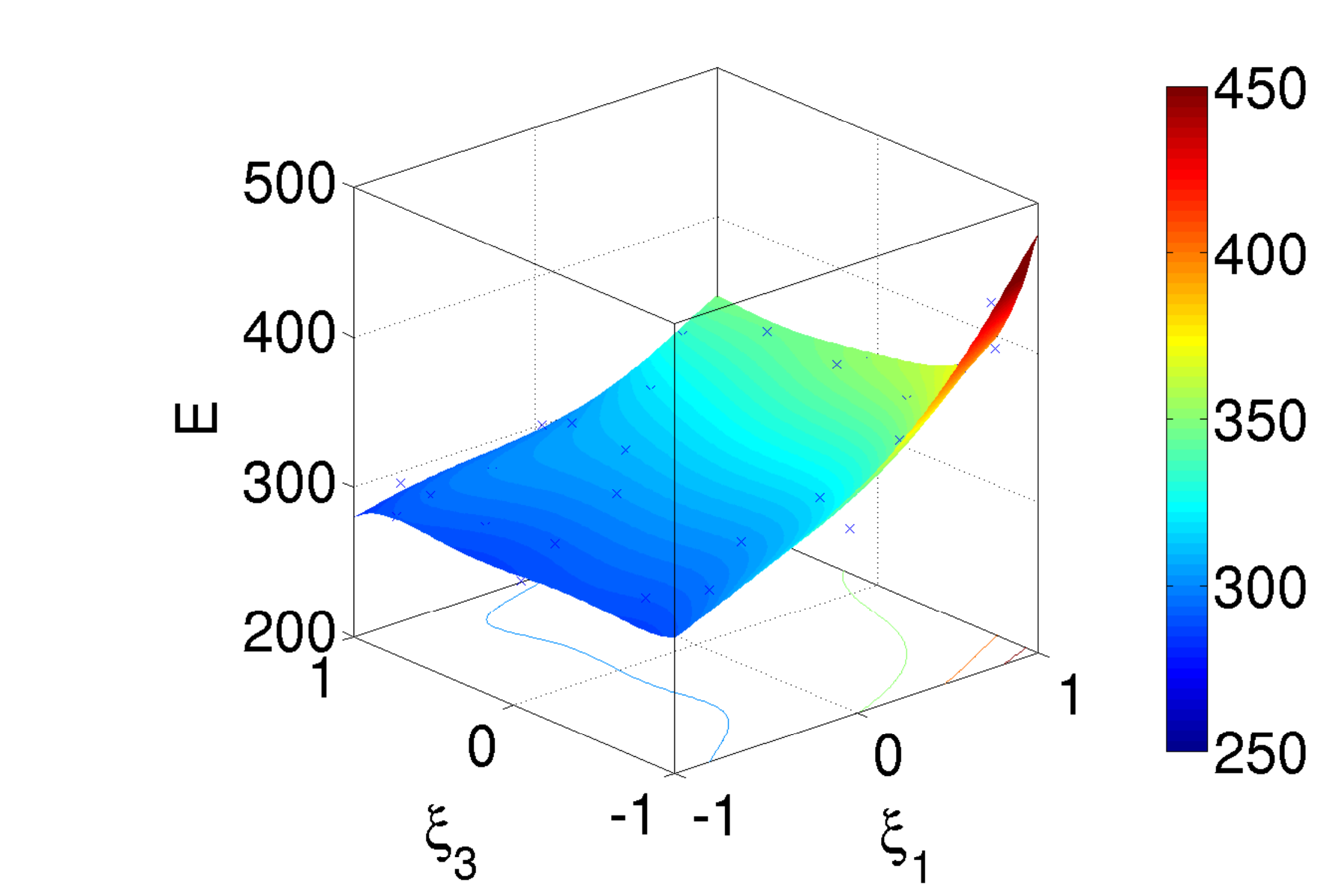}
\end{tabular}
\caption{Response surfaces of test statistic ($E$) function of (top) $Ri_c$ and $Ri_g$, (bottom) $Ri_c$ versus $\phi_{m,unst}$
(in canonical space). For each surface, the other three parameters are set to $\xi_j=0$.}  
\label{fig:response_2d}
\end{figure}

In addition to the moments and sensitivities, the PC surrogate 
can be used to construct a response surface for $E$ as a function of
the uncertain input parameters. To this end, we sample the PC 
surrogate for different values of the canonical vector of random 
variables $\xxi$ within the prior range $ [-1 , +1 ] $ as 
illustrated in Figure~\ref{fig:response_1d}. The different plots 
represent the response curves function of single uncertain parameter while 
the other parameters are set to $\xxi=0$. 
The plots show the strong dependance of
$E$ on $Ri_c$, consistent with the sensitivity results shown earlier. Similarly 
$E$ shows some variations function of $Ri_g$ and $\phi_{m,unst}$ but with a relatively less dependance compared to $Ri_c$. 
The remaining curves exhibit lines with a small slope, suggesting 
that $E$ depends only mildly on $\phi_{s,unst}$ and $C^*$.
We also show the 2D response surfaces for $Ri_c$ versus $Ri_g$  in Figure~\ref{fig:response_2d}~(top),
and $Ri_c$ versus $\phi_{m,unst}$ in Figure~\ref{fig:response_2d}~(bottom) (the other parameters are set to $\xxi=0$.).
The quadrature sample $E_q$ are shown on top of the surfaces for comparison.
\section{KPP Inference}
\label{sec:mcmc}
Bayesian inference is now used to estimate the KPP parameters
such that the likelihood based on the test statistic is maximized.
To this end, a random-walk MCMC method is implemented to sample 
the posterior distributions~\citep{Gareth2009,Haario2001} (Equations~\ref{eq:post3} and \ref{eq:postS}) and consequently 
update the KPP parameters' distributions. This sampling requires tens of thousands of 
\MIT runs that are prohibitively expensive as each MCMC 
sample requires an independent \MIT realization. Instead,
the surrogate model created using PC expansions provides a computationally
efficient alternative that requires only evaluating the PC expansion
for different values of the canonical vector of random variables $\xxi$.

\begin{figure}[ht]
\centering
\begin{tabular}{clc}
{\includegraphics[width=0.45\textwidth]{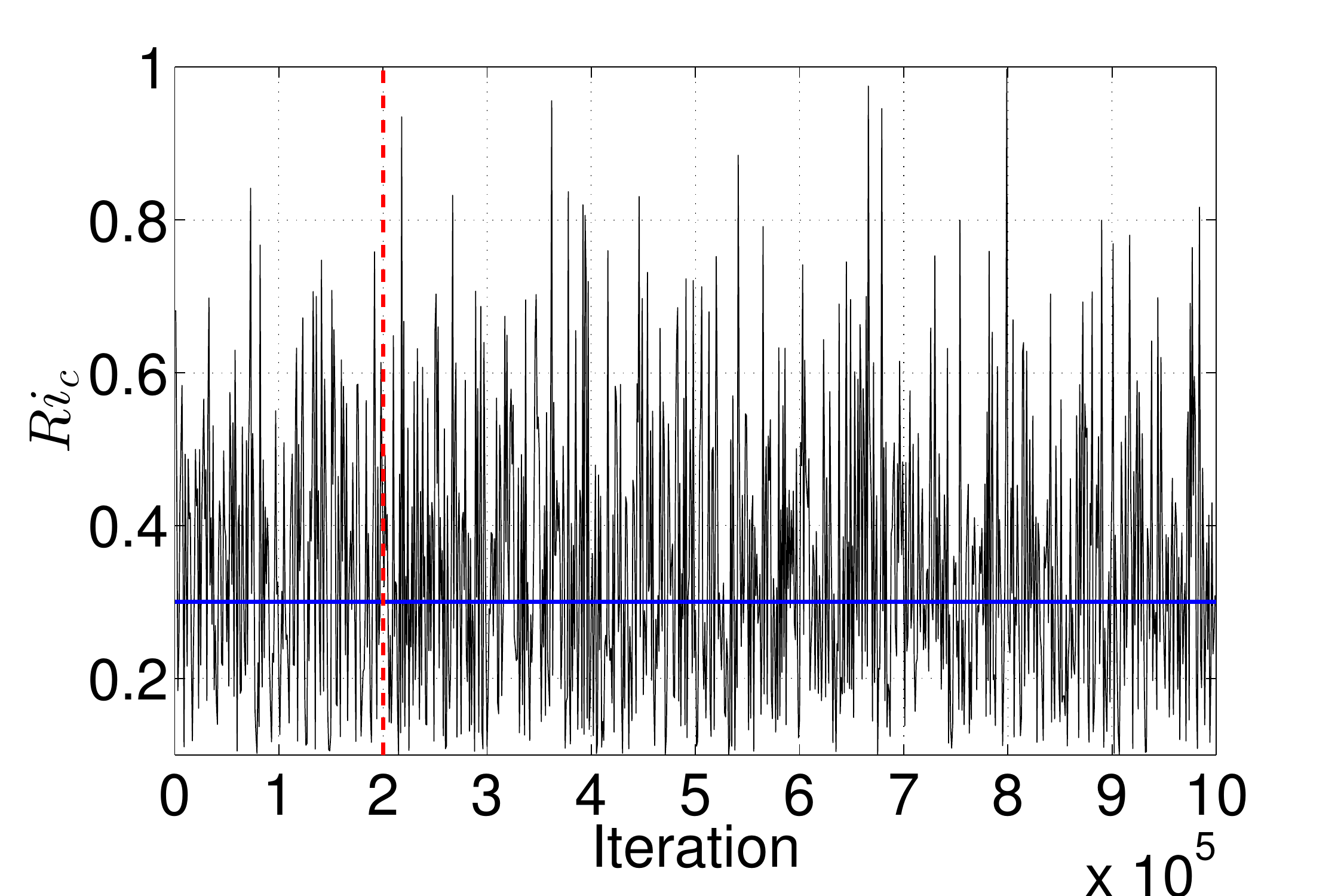}}\,
{\includegraphics[width=0.45\textwidth]{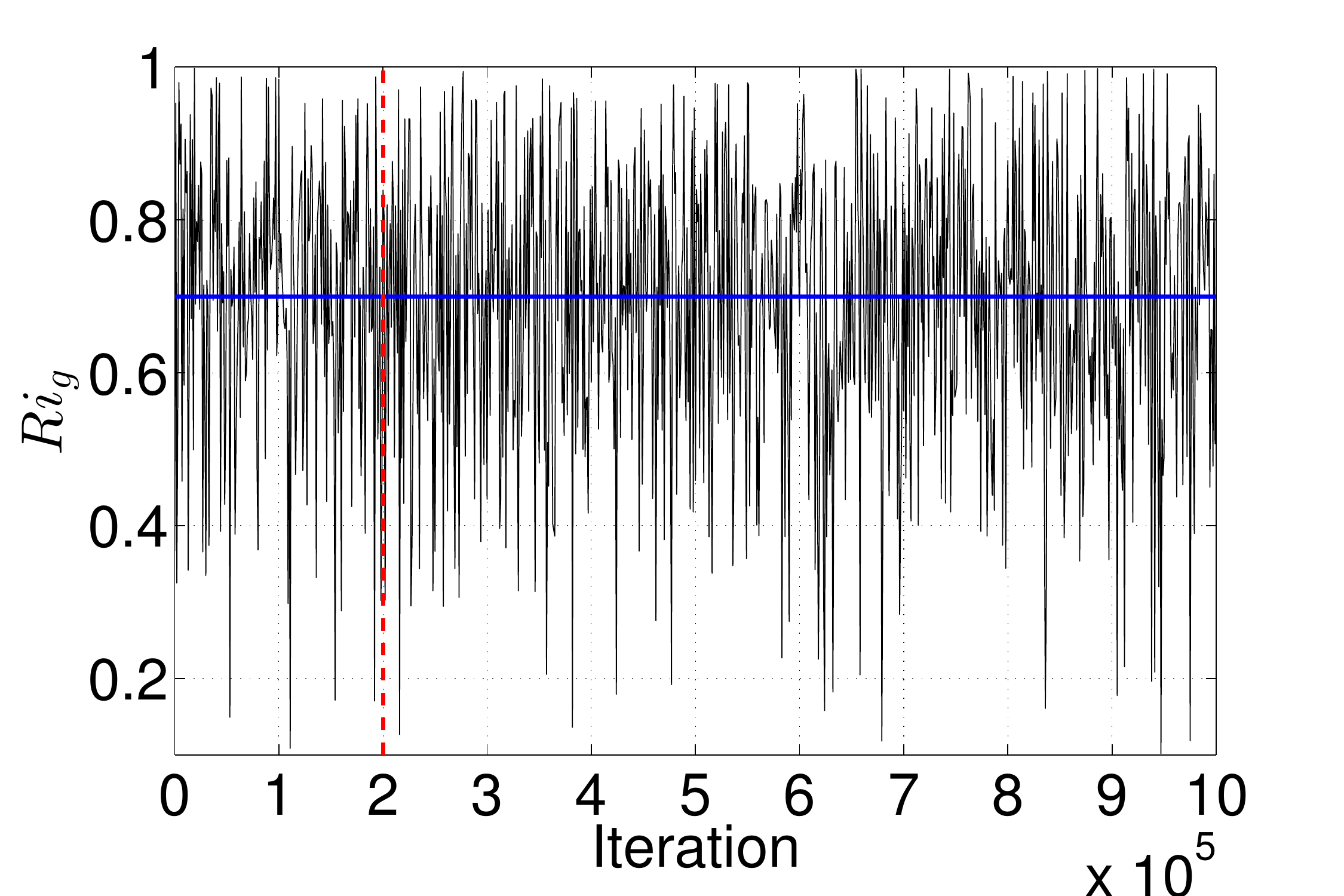}}\,
\\
{\includegraphics[width=0.45\textwidth]{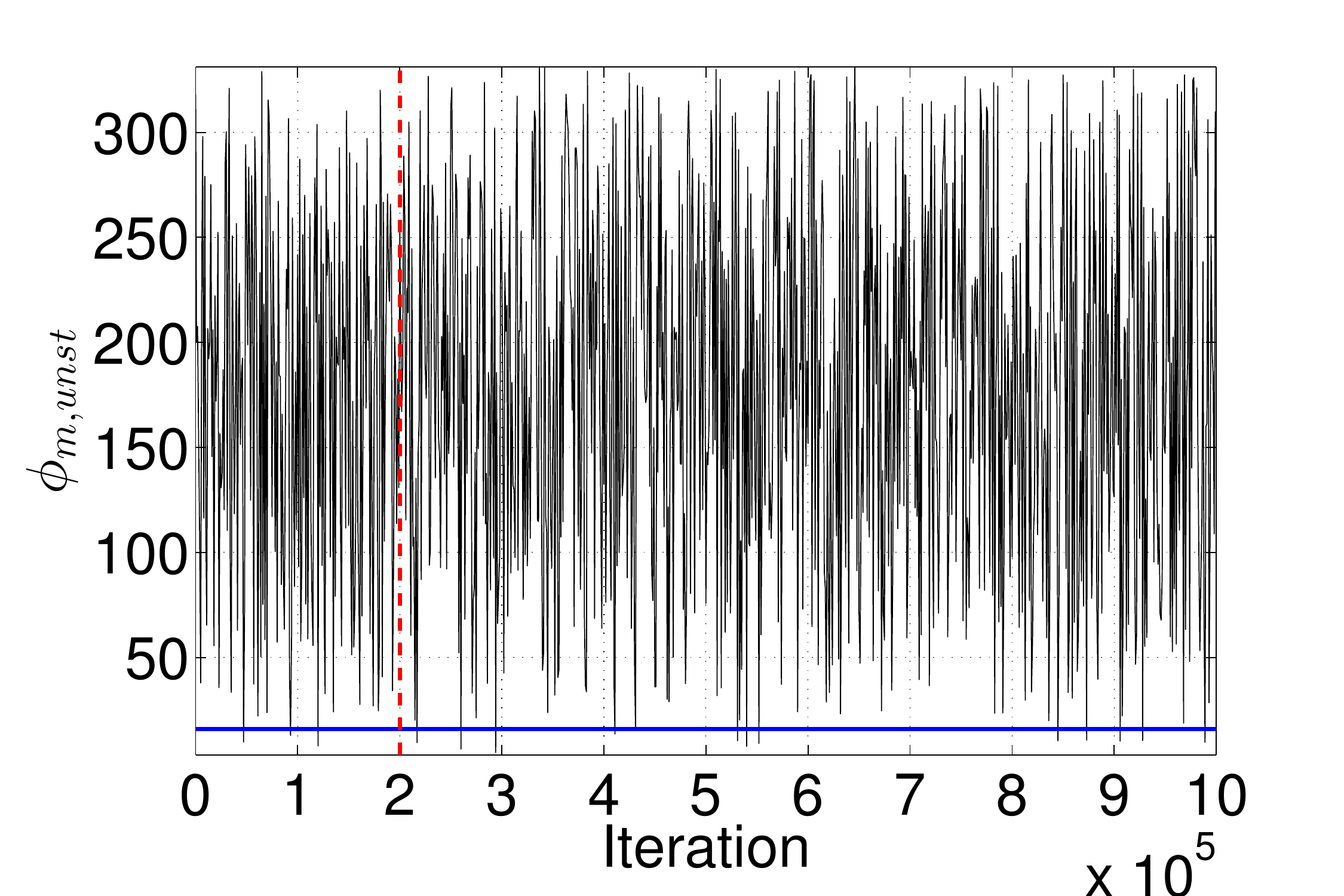}} 
{\includegraphics[width=0.45\textwidth]{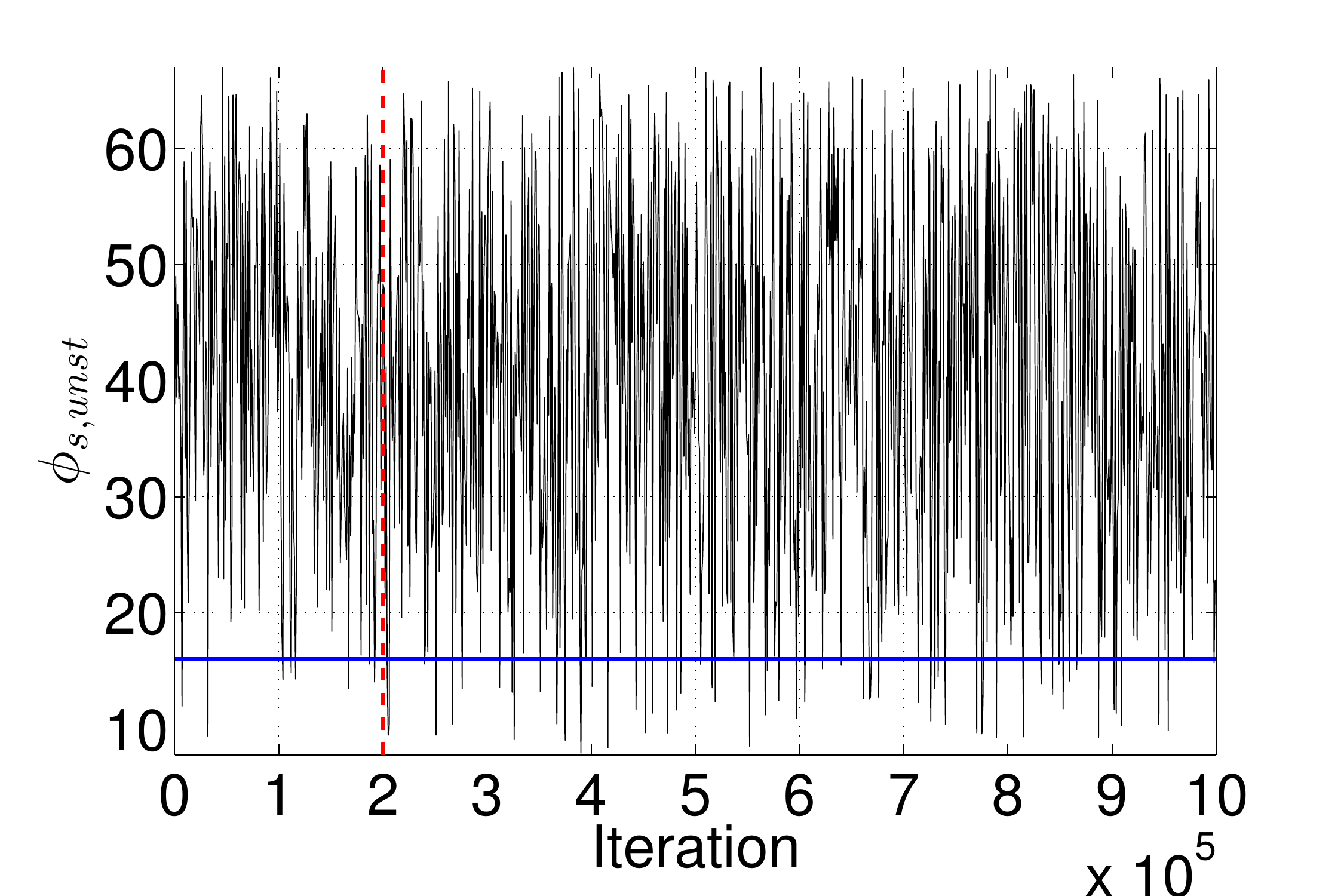}}\,
\\
{\includegraphics[width=0.45\textwidth]{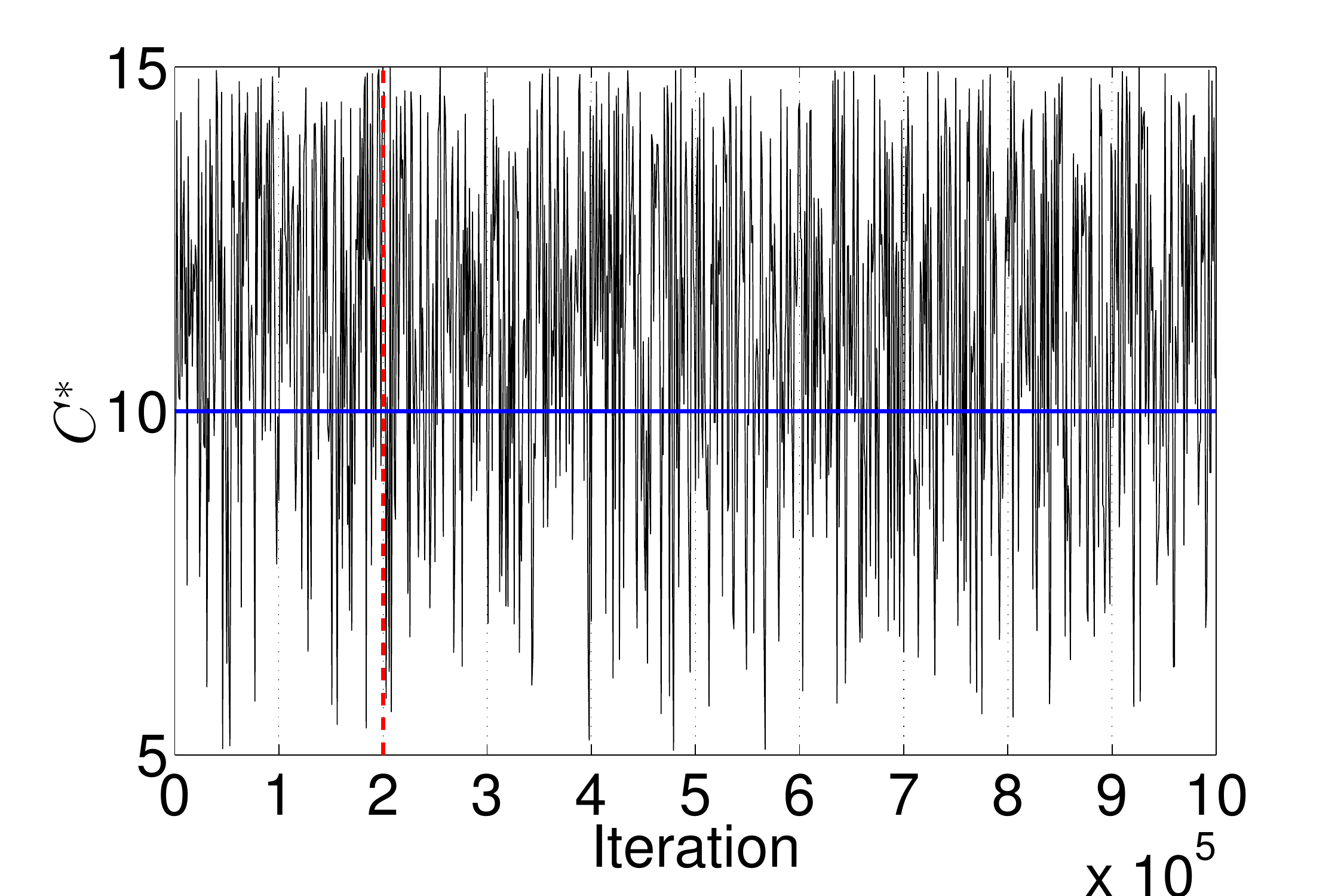}}
{\includegraphics[width=0.45\textwidth]{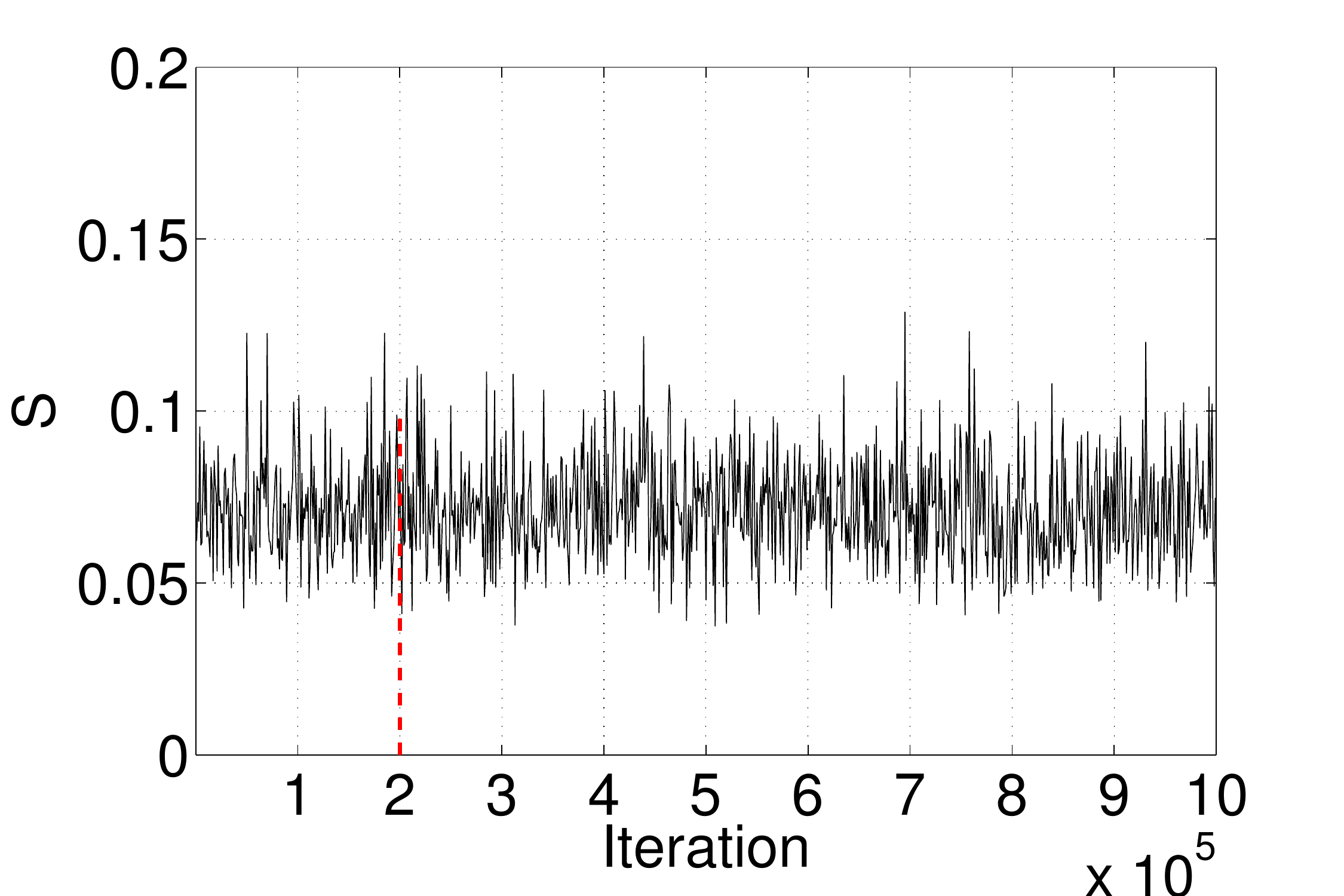}}
\end{tabular}
\caption{Chains of parameters using MCMC sample from PC surrogate constructed. \MIT default values are indicated as horizontal lines. }  
\label{fig:chains}
\end{figure}

\begin{figure}[ht]
\centering
\begin{tabular}{clc}
\includegraphics[width=0.6\textwidth]{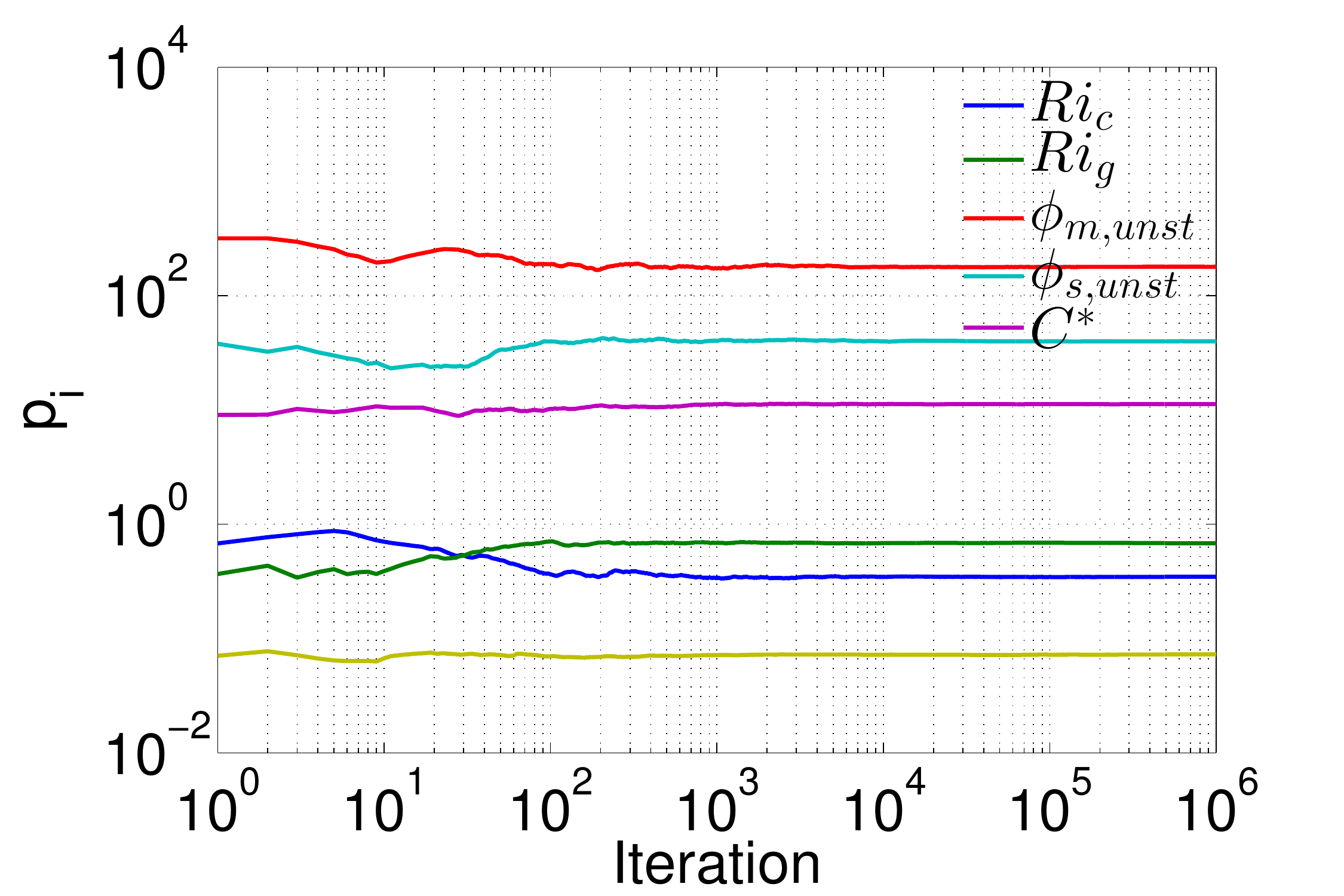}
\end{tabular}
\caption{Running mean of the different chains. 
%using (Left) sparse quadrature (Right) a combination 
%of sparse quadrature and $89$ runs.
}  
\label{fig:rmean}
\end{figure}

All of the results presented here are based on $10^6$ MCMC samples; we find 
negligible changes in the obtained posteriors of the KPP parameters with further iterations.
Figure \ref{fig:chains} shows the sample chains for the input parameters and scaling parameter $S$ for different 
iterations of the MCMC algorithm. The different panels suggest well-mixed chains for all input parameters
where the chains of all parameters appear to be concentrated in an area of
the parameter prior range. The running mean plotted in Figure~\ref{fig:rmean}
is an indication of the convergence of the MCMC.
For $Ri_c$, the chain appears to be concentrated in the  lower end of the 
parameter range with values between $0.1$ and $0.4$
while for $Ri_g$, the chain appears to be concentrated in the  upper end of the parameter range with values between $0.6$ and $1.0$.
These values align well with what is commonly considered physically relevant for such parameters shown as horizontal lines in each panel.
For $\phi_{m,unst}$ and $\phi_{s,unst}$, the chains appear to cover almost all the prior range.
This is an indication of a non-informative posterior due to lack of data to infer those parameters.
As for $C^*$, the chain is concentrated in the  upper end of the parameter range between $10$ and $15$ as shown.
Finally, the hyper-parameter $S$ is well-mixed with values ranging between $0.03$ and $0.12$.

\begin{figure}[h]
\centering
\begin{tabular}{clc}
{\includegraphics[width=0.45\textwidth]{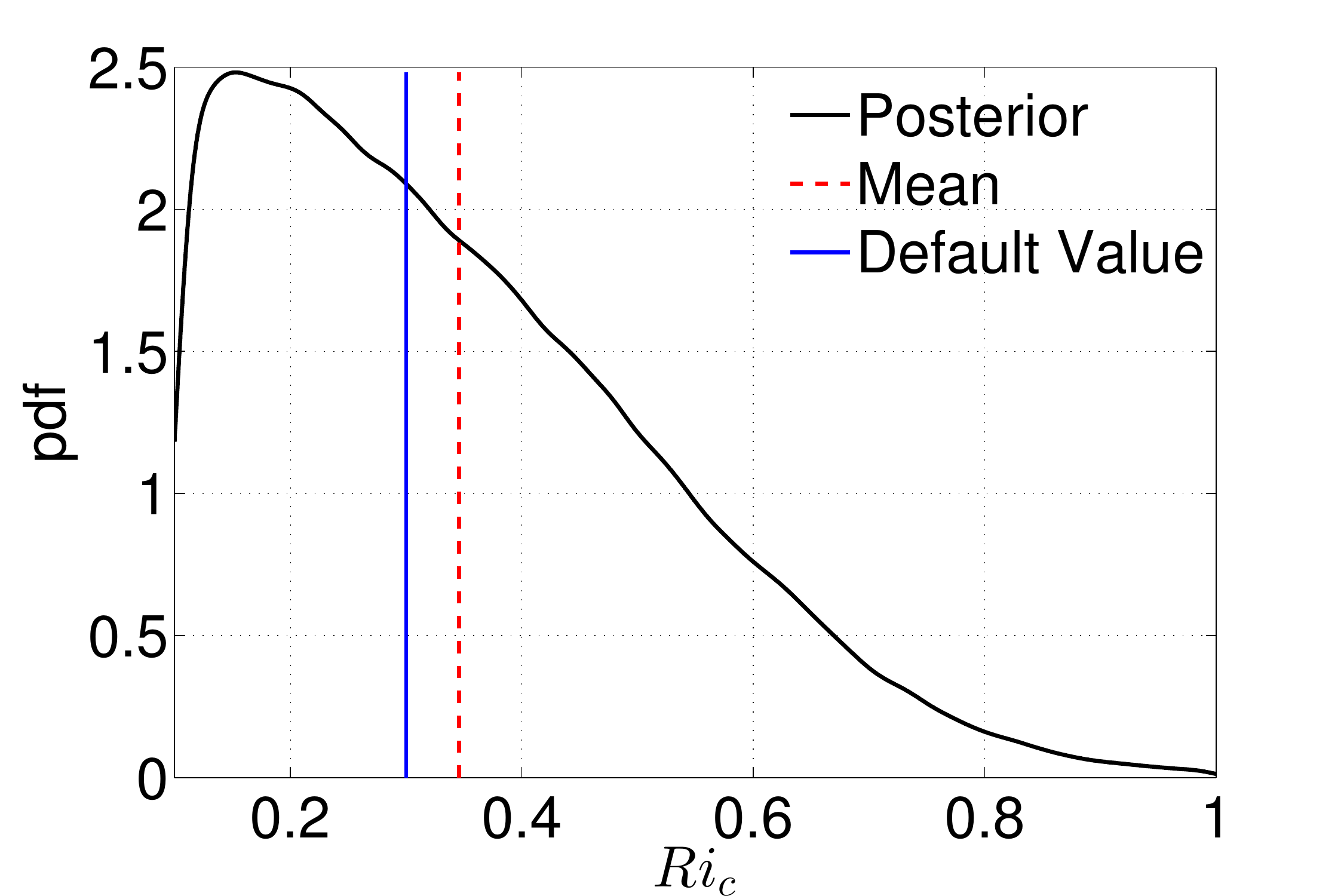}}\,
{\includegraphics[width=0.45\textwidth]{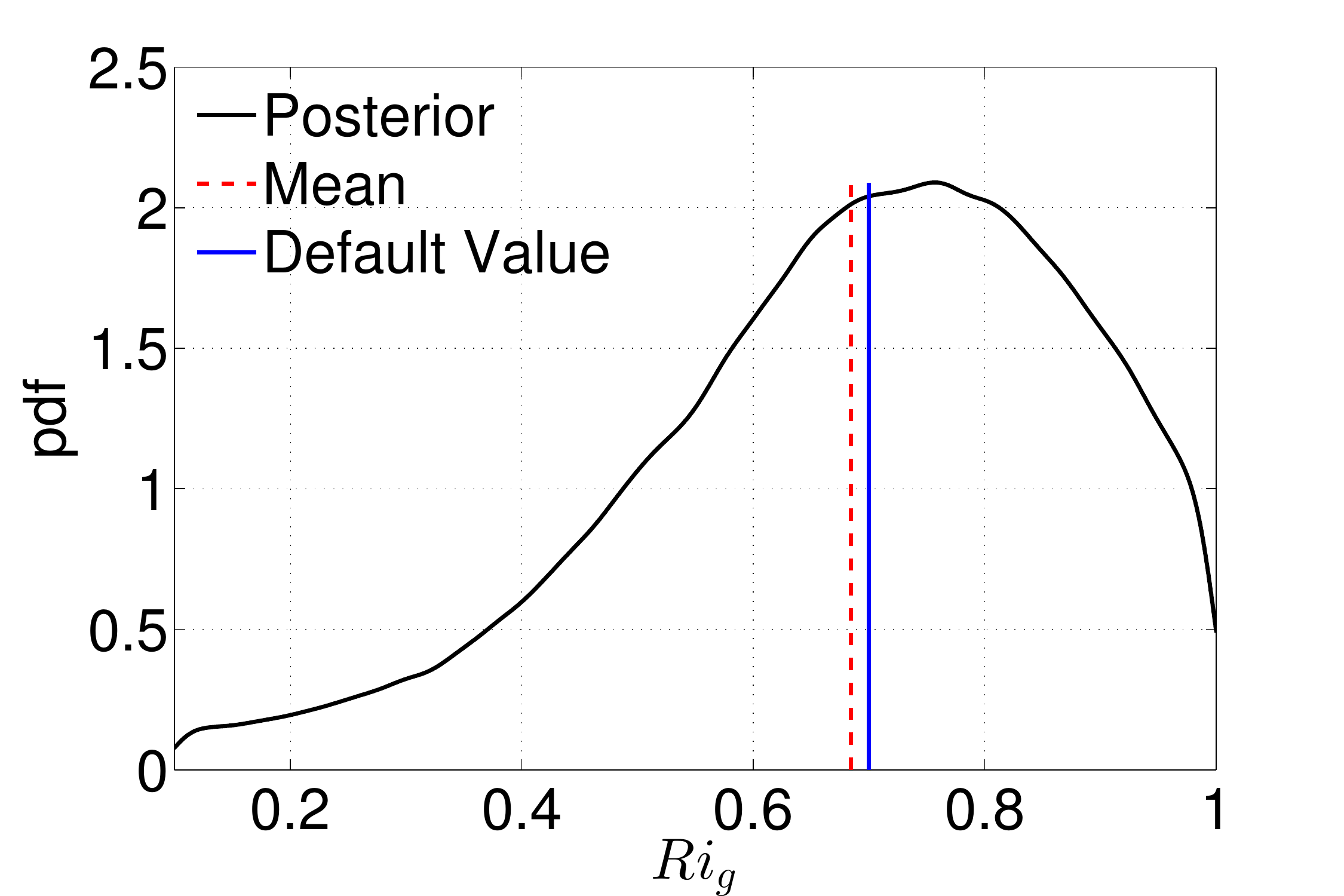}}\,
\\
{\includegraphics[width=0.45\textwidth]{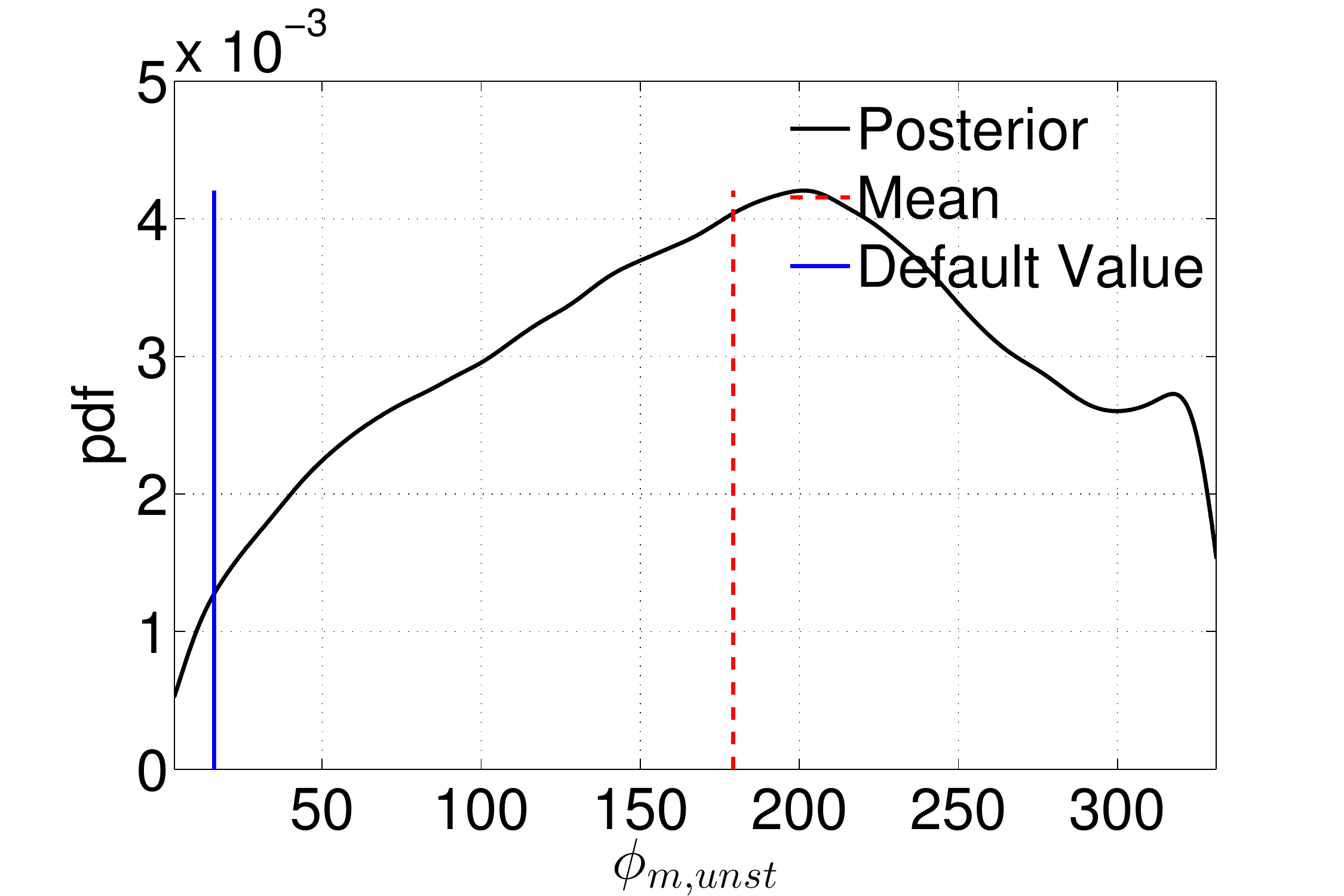}} 
{\includegraphics[width=0.45\textwidth]{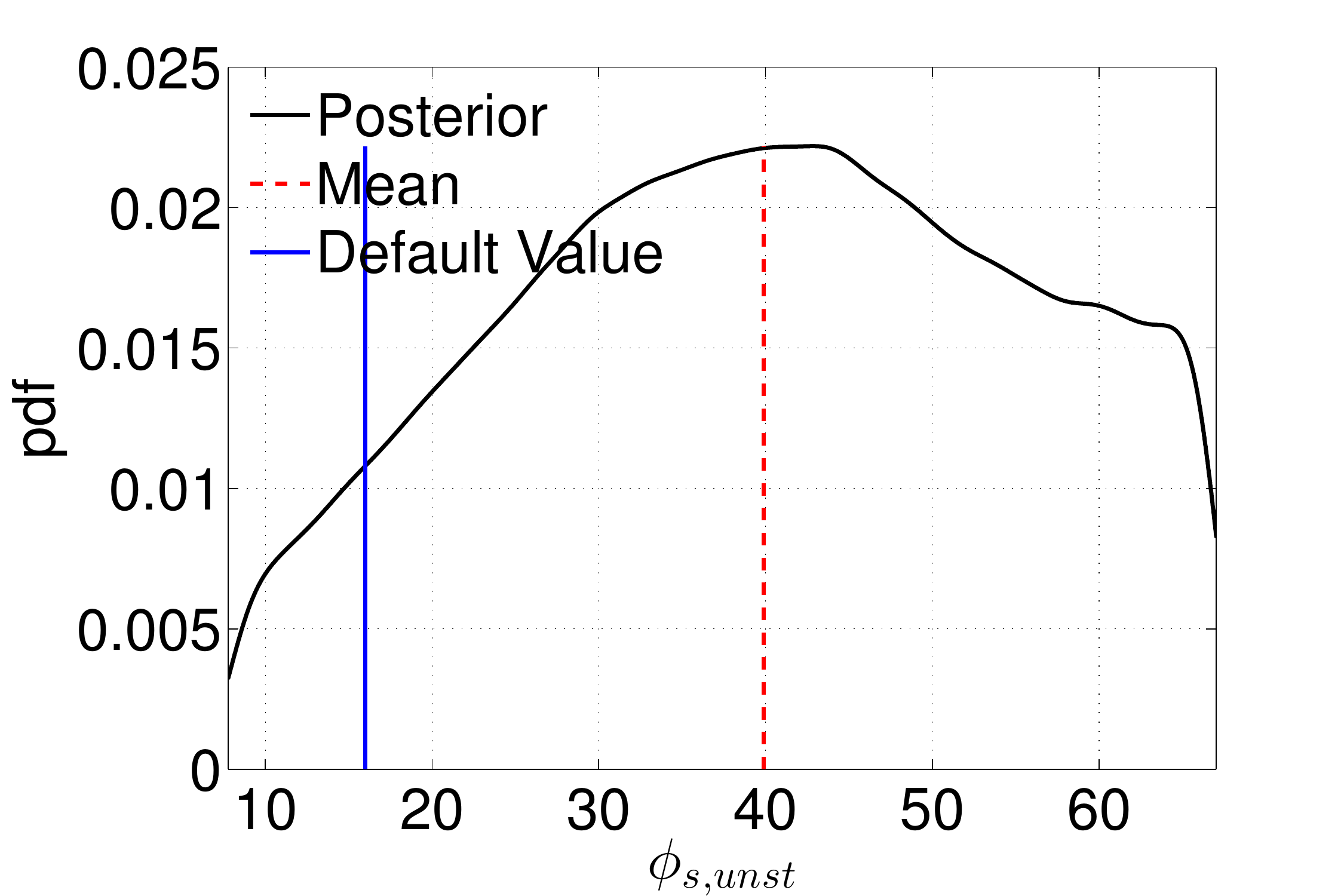}}\,
\\
{\includegraphics[width=0.45\textwidth]{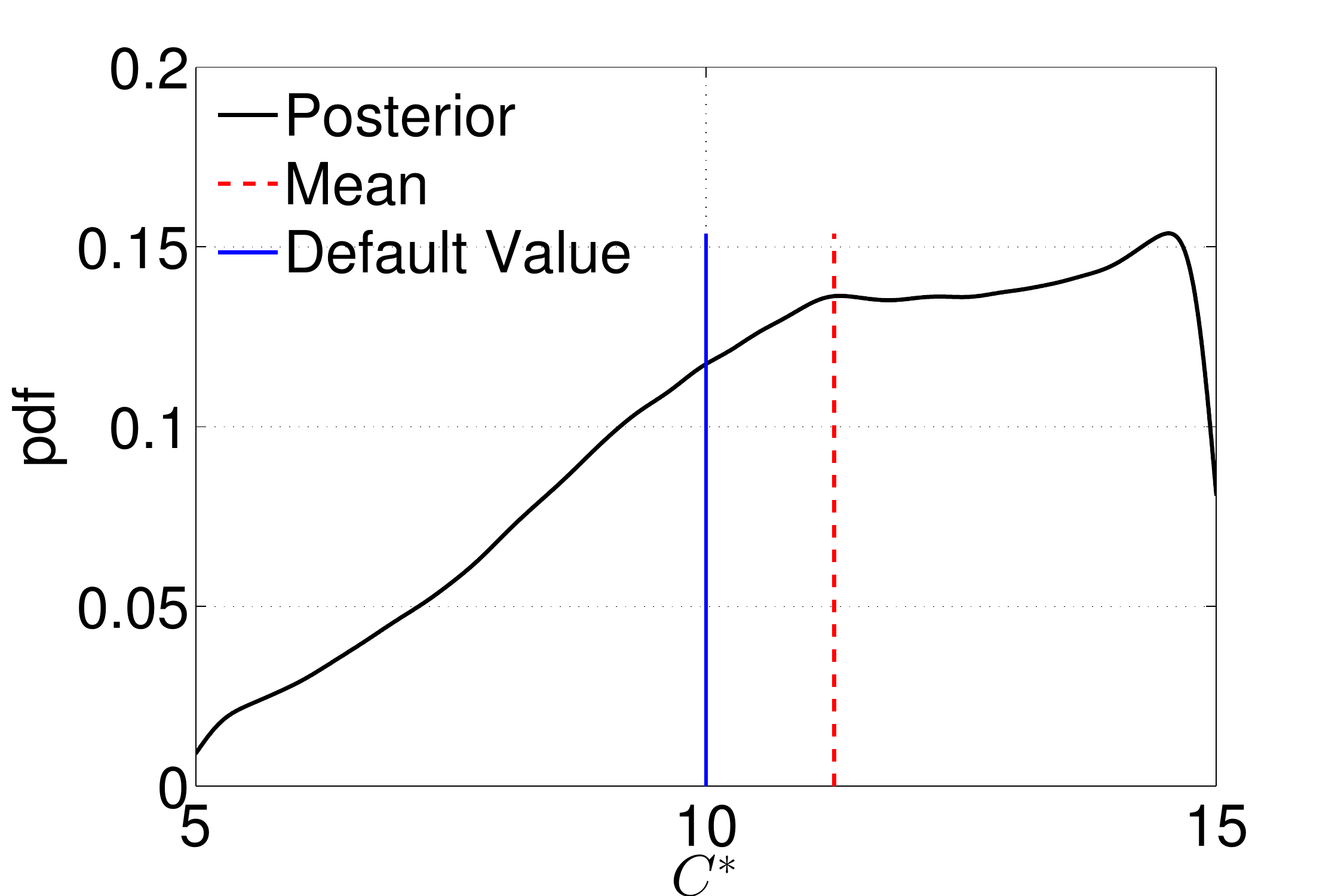}}
{\includegraphics[width=0.45\textwidth]{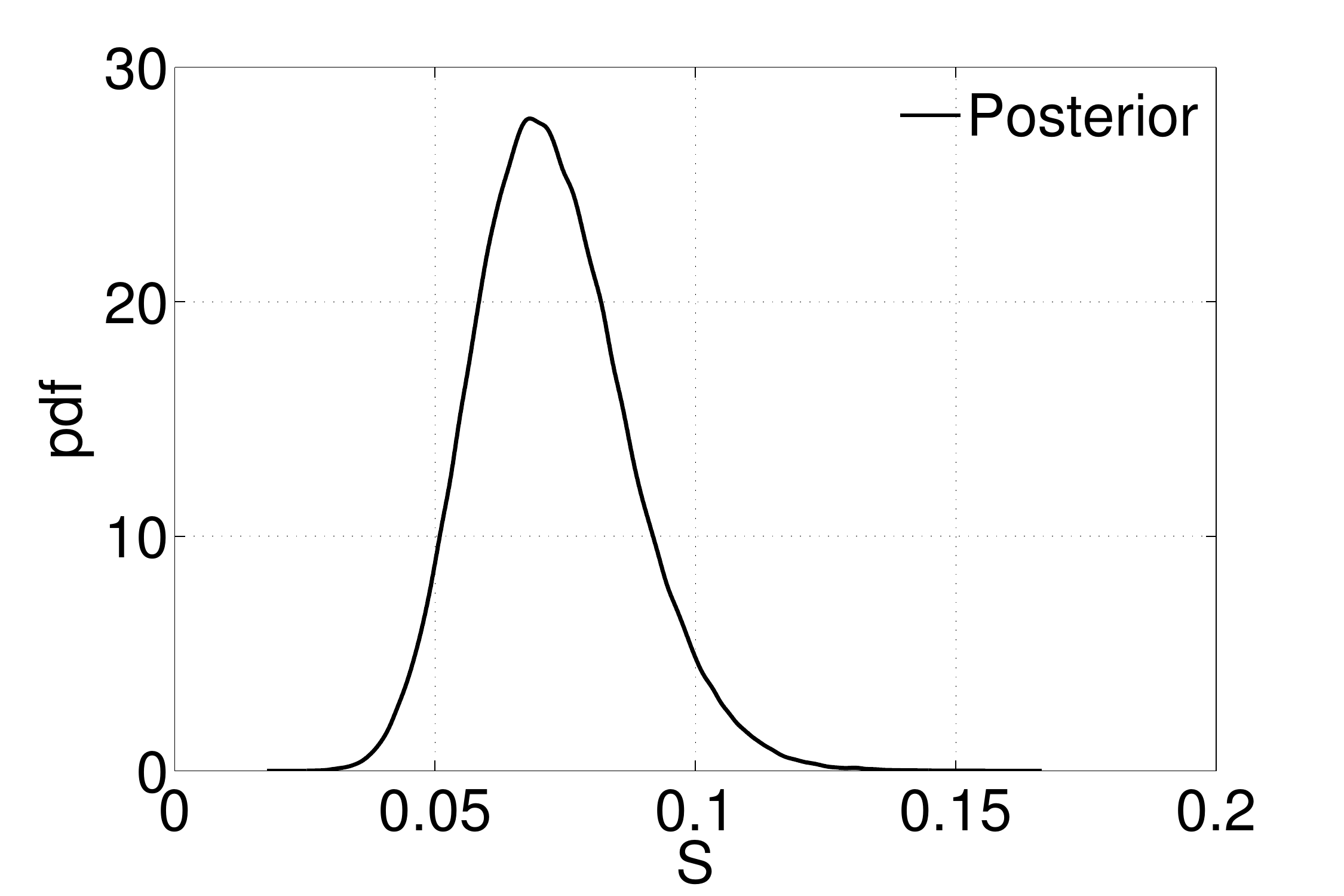}}
\end{tabular}
\caption{\emph{pdfs} of parameters using KDE % compared with pdfs from sarah's study (PC constructed using sparse quadrature)
.}  
\label{fig:posteriors}

\end{figure}

Next, the computed MCMC chains are used to determine the marginalized posterior 
distributions using Kernel Density Estimation (KDE)~\citep{Parzen1962,Silverman1986}.  
The resulting marginalized posterior \emph{pdfs} of the KPP are shown in Figure~\ref{fig:posteriors} 
in addition to the scaling parameter $S$. Note that the first $2\times 10^5$ iterations, 
associated with the burn-in period, were discarded.  As expected from the chains shown in 
Figure~\ref{fig:chains}, the posterior \emph{pdf} of $Ri_c$ appears
to be skewed to the right with an extended tail towards the higher $Ri_c$ values.
The pdf exhibits a well-defined peak but due to the skewness, we instead report the mean estimate calculated  as $\sim 0.23$;
it is close to the default \MIT value (shown as a vertical line on the same pdf). For $Ri_g$, we observe a posterior that has a well 
defined peak of $0.68$; it matches the \MIT default value. 
In contrast, the marginalized posterior \emph{pdfs} of $\phi_{m,unst}$ and $\phi_{s,unst}$ appear to be fairly flat, 
and similar to the uniform prior; an indication that the observed data were not informative to refine our prior knowledge of those variables. 
Regarding $C^*$, the pdf indicates that the mean value is in the upper range of the prior.
Finally the scaling parameter $S$ exhibits a Gaussian like shape contrasted to the prior gamma distribution.

The posterior distributions are consistent with results of previous efforts to calibrate the KPP model~\citep{Large:94,lar99}.  They show that the default critical bulk and gradient Richardson numbers and the non-local convective parameter $C^*$ are within the range of acceptable parameter settings.  
In the absence of a detailed analysis of the role of $\phi_{m,unst}$ and $\phi_{s,unst}$ in our calibration, we can remark on possible explanations for their broad posterior distributions.  It must be the case that either the eddy diffusivity and viscosity are weak functions of these parameters, or that there are negative feedbacks in the model that ultimately limit their influence on the ocean state.  Since under equivalent local ocean conditions the functional dependence of eddy diffusivity on $\phi_{s,unst}$ or eddy viscosity on $\phi_{m,unst}$ implies variation in their values by a factor of 2-3 over the range in our priors, we believe that the latter explanation is more likely (Large et al., 1994, see Eq B1).
The fact that the posterior distributions are in line with previous research is encouraging for the application of uncertainty quantification methods such as polynomial chaos to calibrate a model using observations. The question of whether data can be used to build more predictive models is still open.

\section{Summary and conclusions}
\label{sec:conc}
In this work, we presented a polynomial chaos-based method for the inference of five KPP parameter.
below, we remark on the PC construction process and the KPP inference results.

The method relied on building a surrogate model for a newly developed test statistic instead of the model output.
The advantage was  to avoid building surrogate for several model outputs and for different time scales.
The PC construction was implemented using two techniques. First a non-intrusive spectral projection method was used
that required a quadrature of level 5 nodes ($Q=903$ model runs). The computed PC expansion suffered from convergence issues
due to the presence of internal noise in the predictions. This resulted in over-fitting of the data with increasing level
of PC refinement. The resulting PC model was thus unsuitable and discarded.
The second technique used was BPDN that tolerates noise in the model and assumes sparsity in the PC expansion,
requiring a smaller number of model runs. This technique proved to be successful as it filters out the noise from the model
in finding the PC coefficient that lead to a faithful PC surrogate.
Several error metrics were computed to check the validity of the PC model.
%\revised{These confirmed that the BPDN-based PC surrogate model can be used
%in the MCMC sampling step to infer the KPP parameters.}

As a follow-up research, we are seeking to infer all nine KPP parameters in a calibration where the wind is treated as an adjustable parameter.
The additional number of parameters dramatically increases the number of required expensive model runs.
Thus a different approach is needed where we will investigate an adaptive approach to PC~\citep{winokur:2012}.

%%%%%%%%%%%%%%%%%%%%%%%%%%%%%%%%%%%%%%%%%%%%%%%%%%%%%%%%%%%%%%%%%%%%%
% ACKNOWLEDGMENTS
%%%%%%%%%%%%%%%%%%%%%%%%%%%%%%%%%%%%%%%%%%%%%%%%%%%%%%%%%%%%%%%%%%%%%
\acknowledgments
This research made use of the resources of the Supercomputing Laboratory 
and computer clusters at King Abdullah University of Science and Technology (KAUST) in Thuwal, Saudi Arabia.
SZ and OK are supported in part by the US Department of Energy, Office of Advance Scientific Computing Research, under Award number DE-SC0008789.

%%%%%%%%%%%%%%%%%%%%%%%%%%%%%%%%%%%%%%%%%%%%%%%%%%%%%%%%%%%%%%%%%%%%%
% APPENDIXES
%%%%%%%%%%%%%%%%%%%%%%%%%%%%%%%%%%%%%%%%%%%%%%%%%%%%%%%%%%%%%%%%%%%%%

%% If only one appendix, use
%%\appendix%

%% If more than one appendix, use \appendix[<letter>], e.g.,
% \appendix[A] 

%\appendixtitle{Title of Appendix}

%%%%%%%%%%%%%%%%%%%%%%%%%%%%%%%%%%%%%%%%%%%%%%%%%%%%%%%%%%%%%%%%%%%%
% REFERENCES
%%%%%%%%%%%%%%%%%%%%%%%%%%%%%%%%%%%%%%%%%%%%%%%%%%%%%%%%%%%%%%%%%%%%%
 \bibliographystyle{ametsoc2014}
 %\bibliography{references}

\clearpage

%%%%%%%%%%%%%%%%%%%%%%%%%%%%%%%%%%%%%%%%%%%%%%%%%%%%%%%%%%%%%%%%%%%%%
% TABLES
%%%%%%%%%%%%%%%%%%%%%%%%%%%%%%%%%%%%%%%%%%%%%%%%%%%%%%%%%%%%%%%%%%%%%
\begin{table}
\centering
\begin{tabular}[h]{|c|c|c|c|c|}
\hline
Parameter & Parameter Name & Symbol & Default Value & Uniform Prior $[a,b]$\\ \hline 
$p_1$     & Critical Bulk Richardson number & $Ri_c$ & $0.3$ & $[0.1,1.0]$  \\ \hline 
$p_2$     & Critical gradient Richardson number & $Ri_g$ & $0.7$ & $[0.1,1.0]$  \\ \hline 
$p_3$     & Structure function, unstable forcing, momentum & $\phi_{m,unst}$ & $16$ & $[3.60,331.06]$  \\ \hline 
$p_4$     & Structure function, unstable forcing, tracer & $\phi_{s,unst}$ & $16$ & $[7.77,67.02]$  \\ \hline 
$p_5$     & Nonlocal transport & $C^*$ & $10.0$ & $[5.0,15.0]$  \\ \hline
\end{tabular}
\caption{List of KPP parameters to be estimated using Bayesian inference along with their \MIT default values
and assumed uniform prior.}
\label{tab:param}
\end{table}
%%%%%%%%%%%%%%%%%%%%%%%%%%%%%%%%%%%%%%%%%%%%%%%%%%%%%%%%%%%%%%%%%%%%%%%%%%%%%
%%%%%%%%%%%%%%%%%%%%%%%%%%%%%%%%%%%%%%%%%%%%%%%%%%%%%%%%%%%%%%%%%%%%%%%%%%%%%
%%%%%%%%%%%%%%%%%%%%%%%%%%%%%%%%%%%%%%%%%%%%%%%%%%%%%%%%%%%%%%%%%%%%%%%%%%%%%
\begin{table}[ht]
\centering
\begin{tabular}[t]{|c|c|c|c|c|c|c|}
\hline
$\mu_{E}$ & $\sigma_{E}$ & $T_1$ &$T_2$ & $T_3$ & $T_4$ & $T_5$ \\ \hline 
325.62  & 30.97 & 0.6907 & 0.1450  & 0.2079 & 0.0472 & 0.0784 \\ \hline 
\end{tabular}
\caption{Reporting mean and variance of test statistic in addition to sensitivity index of each uncertain parameter.}
\label{tab:sens}
\end{table}
%%%%%%%%%%%%%%%%%%%%%%%%%%%%%%%%%%%%%%%%%%%%%%%%%%%%%%%%%%%%%%%%%%%%%
%\subsection{Figures}
%%%%%%%%%%%%%%%%%%%%%%%%%%%%%%%%%%%%%%%%%%%%%%%%%%%%%%%%%%%%%%%%%%%%%
\clearpage
%%%%%%%%%%%%%%%%%%%%%%%%%%%%%%%%%%%%%%%%%%%%%%%%%%%%%%%%%%%%%%%%%%%%%%%%%%%%%
%%%%%%%%%%%%%%%%%%%%%%%%%%%%%%%%%%%%%%%%%%%%%%%%%%%%%%%%%%%%%%%%%%%%%%%%%%%%%
%%%%%%%%%%%%%%%%%%%%%%%%%%%%%%%%%%%%%%%%%%%%%%%%%%%%%%%%%%%%%%%%%%%%%%%%%%%%%
%%%%%%%%%%%%%%%%%%%%%%%%%%%%%%%%%%%%%%%%%%%%%%%%%%%%%%%%%%%%%%%%%%%%%%%%%%%%%
%%%%%%%%%%%%%%%%%%%%%%%%%%%%%%%%%%%%%%%%%%%%%%%%%%%%%%%%%%%%%%%%%%%%%%%%%%%%%
%%%%%%%%%%%%%%%%%%%%%%%%%%%%%%%%%%%%%%%%%%%%%%%%%%%%%%%%%%%%%%%%%%%%%%%%%%%%%
\clearpage
%%%%%%%%%%%%%%%%%%%%%%%%%%%%%%%%%%%%%%%%%%%%%%%%%%%%%%%%%%%%%%%%%%%%%%%%%%%%%
%%%%%%%%%%%%%%%%%%%%%%%%%%%%%%%%%%%%%%%%%%%%%%%%%%%%%%%%%%%%%%%%%%%%%%%%%%%%%
%%%%%%%%%%%%%%%%%%%%%%%%%%%%%%%%%%%%%%%%%%%%%%%%%%%%%%%%%%%%%%%%%%%%%%%%%%%%%
\clearpage
%%%%%%%%%%%%%%%%%%%%%%%%%%%%%%%%%%%%%%%%%%%%%%%%%%%%%%%%%%%%%%%%%%%%%%%%%%%%%
%%%%%%%%%%%%%%%%%%%%%%%%%%%%%%%%%%%%%%%%%%%%%%%%%%%%%%%%%%%%%%%%%%%%%%%%%%%%%
%%%%%%%%%%%%%%%%%%%%%%%%%%%%%%%%%%%%%%%%%%%%%%%%%%%%%%%%%%%%%%%%%%%%%%%%%%%%%
%%%%%%%%%%%%%%%%%%%%%%%%%%%%%%%%%%%%%%%%%%%%%%%%%%%%%%%%%%%%%%%%%%%%%%%%%%%%%
%%%%%%%%%%%%%%%%%%%%%%%%%%%%%%%%%%%%%%%%%%%%%%%%%%%%%%%%%%%%%%%%%%%%%%%%%%%%%
%%%%%%%%%%%%%%%%%%%%%%%%%%%%%%%%%%%%%%%%%%%%%%%%%%%%%%%%%%%%%%%%%%%%%%%%%%%%%
%%%%%%%%%%%%%%%%%%%%%%%%%%%%%%%%%%%%%%%%%%%%%%%%%%%%%%%%%%%%%%%%%%%%%%%%%%%%%
%%%%%%%%%%%%%%%%%%%%%%%%%%%%%%%%%%%%%%%%%%%%%%%%%%%%%%%%%%%%%%%%%%%%%%%%%%%%%
%%%%%%%%%%%%%%%%%%%%%%%%%%%%%%%%%%%%%%%%%%%%%%%%%%%%%%%%%%%%%%%%%%%%%%%%%%%%%
%%%%%%%%%%%%%%%%%%%%%%%%%%%%%%%%%%%%%%%%%%%%%%%%%%%%%%%%%%%%%%%%%%%%%%%%%%%%%
%%%%%%%%%%%%%%%%%%%%%%%%%%%%%%%%%%%%%%%%%%%%%%%%%%%%%%%%%%%%%%%%%%%%%%%%%%%%%
%%%%%%%%%%%%%%%%%%%%%%%%%%%%%%%%%%%%%%%%%%%%%%%%%%%%%%%%%%%%%%%%%%%%%%%%%%%%%
%%%%%%%%%%%%%%%%%%%%%%%%%%%%%%%%%%%%%%%%%%%%%%%%%%%%%%%%%%%%%%%%%%%%%%%%%%%%%
%%%%%%%%%%%%%%%%%%%%%%%%%%%%%%%%%%%%%%%%%%%%%%%%%%%%%%%%%%%%%%%%%%%%%%%%%%%%%
%%%%%%%%%%%%%%%%%%%%%%%%%%%%%%%%%%%%%%%%%%%%%%%%%%%%%%%%%%%%%%%%%%%%%%%%%%%%%
\clearpage
%%%%%%%%%%%%%%%%%%%%%%%%%%%%%%%%%%%%%%%%%%%%%%%%%%%%%%%%%%%%%%%%%%%%%%%%%%%%%
%%%%%%%%%%%%%%%%%%%%%%%%%%%%%%%%%%%%%%%%%%%%%%%%%%%%%%%%%%%%%%%%%%%%%%%%%%%%%
%%%%%%%%%%%%%%%%%%%%%%%%%%%%%%%%%%%%%%%%%%%%%%%%%%%%%%%%%%%%%%%%%%%%%%%%%%%%%
\clearpage
%%%%%%%%%%%%%%%%%%%%%%%%%%%%%%%%%%%%%%%%%%%%%%%%%%%%%%%%%%%%%%%%%%%%%%%%%%%%%
%%%%%%%%%%%%%%%%%%%%%%%%%%%%%%%%%%%%%%%%%%%%%%%%%%%%%%%%%%%%%%%%%%%%%%%%%%%%%
%%%%%%%%%%%%%%%%%%%%%%%%%%%%%%%%%%%%%%%%%%%%%%%%%%%%%%%%%%%%%%%%%%%%%%%%%%%%%
\clearpage
\clearpage

%%%%%%%%%%%%%%%%%%%%%%%%%%%%%%%%%%%%%%%%%%%%%%%%%%%%%%%%%%%%%%%%%%%%%%%%%%%%%
%%%%%%%%%%%%%%%%%%%%%%%%%%%%%%%%%%%%%%%%%%%%%%%%%%%%%%%%%%%%%%%%%%%%%%%%%%%%%
%%%%%%%%%%%%%%%%%%%%%%%%%%%%%%%%%%%%%%%%%%%%%%%%%%%%%%%%%%%%%%%%%%%%%%%%%%%%%
%%%%%%%%%%%%%%%%%%%%%%%%%%%%%%%%%%%%%%%%%%%%%%%%%%%%%%%%%%%%%%%%%%%%%%%%%%%%%
%%%%%%%%%%%%%%%%%%%%%%%%%%%%%%%%%%%%%%%%%%%%%%%%%%%%%%%%%%%%%%%%%%%%%%%%%%%%%
%%%%%%%%%%%%%%%%%%%%%%%%%%%%%%%%%%%%%%%%%%%%%%%%%%%%%%%%%%%%%%%%%%%%%%%%%%%%%
%%%%%%%%%%%%%%%%%%%%%%%%%%%%%%%%%%%%%%%%%%%%%%%%%%%%%%%%%%%%%%%%%%%%%%%%%%%%%
%%%%%%%%%%%%%%%%%%%%%%%%%%%%%%%%%%%%%%%%%%%%%%%%%%%%%%%%%%%%%%%%%%%%%%%%%%%%%
%%%%%%%%%%%%%%%%%%%%%%%%%%%%%%%%%%%%%%%%%%%%%%%%%%%%%%%%%%%%%%%%%%%%%%%%%%%%%
%%%%%%%%%%%%%%%%%%%%%%%%%%%%%%%%%%%%%%%%%%%%%%%%%%%%%%%%%%%%%%%%%%%%%%%%%%%%%
%%%%%%%%%%%%%%%%%%%%%%%%%%%%%%%%%%%%%%%%%%%%%%%%%%%%%%%%%%%%%%%%%%%%%%%%%%%%%
%%%%%%%%%%%%%%%%%%%%%%%%%%%%%%%%%%%%%%%%%%%%%%%%%%%%%%%%%%%%%%%%%%%%%%%%%%%%%
%%%%%%%%%%%%%%%%%%%%%%%%%%%%%%%%%%%%%%%%%%%%%%%%%%%%%%%%%%%%%%%%%%%%%%%%%%%%%
%%%%%%%%%%%%%%%%%%%%%%%%%%%%%%%%%%%%%%%%%%%%%%%%%%%%%%%%%%%%%%%%%%%%%%%%%%%%%
%%%%%%%%%%%%%%%%%%%%%%%%%%%%%%%%%%%%%%%%%%%%%%%%%%%%%%%%%%%%%%%%%%%%%%%%%%%%%
\clearpage
%%%%%%%%%%%%%%%%%%%%%%%%%%%%%%%%%%%%%%%%%%%%%%%%%%%%%%%%%%%%%%%%%%%%%%%%%%%%%
%%%%%%%%%%%%%%%%%%%%%%%%%%%%%%%%%%%%%%%%%%%%%%%%%%%%%%%%%%%%%%%%%%%%%%%%%%%%%
%%%%%%%%%%%%%%%%%%%%%%%%%%%%%%%%%%%%%%%%%%%%%%%%%%%%%%%%%%%%%%%%%%%%%%%%%%%%%
\clearpage

%%%%%%%%%%%%%%%%%%%%%%%%%%%%%%%%%%%%%%%%%%%%%%%%%%%%%%%%%%%%%%%%%%%%%%%%%%%%%
%%%%%%%%%%%%%%%%%%%%%%%%%%%%%%%%%%%%%%%%%%%%%%%%%%%%%%%%%%%%%%%%%%%%%%%%%%%%%
%%%%%%%%%%%%%%%%%%%%%%%%%%%%%%%%%%%%%%%%%%%%%%%%%%%%%%%%%%%%%%%%%%%%%%%%%%%%%
\end{document}